\documentclass[pdflatex]{article}[14pt]
\usepackage{amssymb,amsmath}
\usepackage{slashed}
\usepackage{hyperref}
\usepackage{cite}
\usepackage{graphicx}
\usepackage[title]{appendix}
\usepackage{ulem}
\numberwithin{equation}{section}
\usepackage[top=25truemm,bottom=25truemm,left=10truemm,right=10truemm]{geometry}
\begin{document}
\title{
	\vbox{
		\baselineskip 14pt
		\hfill \hbox{\normalsize KUNS-2727
	}} \vskip 1cm
	\bf \Large Universality of soft theorem from locality of soft vertex operators
	\vskip 0.5cm
}
\author{
	Sho Higuchi\thanks{E-mail: \tt sho.h@gauge.scphys.kyoto-u.ac.jp}~ and
	Hikaru~Kawai\thanks{E-mail: \tt hkawai@gauge.scphys.kyoto-u.ac.jp}
	\bigskip\\
	\it \normalsize
	Department of Physics, Kyoto University, Kyoto 606-8502, Japan\\
	\smallskip
}
\date{}

\setcounter{page}{0}
\maketitle\thispagestyle{empty}
\abstract{\normalsize
	The universal behavior of the soft theorem at the tree level is explained by considering the operator product expansion of the soft and hard vertex operators.
	We find that the world-sheet integral for the soft vertex is determined only by the regions that are close to the hard vertices after eliminating total derivative terms.  
	This analyses can be applied to massless particles in various theories such as bosonic closed string, closed superstring and heterotic string.
}

\section{Introduction}	
\begin{figure}[h] \label{fig:diagram}
	\centering
	\includegraphics[width=15cm,bb=0 0 940 241]{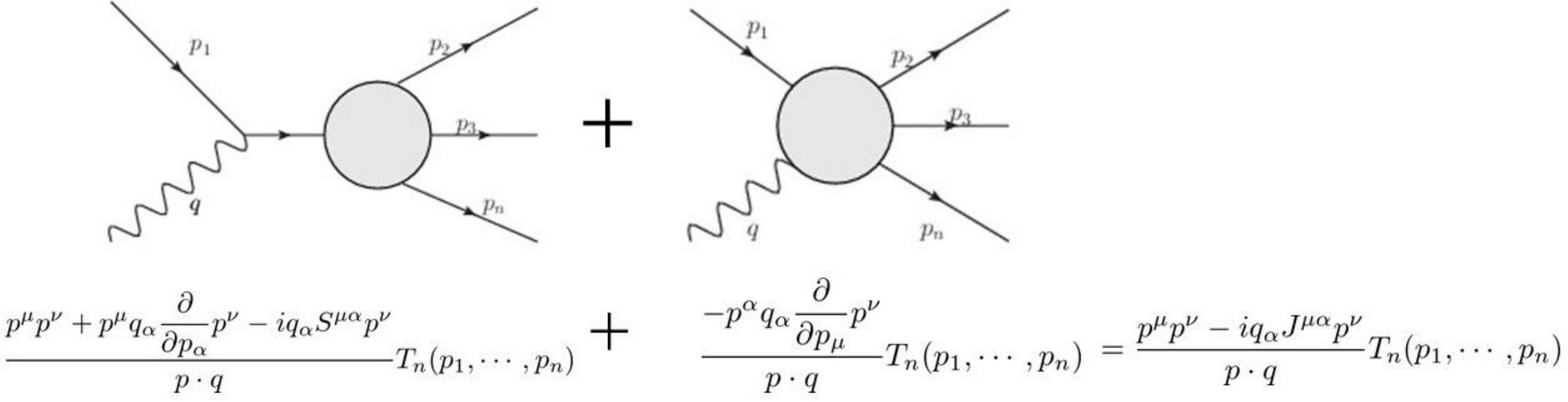}
	\caption{Soft graviton theorem in terms of Feynman diagrams through subleading order}
\end{figure}

$ \qquad $ In recent years much progress has been made in understanding the origin of the universality of the soft theorems \cite{White:2011yy}-\cite{Broedel:2014bza}.
For example, the universal behavior of soft graviton is given by
\begin{align}
M_{n+1}(q;p_{1},\cdots,p_{n})=\left[S^{(0)}+S^{(1)}+S^{(2)}\right]M_{n}(p_{1},\cdots,p_{n}),
\end{align}
where
\begin{align}
&S^{(0)}\equiv \sum_{k=1}^{n}\dfrac{h_{\mu \nu}p^{\mu}_{k}p^{\nu}_{k}}{p_{k}\cdot q}, \nonumber\\
&S^{(1)}\equiv -i\sum_{k=1}^{n}\dfrac{h_{\mu \nu}p^{\mu}_{k}q_{\alpha}J^{\nu \alpha}_{k}}{p_{k}\cdot q}, \nonumber\\
&S^{(2)}\equiv -\dfrac{1}{2}\sum_{k=1}^{n}\dfrac{h_{\mu \nu}q_{\alpha}J^{\mu \alpha}_{k}q_{\beta}J^{\nu \beta}_{k}}{p_{k}\cdot q},\nonumber\\ 
& J^{\mu \alpha}_{k} \equiv L^{\mu \alpha}_{k}+S^{\mu \alpha}_{k}, \nonumber\\
&L^{\mu \alpha}_{k} \equiv i\left(p^{\mu}_{k}\dfrac{\partial}{\partial p_{k\alpha}}-p^{\alpha}_{k}\dfrac{\partial}{\partial p_{k\mu}}\right).
\end{align}
Here $ J_{k}^{\mu\alpha} $,  $ L_{k}^{\mu\alpha} $ and $ S^{\mu\alpha}_{k} $  are the total, orbital and spin angular momenta of the k-th particle respectively.
From the viewpoint of field theory, the soft theorems are beautifully derived by using the Ward identity\cite{Bern:2014vva}, although it is not clear why the total angular momentum comes out from the Feynman diagrams when a soft particle is added (Fig.\ref{fig:diagram}).

The soft theorems can also be understood in string theory \cite{Bianchi:2014gla}-\cite{Laddha:2018myi}. The scattering amplitude is expressed by the insertions of the vertex operators on the world-sheet that correspond to in and out states. The soft theorems are obtained by considering the operator product expansion(OPE) of the soft vertex operator with the hard ones. In this paper we give a simple explanation for the universality of the soft theorems in string theory. For a while we focus on the tree amplitudes of bosonic string. 

\begin{figure}[h]\label{superlocal}
	\centering
	\includegraphics[width=15cm,bb=0 0 951 240]{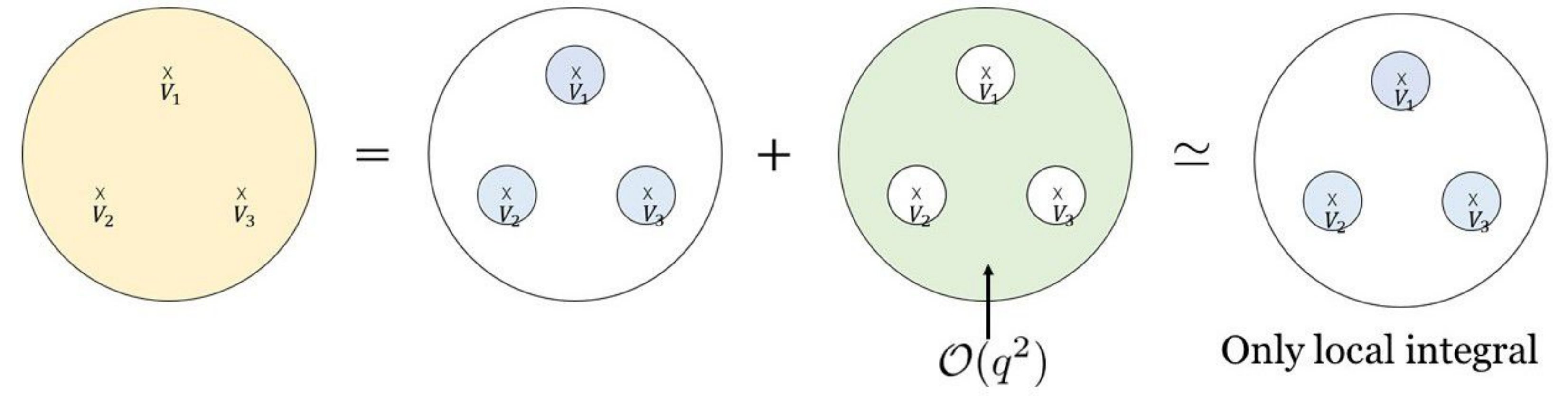}
	\caption{Contributions from $ \frac{1}{2}:X^{\mu}(z,\bar{z})X^{\nu}(z,\bar{z})\partial_{z}\bar{\partial_{z}} \exp\left(iq\cdot X(z,\bar{z})\right) $}
\end{figure}

The soft massless vertex operator in bosonic closed string is given by
\begin{align}\label{math;massless}
V_{soft}(z,\bar{z})=h_{\mu\nu}:\partial X^{\mu}(z)\bar{\partial}X^{\nu}\exp\left(iq\cdot X(z,\bar{z})\right):,
\end{align}
where q is the momentum of the soft particle. As we will see in Section \ref{local}, by dropping the surface terms at the infinity, eq.(\ref{math;massless}) can be replaced by
\begin{align}
\left \{ 
\begin{aligned}
&V_{s}\equiv\frac{1}{2}h_{\mu\nu}:X^{\mu}(z,\bar{z})X^{\nu}(z,\bar{z})\partial_{z}\bar{\partial_{z}} \exp\left(iq\cdot X(z,\bar{z})\right) \ \text{for symmetric} \  \mathrm{h}_{\mu\nu} \\
&V_{a}\equiv-h_{\mu\nu}:X^{\mu}(z,\bar{z})\bar{\partial}X^{\nu}(\bar{z})\partial_{z} \exp\left(iq\cdot X(z,\bar{z})\right) \   \ \text{for antisymmetric} \ \mathrm{h}_{\mu\nu} \end{aligned}
\right. .\label{eq.softvertex}
\end{align}
For the soft dilaton or graviton, where $ h_{\mu\nu}$ is symmetric, this operator is superlocal through the linear order in q, while for the B field, where $ h_{\mu\nu}$ is antisymmetric, through the 0-th order.  A superlocal operator is highly local in the sense that it takes a nonzero value only when its position coincides with the other operators' positions (see Section \ref{superlocal}).
In this paper we will show that the universality of the soft theorems is a direct consequence of this superlocality. 
We can apply the same analysis for superstring and heterotic string theory.\\

The structure of this paper is as follows. In Section \ref{convention} we review the calculation of the scattering amplitudes in string theory and see how the leading soft theorem arises from the OPE.
In Section \ref{local} we introduce the concept of the superlocal operator and see that the soft graviton/dilaton vertex operator is  superlocal through the linear order in q. In section \ref{sec:graviton,dilaton} and \ref{sec;bfield} we give a unified explanation for the universality of the soft theorems for graviton, dilaton and B field.
In Section \ref{superstring} we apply this idea for superstring and heterotic string theory.
The details of the calculations are given in the Appendixes.

\section{Soft graviton/dilaton theorems from OPE}\label{convention}
\qquad In this section we review the calculation of the scattering amplitudes in string theory and explain how to derive the leading soft graviton or dilaton theorem by using the OPE. 

The tree level amplitudes are represented as the insertions of the vertex operators on a complex plane.
\begin{align}\label{scattering}
	M_{N+1}(p_{1},\cdots,p_{N},q)
	\sim 
	\int d^2z \int d^2w_{i}
	\langle V_{soft}(q,z)\prod_{i=1}^{N}V_{i}(p_{i},w_{i}) \rangle .
\end{align}
Here we use the following normalization:
\begin{align}
X^{\mu}(z,\bar{z})X^{\nu}(w,\bar{w})\sim -\dfrac{\alpha'}{2}\eta^{\mu\nu}\ln|z-w|^2 .
\end{align}
For convenience we define a disk $ D_{i} $ of radius $ \epsilon $ around each vertex $w_{i}$, and denote the rest bulk region by B, $ \mathrm{B}=\mathbb{C}-\bigcup_{i=1}^{n}D_{i} $.  We evaluate the integration over each of these regions.

First we calculate the contribution from the disk $ D_{i}$ by using the OPE
\begin{align}
:V_{soft}(q,z)::V_{i}(p,w):
=&
\cdots 
+|z-w|^{\alpha'p\cdot q-4}\times \cdots\nonumber\\ 
&+|z-w|^{\alpha'p\cdot q-2}\left  [\left(-\dfrac{i\alpha'}{2}\right)^2p^{\mu}p^{\nu}+\mathcal{O}(q)\right ]:V_{i}(p,w):\\
&+|z-w|^{\alpha'p\cdot q}\times \cdots \nonumber\\ 
&+\cdots .\nonumber
\end{align}
We then perform the z integration
\begin{align}
\int_{|z|<\epsilon} d^2z |z-w|^{\alpha'q\cdot p+m-2}
=\dfrac{2\pi}{\alpha'p\cdot q+m}\epsilon^{\alpha'p\cdot q+m}.
\end{align}
If we pick up the most singular terms $ (m=0)  $ for q, the leading soft graviton theorem is reproduced:
\begin{align}
\int d^2z :V_{soft}(q,z)::V_{i}(p,w):= -\dfrac{\pi\alpha'p^{\mu}p^{\nu}}{2p\cdot q}:V_{i}(p,w):+\mathcal{O}(1) .
\end{align}

\begin{figure}[h]\label{analogy}
	\centering
	\includegraphics[width=15cm,bb=0 0 960 540]{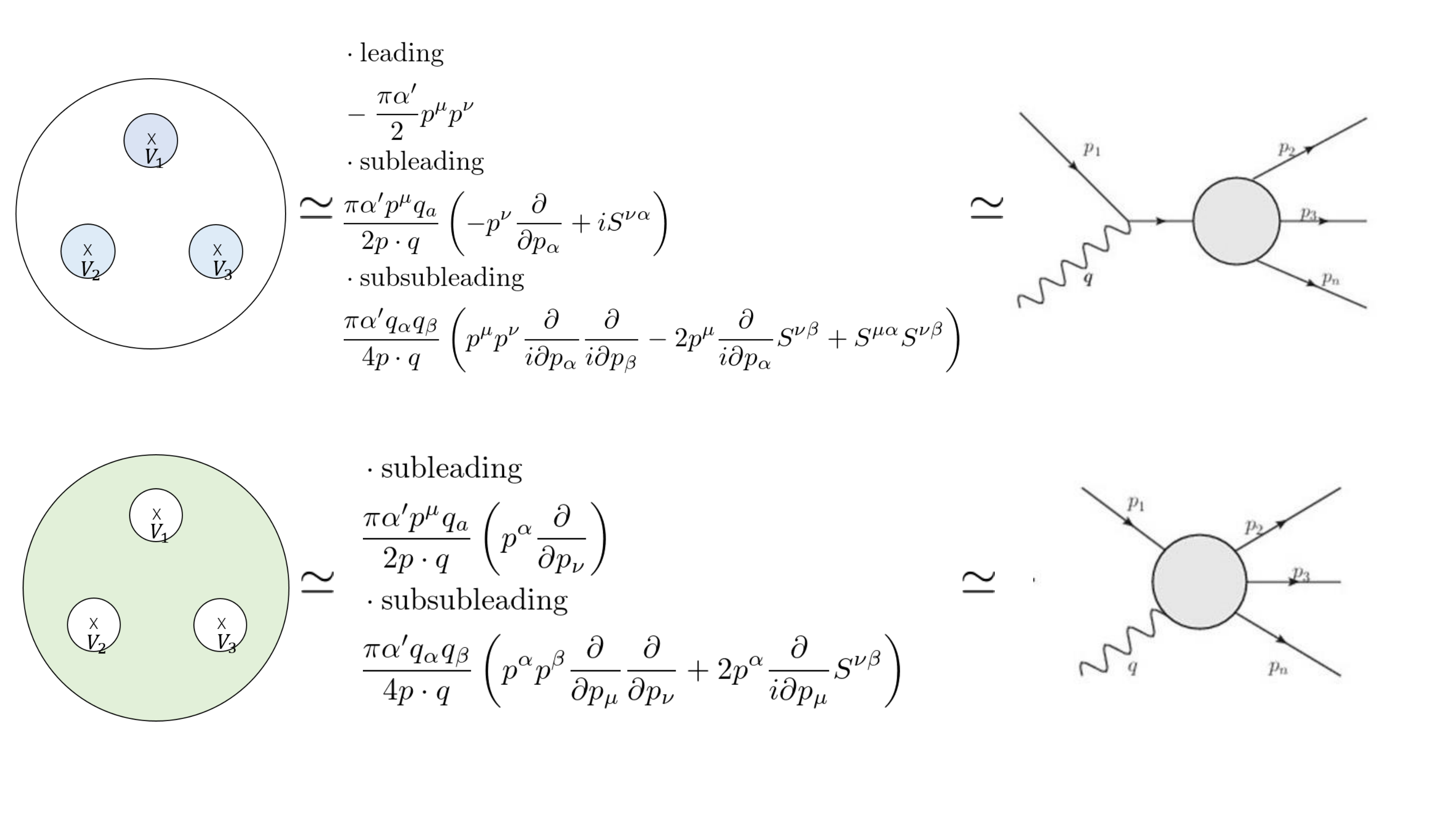}
	\caption{Analogy with field theory}
\end{figure}

We also evaluate the contributions from the bulk B by some partial integration. The subleading and subsubleading soft graviton or dilaton theorems are reproduced from the disks $ D_{i} $ and bulk B, if we ignore the higher order corrections in $\alpha'$ and mixing with different vertex operators. The detail of the calculation for massless hard particles is given in Appendix \ref{simple}. The structure is the same as in field theory. The singular terms in q arise from diagrams with a soft particle attached to the external lines, while the regular terms arise from the the internal lines(See Figure \ref{analogy}.).

In the above calculation there are three problems.
First, we cannot control the higher order corrections in $ \alpha' $ because $ \epsilon $ dependence remains in the expansion of $ \epsilon^{\alpha'p\cdot q} $.  We will calculate these corrections by a better method in the next section. 
Second, we cannot determine through which order the universality of the soft theorems holds. This also becomes clear in the next section.
Third, it seems that the hard particles are mixed to other levels. For example, we assume that the hard particle is massless, $ V_{i}=h_{\rho\sigma}\partial X^{\rho}\bar{\partial}X^{\sigma}\mathrm{e}^{ip\cdot X} $. Then from the contractions between the exponential functions in the soft and hard part, we would have the following term:
\begin{align}
\int_{|z|<\epsilon} d^2z V_{soft}(q,z)V_{i}(p,w)
&\sim \int_{|z|<\epsilon}d^2z |z-w|^{\alpha'p\cdot q}:\partial X^{\mu}(w)\partial X^{\rho}(w)\bar{\partial}X^{\nu}(\bar{w})\bar{\partial}X^{\sigma}(\bar{w})\exp(i(p+q)\cdot X)(w): \nonumber\\
&\sim \dfrac{2\pi}{\alpha'p\cdot q+2}\epsilon^{\alpha'p\cdot q} \partial X^{\mu}(w)\bar{\partial}X^{\nu}(\bar{w})\partial X^{\rho}(w)\bar{\partial}X^{\sigma}\exp(ip\cdot X(w,\bar{w})).
\end{align}
We will show that such term does not appear in the next section. It should be canceled if we correctly evaluate the integration over the bulk region B.

\section{Superlocal operator and universality of soft theorems}\label{local}
\qquad We introduce the concept of superlocal operator in this section.
A superlocal operator is highly local in the sense that it takes a nonzero value only when its position coincides with the other operators. More precisely, an operator $ O(z) $ is said to be superlocal when the following equation holds for any operators $ \phi_{i}(w_{i}) $:
\begin{align}
\int d^2z \langle O(z)\phi_{1}(w_{1})\cdots \phi_{n}(w_{n})\rangle= \sum_{a}\sum_{i=1}^{n}c_{a}^{i}\langle O_{a}^{i}(w_{i})\phi_{1}(w_{1})\cdots \check{\phi_{i}}(w_{i})\cdots \phi_{n}(w_{n})\rangle.
\end{align}
where $ c_ {a} ^ {i} $ is a constant, $ O_{a}^{i} $ are local operators and $ \check{\phi_{i}} (w_{i}) $ means that i-th operator $ \phi_{i} $ is removed. 

The simplest example of superlocal operator is $ \partial \bar {\partial} X^{\mu} (z) $:\\
\begin{align}
\int d^2z \langle \partial \bar{\partial}X^{\mu}(z)X^{\nu_{1}}(w_{1})\cdots X^{\nu_{n}}(w_{n})\rangle =-\pi\alpha'\sum_{i=1}^{n}\eta^{\mu\nu_{i}}\langle X^{\nu_{1}}(w_{1})\cdots \check{X}^{\nu_{i}}(w_{i})\cdots X^{\nu_{n}}(w_{n})\rangle ,
\end{align}
which is nothing but the Schwinger-Dyson equation.

As we discuss now, the universality of soft theorem is a direct consequence of the fact that the soft vertex is superlocal through the linear order in q. We rewrite the emission vertex in eq.(\ref{math;massless}) for the symmetric $ h_{\mu\nu} $:
\begin{align}
&h_{\mu\nu}:\partial X^{\mu}(z)\bar{\partial}X^{\nu}\exp\left(iq\cdot X(z,\bar{z})\right):\nonumber\\
&=h_{\mu\nu}\partial:X^{\mu}(z,\bar{z})\bar{\partial}X^{\nu}(\bar{z})\exp\left(iq\cdot X(z,\bar{z})\right):
-h_{\mu\nu}:X^{\mu}(z,\bar{z})\bar{\partial}X^{\nu}(z,\bar{z})\partial_{z} \exp\left(iq\cdot X(z,\bar{z})\right):\nonumber\\
&=h_{\mu\nu}\partial:X^{\mu}(z,\bar{z})\bar{\partial}X^{\nu}(\bar{z})\exp\left(iq\cdot X(z,\bar{z})\right):
-h_{\mu\nu}\frac{\bar{\partial}}{2}:X^{\mu}(z,\bar{z})X^{\nu}(z,\bar{z})\partial_{z} \exp\left(iq\cdot X(z,\bar{z})\right):\nonumber\\
&+\frac{h_{\mu\nu}}{2}:X^{\mu}(z,\bar{z})X^{\nu}(z,\bar{z})\partial_{z}\bar{\partial_{z}} \exp\left(iq\cdot X(z,\bar{z})\right):\label{symmetric}.
\end{align}
Here in the last equation we have used the symmetry of $ h_{\mu\nu} $. Because the world-sheet is a complex plane for the tree level amplitudes, we can drop the total derivatives, and we obtain
\begin{align}
\int d^2z \langle:\partial X^{\mu}(z)\bar{\partial}X^{\nu}\exp\left(iq\cdot X(z,\bar{z})\right):\cdots\rangle
=\int d^2z \langle\frac{1}{2}:X^{\mu}(z,\bar{z})X^{\nu}(z,\bar{z})\partial_{z}\bar{\partial_{z}} \exp\left(iq\cdot X(z,\bar{z})\right):\cdots\rangle.
\end{align}
As in the previous section, we divide the complex plane into the disks $ D_{i} $ and the bulk region B.

In the bulk B, because there is no singularity in $ z-w $, we can simply expand the soft exponential $ \mathrm{e}^{iq\cdot X} $. By using the equation of motion we find that the soft vertex operator is 0 through the linear order in q on B:
\begin{align}
\partial \bar{\partial}\exp(iq\cdot X(z))
=\partial \bar{\partial}\left (1+iq\cdot X(z)+\mathcal{O}(q^2)\right )
=\mathcal{O}(q^2).
\end{align}
 Because there is no contribution from B for any radius $ \epsilon $, we can conclude that the soft vertex operator is superlocal through the linear order in q.

 In the disk $ D_{i} $, however, we cannot simply expand $ \mathrm{e}^{iq\cdot X} $ because of the singularity in $ z-w $. We should consider the contraction of the exponential functions $ \mathrm{e}^{iq\cdot X} $ and $ \mathrm{e}^{ip\cdot X} $ before expanding in q:
\begin{align}
&\int_{|z|<\epsilon} \langle\frac{1}{2}:X^{\mu}(z,\bar{z})X^{\nu}(z,\bar{z})\partial_{z}\bar{\partial_{z}} \exp\left(iq\cdot X(z,\bar{z})\right)::\partial X^{\rho}(w)\exp(ip\cdot X(w,\bar{w})\cdots):\rangle\nonumber\\
=&\int_{|z|<\epsilon}  \langle\frac{1}{2}:X^{\mu}(z,\bar{z})X^{\nu}(z,\bar{z}):\partial_{z}\bar{\partial_{z}}\left ( |z-w|^{\alpha'p\cdot q}\partial X^{\rho}(w):\exp(iq\cdot X(z,\bar{z})+ip\cdot X(w,\bar{w})):\cdots \right )\rangle,
\end{align}
where operators at the same position, z or w, must not be contracted. Then if we expand $ \mathrm{e}^{iq\cdot X} $ with respect to q, we can get the leading, subleading, subsubleading soft theorem. The details are given in the following sections. 

\section{Soft graviton/dilaton theorem }\label{sec:graviton,dilaton}
\subsection{General formula}
\qquad In this section we consider a hard vertex operator of the form, $V_{hard}=\partial^{m_{1}}X\cdots \bar{\partial}^{n_{1}}X\cdots \exp(ip\cdot X)$, and we refer to the factors before the exponential function, $ \partial^{m_{1}}X\cdots \bar{\partial}^{n_{1}}X\cdots $, as prefactors. Then we can evaluate the OPE of the soft vertex operator $ V_{s} $ (eq.(\ref{eq.softvertex})) and $ V_{hard} $ as follows. 

First in order to simplify the contractions we rewrite the soft and hard vertex operators as
\begin{align}
&
V_{s}=\frac{h_{\mu\nu}}{2}:X^{\mu}(z,\bar{z})X^{\nu}(z,\bar{z})\partial_{z}\bar{\partial_{z}} \exp\left(iq\cdot X(z,\bar{z})\right):
=-\dfrac{h_{\mu\nu}}{2}\lim_{\xi \to 0}
\partial^{\mu}_{\xi}\partial^{\nu}_{\xi}\lim_{z'\to z}\partial_{z}\bar{\partial}_{z}
:\exp(iq\cdot X(z,\bar{z})+i\xi\cdot X(z',\bar{z'}):,\label{softvertex}\\
&\begin{aligned}
V_{hard}&=A_{\rho_{1}\cdots \sigma_{1}\cdots }\partial^{m_{1}}X^{\rho_{1}}\cdots\bar{\partial}^{n_{1}}X^{\sigma_{1}}\cdots \exp(ip\cdot X(w))\nonumber\\
&\left .=A_{\rho_{1}\cdots \sigma_{1}\cdots }\exp\left(ip\cdot X(w)+\sum_{i}i\zeta_{i}\cdot \partial^{m_{i}}X+\sum_{i}i\lambda_{i}\cdot \bar{\partial}^{n_{i}}X(\bar{w}) \right)\right|_{\text{multilinear in }  i\zeta_{i},i\lambda_{i} }.
\end{aligned}
\end{align}
Here we keep only the multilinear terms in $ i\zeta_{i}$ and $i\lambda_{i} $, and perform the replacement, $ i\zeta_{1\rho_{1}}\cdots i\lambda_{1\sigma_{1}}\cdots \rightarrow A_{\rho_{1}\cdots \sigma_{1}\cdots }$, at the end.

The OPE of these operators is given by
\begin{align}
&:V_{s}(z)::V_{hard}(w):\nonumber\\
=&\lim_{\xi \to 0}
\partial^{\mu}_{\xi}\partial^{\nu}_{\xi}\lim_{z'\to z}\left .-\dfrac{1}{2}h_{\mu\nu}\partial_{z}\bar{\partial}_{z}:\exp(iq\cdot X(z)+i\xi\cdot X(z')):
:\exp\left (ip\cdot X(w)+i\sum_{i}\zeta_{i}\cdot \partial_{w}^{m_{i}}X(w)+i\sum_{i}\lambda_{i}\cdot \bar{\partial}^{n_{i}}X(\bar{w})\right ):\right|_{\text{multilinear in }  i\zeta_{i},i\lambda_{i} }\nonumber\\
=&-\lim_{\xi \to 0}
\partial^{\mu}_{\xi}\partial^{\nu}_{\xi}\lim_{z'\to z}\dfrac{1}{2}h_{\mu\nu}\partial_{z}\bar{\partial}_{z}
: |z-w|^{\alpha'p\cdot q}|z'-w|^{\alpha'p\cdot \xi}\nonumber\\
&\exp\left(-\dfrac{\alpha'}{2}\sum_{i}\dfrac{(m_{i}-1)!q\cdot \zeta_{i}}{(z-w)^{m_{i}}} -\dfrac{\alpha'}{2}\sum_{i}\dfrac{(n_{i}-1)!q\cdot \lambda_{i}}{(\bar{z}-\bar{w})^{n_{i}}} \right) \nonumber\\
\times&\exp\left(-\dfrac{\alpha'}{2}\sum_{i}\dfrac{(m_{i}-1)!\xi\cdot \zeta_{i}}{(z'-w)^{m_{i}}} -\dfrac{\alpha'}{2}\sum_{i}\dfrac{(n_{i}-1)!\xi\cdot \lambda_{i}}{(\bar{z'}-\bar{w})^{n_{i}}} \right) \nonumber\\
\times&\left .:\exp\left(iq\cdot X(z)+i\xi\cdot X(z')+ip\cdot X(w)+i\sum_{i}\zeta_{i}\cdot
\partial_{w}^{m_{i}}X(w)+i\sum_{i}\lambda_{i}\cdot \bar{\partial}^{n_{i}}X(\bar{w})\right):\right|_{\text{multilinear in }  i\zeta_{i},i\lambda_{i} }\nonumber\\
=&-\lim_{\xi \to 0}
\partial^{\mu}_{\xi}\partial^{\nu}_{\xi}\dfrac{1}{2}h_{\mu\nu}
: |z-w|^{\alpha'p\cdot (q+\xi)}\nonumber\\
\times&\exp\left(-\dfrac{\alpha'}{2}\sum_{i}\dfrac{(m_{i}-1)!(q+\xi)\cdot \zeta_{i}}{(z-w)^{m_{i}}} -\dfrac{\alpha'}{2}\sum_{i}\dfrac{(n_{i}-1)!(q+\xi)\cdot \lambda_{i}}{(\bar{z}-\bar{w})^{n_{i}}} \right) \nonumber\\
\times&:\exp\left(i(q+\xi)\cdot X(z)+ip\cdot X(w)+i\sum_{i}\zeta_{i}\cdot
\partial_{w}^{m_{i}}X(w)+i\sum_{i}\lambda_{i}\cdot \bar{\partial}^{n_{i}}X(\bar{w})\right)\nonumber\\
\times&\left .\left(\dfrac{\alpha'p\cdot q}{2(z-w)}+\sum_{i}\dfrac{\alpha'm_{i}!q\cdot \zeta_{i}}{2(z-w)^{m_{i}+1}}+iq\cdot \partial X(z)  \right)\left(\dfrac{\alpha'p\cdot q}{2(\bar{z}-\bar{w})}+\sum_{i}\dfrac{\alpha'n_{i}!q\cdot \lambda_{i}}{2(\bar{z}-\bar{w})^{n_{i}+1}}+iq\cdot \bar{\partial} X(\bar{z})  \right):\right|_{\text{multilinear in }  i\zeta_{i},i\lambda_{i} }.
\label{eq.before}
\end{align}
The following two facts are crucial.
\begin{itemize}
	\item First, we focus on the powers of $ z-w $. Because the last line in eq.(\ref{eq.before}) is quadratic in q, the z integration needs to yield a singular behavior in q in order to obtain a nonzero result through the linear order in q. Therefore it is sufficient to consider the coefficients of  $ |z-w|^{\alpha'p\cdot (q+\xi)-2} $.
	\item The singular factor that emerges after the z-integration of $ |z-w|^{\alpha'p\cdot (q+\xi)-2} $ is expanded as
	\begin{align}
	\dfrac{1}{p\cdot(q+\xi)}=\dfrac{1}{p\cdot q}\left(1-\dfrac{p\cdot\xi}{p\cdot q}+\left(\dfrac{p\cdot\xi}{p\cdot q}\right)^2+\cdots    \right).\label{eq.expand}
	\end{align}
	We are interested in quadratic terms in $ \xi $, and the last line of eq.(\ref{eq.before}) is quadratic in q. Therefore the leading, subleading and subsubleading terms come from the 0-th, first and second order of the expansion of the exponential functions with respect to $ (q+\xi) $, respectively.
\end{itemize}
In the following we label the degrees of $ (q+\xi) $ in each expansion of $ \exp\left(-\dfrac{\alpha'}{2}\sum_{i}\dfrac{(m_{i}-1)!(q+\xi)\cdot \zeta_{i}}{(z-w)^{m_{i}}} \right) $, \\$ \exp\left( -\dfrac{\alpha'}{2}\sum_{i}\dfrac{(n_{i}-1)!(q+\xi)\cdot \lambda_{i}}{(\bar{z}-\bar{w})^{n_{i}}} \right) $ and $ \exp(i(q+\xi)\cdot X(z) $, as (a,b;c), for example (0,0;1).\\

\paragraph{The leading soft theorem}\quad

 We keep the 0-th order in $q+\xi$ in eq.(\ref{eq.before}). We have only one contribution.\\
	\underline{(0,0;0) terms}
	\begin{align}
	&-\lim_{\xi \to 0}\dfrac{1}{2}h_{\mu\nu}\partial_{\xi}^{\mu}\partial_{\xi}^{\nu}
	: |z-w|^{\alpha'p\cdot (q+\xi)}:\exp\left(ip\cdot X(w)+i\sum_{i}\zeta_{i}\cdot
	\partial_{w}^{m_{i}}X(w)+i\sum_{i}\lambda_{i}\cdot \bar{\partial}^{n_{i}}X(\bar{w})\right):\nonumber\\
	\times&\left .\left(\dfrac{\alpha'p\cdot q}{2(z-w)}+\sum_{i}\dfrac{\alpha'm_{i}!q\cdot \zeta_{i}}{2(z-w)^{m_{i}+1}}+iq\cdot \partial X(z)  \right)\left(\dfrac{\alpha'p\cdot q}{2(\bar{z}-\bar{w})}+\sum_{i}\dfrac{\alpha'n_{i}!q\cdot \lambda_{i}}{2(\bar{z}-\bar{w})^{n_{i}+1}}+iq\cdot \bar{\partial} X(\bar{z})  \right)\right|_{\text{multilinear in }  i\zeta_{i},i\lambda_{i} }.
	\end{align}
	In the last line only the product of the first terms in each bracket yields the factor $ |z-w|^{-2} $, and the z integration becomes
	\begin{align}
	&-\lim_{\xi \to 0}\dfrac{1}{2}h_{\mu\nu}\partial_{\xi}^{\mu}\partial_{\xi}^{\nu}
	|z-w|^{\alpha'p\cdot (q+\xi)}	\left(\dfrac{\alpha'p\cdot q}{2(z-w)}  \right)\left(\dfrac{\alpha'p\cdot q}{2(\bar{z}-\bar{w})} \right)\nonumber\\
	&\times \left .:\exp\left(ip\cdot X(w)+i\sum_{i}\zeta_{i}\cdot
	\partial_{w}^{m_{i}}X(w)+i\sum_{i}\lambda_{i}\cdot \bar{\partial}^{n_{i}}X(\bar{w})\right):
	\right|_{\text{multilinear in }  i\zeta_{i},i\lambda_{i} }\nonumber\\
	\rightarrow&\left .-\lim_{\xi \to 0}\dfrac{1}{2}h_{\mu\nu}\partial_{\xi}^{\mu}\partial_{\xi}^{\nu}\dfrac{2\pi}{\alpha'p\cdot(q+\xi)}\left(\dfrac{\alpha'p\cdot q}{2}\right)^2 :\exp\left(ip\cdot X(w)+i\sum_{i}\zeta_{i}\cdot
	\partial_{w}^{m_{i}}X(w)+i\sum_{i}\lambda_{i}\cdot \bar{\partial}^{n_{i}}X(\bar{w})\right):\right|_{\text{multilinear in }  i\zeta_{i},i\lambda_{i} }.
	\label{eq.(0,0;0)}
	\end{align}
	Here the arrow stands for the integration over z.
	By using eq.(\ref{eq.expand}) and taking terms that are quadratic in $ \xi $ and multilinear in $ i\zeta_{i} $ and $i\lambda_{i} $, eq.(\ref{eq.(0,0;0)}) becomes
	\begin{align}
	&-\dfrac{1}{2}h_{\mu\nu}\dfrac{2\pi 2p^{\mu}p^{\nu}}{\alpha'(p\cdot q)^3}\left(\dfrac{\alpha'p\cdot q}{2}\right)^2 :\exp(ip\cdot X(w))\partial^{m_{1}}X\cdots \bar{\partial}^{n_{1}}X\cdots :\nonumber\\
	=&-\dfrac{\pi\alpha' h_{\mu\nu}p^{\mu}p^{\nu}}{2p\cdot q} :\exp(ip\cdot X(w))\partial^{m_{1}}X\cdots \bar{\partial}^{n_{1}}X\cdots .\label{graviton leading}
	\end{align}
	Thus the leading soft theorem is obtained for any hard vertex.\\
	
\paragraph{The subleading soft theorem}\quad

 We keep the first order in $ (q+\xi) $ in eq.(\ref{eq.before}). We have the following three contributions.
	\begin{enumerate}
		\item \underline{(0,0;1) terms }
		\begin{align}
		&-\lim_{\xi \to 0}\dfrac{1}{2}h_{\mu\nu}\partial_{\xi}^{\mu}\partial_{\xi}^{\nu}
		: |z-w|^{\alpha'p\cdot (q+\xi)}
		i(q+\xi)\cdot X(z):\exp\left(ip\cdot X(w)+i\sum_{i}\zeta_{i}\cdot
		\partial_{w}^{m_{i}}X(w)+i\sum_{i}\lambda_{i}\cdot \bar{\partial}^{n_{i}}X(\bar{w})\right):\nonumber\\
		\times&\left .\left(\dfrac{\alpha'p\cdot q}{2(z-w)}+\sum_{i}\dfrac{\alpha'm_{i}!q\cdot \zeta_{i}}{2(z-w)^{m_{i}+1}}+iq\cdot \partial X(z)  \right)\left(\dfrac{\alpha'p\cdot q}{2(\bar{z}-\bar{w})}+\sum_{i}\dfrac{\alpha'n_{i}!q\cdot \lambda_{i}}{2(\bar{z}-\bar{w})^{n_{i}+1}}+iq\cdot \bar{\partial} X(\bar{z})  \right)\right|_{\text{multilinear in }  i\zeta_{i},i\lambda_{i} }.
		\end{align}
	    We consider the expansion of the last line. If we pick up one of the third terms in each bracket, we cannot have a pole for $ z-w $ or $ \bar{z}-\bar{w} $. Furthermore the product of the second terms in each bracket cannot become $ |z-w|^{-2} $ because the derivatives of $ X $ with respect to both $ z $ and $ \bar{z} $, $ \partial^{a}\bar{\partial}^{b}X(a,b\geq 1) $, are zero.
		
		The remaining terms are
		\begin{align}
		&-\lim_{\xi \to 0}\dfrac{1}{2}h_{\mu\nu}\partial_{\xi}^{\mu}\partial_{\xi}^{\nu}
		: |z-w|^{\alpha'p\cdot (q+\xi)}i(q+\xi)\cdot X(z):\exp\left(ip\cdot X(w)+i\sum_{i}\zeta_{i}\cdot
		\partial_{w}^{m_{i}}X(w)+i\sum_{i}\lambda_{i}\cdot \bar{\partial}^{n_{i}}X(\bar{w})\right):\nonumber\\
		\times&\left .\left [\left(\dfrac{\alpha'p\cdot q}{2(z-w)}  \right)\left(\dfrac{\alpha'p\cdot q}{2(\bar{z}-\bar{w})} \right)+\left(\dfrac{\alpha'p\cdot q}{2(z-w)}\right)\left (\sum_{i}\dfrac{\alpha'n_{i}!q\cdot \lambda_{i}}{2(\bar{z}-\bar{w})^{n_{i}+1}}\right )+\left (\sum_{i}\dfrac{\alpha'm_{i}!q\cdot \zeta_{i}}{2(z-w)^{m_{i}+1}}\right )\left(\dfrac{\alpha'p\cdot q}{2(\bar{z}-\bar{w})} \right)   \right ]\right|_{\text{multilinear in }  i\zeta_{i},i\lambda_{i} }\nonumber\\
		\rightarrow&\lim_{\xi \to 0}-\dfrac{1}{2}h_{\mu\nu}\partial_{\xi}^{\mu}\partial_{\xi}^{\nu}\dfrac{2\pi}{\alpha'p\cdot(q+\xi)}\left(\dfrac{\alpha'p\cdot q}{2}\right)
		\left [ \left (\dfrac{\alpha'p\cdot q}{2}\right)i(q+\xi)\cdot X(w)
		+\left(\sum_{i}\dfrac{\alpha'q\cdot \zeta_{i}}{2}\right)i(q+\xi)\cdot \partial^{m_{i}}X(w) \right .\nonumber\\
		&\left .+\left .\left (\sum_{i}\dfrac{\alpha'q\cdot \lambda_{i}}{2}\right ) :i(q+\xi)\cdot \bar{\partial}^{n_{i}}X(w)\right ] 
		\exp\left(ip\cdot X(w)+i\sum_{i}\zeta_{i}\cdot
		\partial_{w}^{m_{i}}X(w)+i\sum_{i}\lambda_{i}\cdot \bar{\partial}^{n_{i}}X(\bar{w})\right)
		\right|_{\text{multilinear in }  i\zeta_{i},i\lambda_{i} }.
		\label{eq.(0,0;1)}
		\end{align}

		\item \uline{(0,1;0) terms }
		\begin{align}
		&-\lim_{\xi \to 0}\dfrac{1}{2}h_{\mu\nu}\partial_{\xi}^{\mu}\partial_{\xi}^{\nu}
		: |z-w|^{\alpha'p\cdot (q+\xi)}
		\left (-\dfrac{\alpha'}{2}\sum_{i}\dfrac{(n_{i}-1)!(q+\xi)\cdot \lambda_{i}}{(\bar{z}-\bar{w})^{n_{i}}} \right) \nonumber\\
		\times&:\exp\left(ip\cdot X(w)+i\sum_{i}\zeta_{i}\cdot
		\partial_{w}^{m_{i}}X(w)+i\sum_{i}\lambda_{i}\cdot \bar{\partial}^{n_{i}}X(\bar{w})\right):\nonumber\\
		\times&\left .\left(\dfrac{\alpha'p\cdot q}{2(z-w)}+\sum_{i}\dfrac{\alpha'm_{i}!q\cdot \zeta_{i}}{2(z-w)^{m_{i}+1}}+iq\cdot \partial X(z)  \right)\left(\dfrac{\alpha'p\cdot q}{2(\bar{z}-\bar{w})}+\sum_{i}\dfrac{\alpha'n_{i}!q\cdot \lambda_{i}}{2(\bar{z}-\bar{w})^{n_{i}+1}}+iq\cdot \bar{\partial} X(\bar{z})  \right)
		\right|_{\text{multilinear in }  i\zeta_{i},i\lambda_{i} }.
		\end{align}
		In the last line only the product of $ \dfrac{\alpha'p\cdot q}{2(z-w)} $ and $ iq\cdot \bar{\partial} X(\bar{z})$ can give the form $ |z-w|^{-2} $, after combining with the Taylor expansion of $ X(z) $. Then we have
		\begin{align}
		&\lim_{\xi \to 0}-\dfrac{1}{2}h_{\mu\nu}\partial_{\xi}^{\mu}\partial_{\xi}^{\nu}
		|z-w|^{\alpha'p\cdot (q+\xi)}
		\left (-\dfrac{\alpha'}{2}\sum_{i}\dfrac{(n_{i}-1)!(q+\xi)\cdot \lambda_{i}}{(\bar{z}-\bar{w})^{n_{i}}} \right) \nonumber\\
		\times&\left .:\exp\left(ip\cdot X(w)+i\sum_{i}\zeta_{i}\cdot
		\partial_{w}^{m_{i}}X(w)+i\sum_{i}\lambda_{i}\cdot \bar{\partial}^{n_{i}}X(\bar{w})\right):
		\left(\dfrac{\alpha'p\cdot q}{2}\dfrac{1}{z-w})\right)\left(iq\cdot \bar{\partial} X(\bar{z})  \right)\right|_{\text{multilinear in }  i\zeta_{i},i\lambda_{i} }\nonumber\\
		\rightarrow&\lim_{\xi \to 0}-\dfrac{1}{2}h_{\mu\nu}\partial_{\xi}^{\mu}\partial_{\xi}^{\nu}\left (-\dfrac{\alpha'}{2}\sum_{i}(q+\xi)\cdot \lambda_{i} \right)\dfrac{2\pi}{\alpha'p\cdot(q+\xi)}
		\dfrac{\alpha'p\cdot q}{2}\nonumber\\
		&\left .\times:q\cdot \bar{\partial}^{n_{i}}X(\bar{w})
		\exp\left(ip\cdot X(w)+i\sum_{i}\zeta_{i}\cdot
		\partial_{w}^{m_{i}}X(w)+i\sum_{i}\lambda_{i}\cdot \bar{\partial}^{n_{i}}X(\bar{w})\right):
		\right|_{\text{multilinear in }  i\zeta_{i},i\lambda_{i} } .
		\label{eq.(0,1;0)}
		\end{align}
		\item \uline{(1,0;0) terms}\\
		We can get the result by replacing $ \lambda_{i}, n_{i},$ and $ \bar{X} $ with $ \zeta_{i}, m_{i},$ and $ X $ in eq.(\ref{eq.(0,1;0)}).
		\begin{align}
		&\lim_{\xi \to 0}\partial_{\xi}^{\mu}\partial_{\xi}^{\nu}-\dfrac{1}{2}h_{\mu\nu}\left (-\dfrac{\alpha'}{2}\sum_{i}(q+\xi)\cdot \zeta_{i} \right)\dfrac{2\pi}{\alpha'p\cdot(q+\xi)}\dfrac{\alpha'p\cdot q}{2}\nonumber\\
		&\left .\times:q\cdot \partial^{m_{i}}X(w) \exp\left(ip\cdot X(w)+i\sum_{i}\zeta_{i}\cdot
		\partial_{w}^{m_{i}}X(w)+i\sum_{i}\lambda_{i}\cdot \bar{\partial}^{n_{i}}X(\bar{w})\right):
		\right|_{\text{multilinear in }  i\zeta_{i},i\lambda_{i} } .
		\label{eq.(1,0;0)}
		\end{align}
		
		Adding eq.(\ref{eq.(0,0;1)}), eq.(\ref{eq.(0,1;0)}) and eq.(\ref{eq.(1,0;0)}), we get
		\begin{align}
		&\lim_{\xi \to 0}-\dfrac{1}{2}h_{\mu\nu}\partial_{\xi}^{\mu}\partial_{\xi}^{\nu}\dfrac{2\pi}{\alpha'p\cdot(q+\xi)}\left(\dfrac{\alpha'p\cdot q}{2}\right)
		\exp\left(ip\cdot X(w)+i\sum_{i}\zeta_{i}\cdot
		\partial_{w}^{m_{i}}X(w)+i\sum_{i}\lambda_{i}\cdot \bar{\partial}^{n_{i}}X(\bar{w})\right)\nonumber\\
		&\left [ \left(\dfrac{\alpha'p\cdot q}{2}\right):i(q+\xi)\cdot X(w)+\left(\sum_{i}\dfrac{\alpha'q\cdot \zeta_{i}}{2}\right)i(q+\xi)\cdot \partial^{m_{i}}X(w)+\left (\sum_{i}\dfrac{\alpha'q\cdot \lambda_{i}}{2}\right )i(q+\xi)\cdot \bar{\partial}^{n_{i}}X(w)\right .\nonumber\\
		&+\left.\left. \left (-\dfrac{\alpha'}{2}\sum_{i}(q+\xi)\cdot \zeta_{i} \right)q\cdot \partial^{m_{i}}X(w)
		+\left (-\dfrac{\alpha'}{2}\sum_{i}(q+\xi)\cdot \lambda_{i} \right)q\cdot \bar{\partial}^{n_{i}}X(\bar{w})
		\right ]\right|_{\text{multilinear in }  i\zeta_{i},i\lambda_{i} }\nonumber\\
		=&\lim_{\xi \to 0}-\dfrac{1}{2}h_{\mu\nu}\partial_{\xi}^{\mu}\partial_{\xi}^{\nu}\dfrac{2\pi}{\alpha'}\left(\dfrac{\alpha'p\cdot q}{2}\right)
		\exp\left(ip\cdot X(w)+i\sum_{i}\zeta_{i}\cdot
		\partial_{w}^{m_{i}}X(w)+i\sum_{i}\lambda_{i}\cdot \bar{\partial}^{n_{i}}X(\bar{w})\right)\nonumber\\
		&\left [ \left(\dfrac{\alpha'p\cdot q}{2}\right):\dfrac{iq\cdot X(w)(p\cdot \xi)^2-i\xi\cdot X(w)(p\cdot \xi)(p\cdot q)}{(p\cdot q)^3}\right .\nonumber\\
		&\left .\left .+\left(\sum_{i}\dfrac{\alpha'q_{a}\xi_{b} \zeta_{ic}\partial^{m_{i}}X_{d}(w)(S^{ab})^{cd}}{2}\right)\dfrac{-p\cdot\xi}{(p\cdot q)^2} +\left (\sum_{i}\dfrac{\alpha'q_{a'}\xi_{b'} \lambda_{ic'}\bar{\partial}^{n_{i}}X_{d'}(\bar{w})(S^{a'b'})^{c'd'}}{2}\right )\dfrac{-p\cdot\xi}{(p\cdot q)^2}
		\right ]\right|_{\text{multilinear in }  i\zeta_{i},i\lambda_{i} }\nonumber\\
		=&-\dfrac{\alpha'}{2}h_{\mu\nu}\pi\exp\left(ip\cdot X(w)+i\sum_{i}\zeta_{i}\cdot
		\partial_{w}^{m_{i}}X(w)+i\sum_{i}\lambda_{i}\cdot \bar{\partial}^{n_{i}}X(\bar{w})\right)\nonumber\\
		&\left [ \dfrac{iq\cdot X(w)p^{\mu}p^{\nu}-iX^{\nu}(w)p^{\mu}(p\cdot q)}{p\cdot q}
		+\left(\sum_{i}q_{a} \zeta_{ic}\partial^{m_{i}}X_{d}(w)(S^{a\mu})^{cd}\right)\dfrac{-p^{\nu}}{(p\cdot q)}\right . \nonumber\\
		&\quad +\left .\left .\left (\sum_{i}q_{a'} \lambda_{ic'}\bar{\partial}^{n_{i}}X_{d'}(\bar{w})(S^{a'\mu})^{c'd'}\right )\dfrac{-p^{\nu}}{(p\cdot q)}
		\right ]\right|_{\text{multilinear in }  i\zeta_{i},i\lambda_{i} }\nonumber\\
		=&-\dfrac{\alpha'h_{\mu\nu}\pi}{2p\cdot q}
		\left [-i p^{\mu}q_{a}L^{\nu a}
		+\left(\sum_{i}q_{a} \zeta_{ic}\partial^{m_{i}}X_{d}(w)(S^{\mu a})^{cd}\right)p^{\nu} +\left (\sum_{i}q_{a'} \lambda_{ic'}\bar{\partial}^{n_{i}}X_{d'}(\bar{w})(S^{\mu a'})^{c'd'}\right )p^{\nu}
		\right ]\nonumber\\
		&\left .\times\exp\left(ip\cdot X(w)+i\sum_{i}\zeta_{i}\cdot
		\partial_{w}^{m_{i}}X(w)+i\sum_{i}\lambda_{i}\cdot \bar{\partial}^{n_{i}}X(\bar{w})\right)
		\right|_{\text{multilinear in }  i\zeta_{i},i\lambda_{i} },\label{eq.subleading graviton theorem}
		\end{align}
		where $ L^{\mu a}=p^{a}X^{\mu} -p^{\mu}X^{a}$, $ (S^{ab})^{cd}=\eta^{ac}\eta^{bd}-\eta^{ad}\eta^{bc} $.\\
		The second and third terms in the square bracket represent a Lorentz transformation for each index of the prefactors. Then we can write eq.(\ref{eq.subleading graviton theorem}) as 
		\begin{align}
		\dfrac{i\alpha'h_{\mu\nu}\pi p^{\mu}q_{a}J^{\nu a}}{2p\cdot q}
		:\partial^{m_{1}}X\cdots\bar{\partial}^{n_{1}}X\cdots\exp\left(ip\cdot X(w)\right ):,
		\end{align}
		where $ J^{\mu a}=L^{\mu a}+S^{\mu a}+\bar{S}^{\mu a} $. $ S^{\mu a} $ and $ \bar{S}^{\nu b} $ are the spin angular momentum operators for the holomorphic and antiholomorphic parts, respectively.
	\end{enumerate}

\paragraph{The subsubleading soft theorem}\quad

We keep the second order terms in $ q+\xi $ in eq.(\ref{eq.before}).

The details are given in Appendix \ref{sec;formula}, and we write only the last expression here:
\begin{align}
	&\dfrac{\pi\alpha'h_{\mu\nu}}{4p\cdot q}
	\left[q_{a}q_{b}\left (L^{\mu a}L^{\nu b}+2S^{\mu a}\bar{S}^{\nu b}\right )
	+2q_{a}L^{\mu a}\left(q_{b}S^{b\nu}+q_{b}\bar{S}^{b\nu}  \right)
	\right]:\partial^{m_{1}}X^{\rho_{1}}\cdots\bar{\partial}^{n_{1}}X^{\sigma_{1}}\cdots \exp\left(ip\cdot X(w)\right ):\nonumber\\
	+&\dfrac{\pi\alpha'^2h_{\mu\nu}}{4p\cdot q}
	:\sum_{i}(S^{\mu a})^{ed}(S^{b\nu})^{\rho_{i}f}\dfrac{q_{a}q_{b}\partial^{m_{i}}X_{d}p_{e}p_{f}}{m_{i}} 
	\cdots\underbrace{\partial^{m_{i}}X^{\rho_{i}}}\cdots\bar{\partial}^{n_{1}}X^{\sigma_{1}}\cdots\exp\left(ip\cdot X(w)\right ):\nonumber\\
	+&\dfrac{\pi\alpha'^2h_{\mu\nu}}{4p\cdot q}:\sum_{i}(S^{\mu a})^{ed}(S^{b\nu})^{\sigma_{i}f}
	\dfrac{q_{a}q_{b}\bar{\partial}^{n_{i}}X_{d}p_{e}p_{f}}{n_{i}}  
	\partial^{m_{1}}X^{\rho_{1}}\cdots\underbrace{\bar{\partial}^{n_{i}}X^{\sigma_{1}}}\cdots\exp\left(ip\cdot X(w)\right ):\nonumber\\
	-&\dfrac{i\pi\alpha'h_{\mu\nu}}{2p\cdot q}
	\left(\sum_{k=1}^{m_{i}-1}\dfrac{(m_{i}-1)! }{k!(m_{i}-k-1)!}  \right)
	:\left(q_{a}p_{c}\partial^{k}X_{d}(S^{a\mu})^{cd}  \right)
	\left(q_{b}\partial^{m_{i}-k}X_{d}(S^{b\nu})^{\rho_{i}d}  \right)
	\cdots\underbrace{\partial^{m_{i}}X^{\rho_{i}}}\cdots\bar{\partial}^{n_{1}}X^{\sigma_{1}}\cdots\exp\left(ip\cdot X(w)\right ):\nonumber\\
	-&\dfrac{i\pi\alpha'h_{\mu\nu}}{2p\cdot q}
	\left(\sum_{k=1}^{n_{i}-1}\dfrac{(n_{i}-1)! }{k!(n_{i}-k-1)!}  \right)
	:\left(q_{a}p_{c}\bar{\partial}^{k}X_{d}(S^{a\mu})^{cd}  \right) 
	\left(q_{a}\bar{\partial}^{n_{i}-k}X_{d}(S^{a\nu})^{\sigma_{i}d}  \right) 
	\partial^{m_{1}}X^{\rho_{1}}\cdots\underbrace{\bar{\partial}^{n_{i}}X^{\sigma_{i}}}\cdots\exp\left(ip\cdot X(w)\right ):\nonumber\\
	+&\dfrac{i\pi\alpha'^2h_{\mu\nu}}{8p\cdot q}
	\left[
	\sum_{i,j}
	\dfrac{(n_{i}-1)!(n_{j}-1)!}{(n_{i}+n_{j}-1)!}
	:\dfrac{q_{a}p_{d}(S^{\mu a})^{\sigma_{i}d}n_{j}q_{e}\bar{\partial}^{n_{i}+n_{j}}X_{h}(S^{\nu e})^{\sigma_{j}h}+(i\leftrightarrow j)}{(n_{i}+n_{j})}\right]
	\partial^{m_{1}}X^{\rho_{1}}\cdots\underbrace{\bar{\partial}^{n_{i}}X^{\sigma_{i}}}\underbrace{\bar{\partial}^{n_{j}}X^{\sigma_{j}}}\cdots\exp\left(ip\cdot X(w)\right ):\nonumber\\
	+&\dfrac{i\pi\alpha'^2h_{\mu\nu}}{8p\cdot q}\left [\sum_{i,j}
	\dfrac{(m_{i}-1)!(m_{j}-1)!}{(m_{i}+m_{j}-1)!}
	:\dfrac{q_{a}p_{d}(S^{\mu a})^{\rho_{i}d}m_{j}q_{e}\partial^{m_{i}+m_{j}}X_{h}(S^{\nu e})^{\rho_{j}h}+(i\leftrightarrow j)}{(m_{i}+m_{j})}
	\right]
	\cdots\underbrace{\partial^{m_{i}}X^{\rho_{i}}}\underbrace{\partial^{m_{j}}X^{\rho_{j}}}\cdots\bar{\partial}^{n_{1}}X^{\sigma_{1}}\cdots\exp\left(ip\cdot X(w)\right ):\label{eq.graviton subsubleading}.
\end{align}
Here the underbrace represents the absence of the factor above it. 
Note that in general we cannot express the subsubleading soft theorem in terms of the total angular momentum.
A compact form of eq.(\ref{eq.graviton subsubleading}) will be given in Section \ref{sec;compact form}.

\subsection{Examples}
We consider the following two examples of eq.(\ref{eq.graviton subsubleading}).
\subsubsection{Example : soft graviton/dilaton theorem for hard massless particles}\quad
 \label{graviton}
\quad We assume that the hard vertex is $ V=A_{\rho\sigma}\partial X^{\rho}(w)\bar{\partial}X^{\sigma}(\bar{w})\exp\left( ip\cdot X(w,\bar{w}) \right) $, where $ A_{\rho\sigma} $ is a polarization tensor of the hard particle, which can be either symmetric or antisymmetric. 
Applying the formulas eq.(\ref{graviton leading}), eq.(\ref{eq.subleading graviton theorem}) and eq.(\ref{eq.graviton subsubleading}) for this case, we obtain
\begin{align}
&\dfrac{\pi\alpha' h_{\mu\nu}p^{\mu}p^{\nu}A_{\rho\sigma}}{2p\cdot q} :\partial X^{\rho} \bar{\partial}X^{\sigma} \exp(ip\cdot X(w)):\nonumber\\
+&\dfrac{\alpha'h_{\mu\nu}\pi p^{\mu}q_{a}A_{\rho\sigma}}{2p\cdot q}
\left [J^{\nu a}:\partial X^{\rho}\bar{\partial}X^{\sigma}\exp\left(ip\cdot X(w)\right ):\right ]\nonumber\\
+&\dfrac{\pi\alpha'h_{\mu\nu}A_{\rho\sigma}}{4p\cdot q}
\left[q_{a}q_{b}\left (L^{\mu a}L^{\nu b}\eta^{\rho\alpha}\eta^{\sigma\beta}+2(S^{\mu a})^{\rho\alpha}(\bar{S}^{\nu b})^{\sigma\beta}\right )
+2q_{a}L^{\mu a}\left(q_{b}(S^{b\nu})^{\sigma\beta}\eta^{\sigma\beta}+q_{b}(\bar{S}^{b\nu})^{\sigma\beta}\eta^{\rho\alpha}  \right)
\right]:\partial X_{\alpha}\bar{\partial}X_{\beta}\exp\left(ip\cdot X(w)\right ):\nonumber\\
+&C_{1}(w)+C_{2}(w),\label{eq.soft graviton theorem for massless}
\end{align}
where $ C_{1}(w) $ and $ C_{2}(w) $ are the higher order terms in $ \alpha' $:
\begin{align}
&C_{1}(w)\equiv \dfrac{\pi\alpha'^2h_{\mu\nu}A_{\rho\sigma}}{4p\cdot q}
:\sum_{i}(S^{\mu a})^{ed}(S^{b\nu})^{\rho f}q_{a}q_{b}\partial X_{d}p_{e}p_{f}
\bar{\partial}X^{\sigma}\exp\left(ip\cdot X(w)\right ):,\nonumber\\
&C_{2}(w)\equiv \dfrac{\pi\alpha'^2h_{\mu\nu}A_{\rho\sigma}}{4p\cdot q}:\sum_{i}(S^{\mu a})^{ed}(S^{b\nu})^{\sigma f}
q_{a}q_{b}\bar{\partial}X_{d}p_{e}p_{f}
\partial X^{\rho}\exp\left(ip\cdot X(w)\right ):.
\end{align}
We have only one prefactor for each of the holomorphic and antiholomorphic part. 
For graviton it is easy to check that the following combination of the spin operators vanishes when it acts on a single vector index:
\begin{align}
h_{\mu\nu}q_{a}q_{b}(S^{\mu a})^{\rho}_{\alpha}(S^{\nu b})^{\alpha}_{\beta}=0.\label{eq.spinspin vanish}
\end{align} 
Here we have used $ h_{\mu\nu}q^{\mu}=0 $ and $ h^{\mu}_{\mu}=0 $.
Then the third line of eq.(\ref{eq.soft graviton theorem for massless}) becomes

\begin{align}
&\dfrac{\pi\alpha'h_{\mu\nu}h_{\rho\sigma}}{4p\cdot q}
\left[q_{a}q_{b}\left (L^{\mu a}L^{\nu b}+2S^{\mu a}\bar{S}^{\nu b}\right )
+2q_{a}L^{\mu a}\left(q_{b}S^{b\nu}+q_{b}\bar{S}^{b\nu}  \right)
\right]:\partial X^{\rho}\bar{\partial}X^{\sigma}\exp\left(ip\cdot X(w)\right ):\nonumber\\
=&\dfrac{\pi\alpha'h_{\mu\nu}A_{\rho\sigma}}{4p\cdot q}
q_{a}q_{b}:\left (J^{\mu a}J^{\nu b}\partial X \bar{\partial}X \exp(ip\cdot X) \right )^{\rho\sigma}:.\label{eq.subsubleading part}
\end{align}   
 This reproduces the subsubleading soft graviton theorem of the field theory at the tree level, if we ignore the higher order terms in $ \alpha' $ in eq.(\ref{eq.soft graviton theorem for massless}).
 
 When we include the $ \alpha' $ corrections, we can find that the polarization tensors of $ C_{1} $ and $ C_{2}(w) $ are not traceless, if we assume that $ A_{\rho\sigma} $ is traceless. Therefore, a mixing between dilaton and graviton occurs in the higher order terms in $ \alpha' $.

	On the other hand, $ C_{1}(w) $ and $ C_{2}(w) $ vanish for soft dilaton as was discussed in \cite{DiVecchia:2016amo}. 
	To see this, we take the polarization tensor of dilaton as 
\begin{align}
h_{\mu\nu}(q)=\dfrac{\eta_{\mu\nu}-q_{\mu}\bar{q}_{\nu}-\bar{q}_{\mu}q_{\nu}}{\sqrt{D-2}},
\end{align}
where $ \bar{q}^2=0, q\cdot \bar{q}=1, $ and D is the dimension of the spacetime.
When we substitute this in $ C_{1}(w) $ and $ C_{2}(w) $, the second and third terms in $ h_{\mu\nu}(q) $, $ -q_{\mu}\bar{q_{\nu}}-\bar{q_{\mu}}q_{\nu} $, vanish because $ S^{\mu a}q_{a}q_{\mu}=0 $. 
Then we can replace $ h_{\mu\nu}$ with $\eta_{\mu\nu} $, and $ C_{1}(w) $ becomes 
\begin{align}
&\dfrac{\pi\alpha'^2\eta_{\mu\nu}A_{\rho\sigma}}{4p\cdot q}
:\sum_{i}(S^{\mu a})^{ed}(S^{b\nu})^{\rho f}q_{a}q_{b}\partial X_{d}p_{e}p_{f}
\bar{\partial}X^{\sigma}\exp\left(ip\cdot X(w)\right ):\nonumber\\
=&\dfrac{\pi\alpha'^2\eta_{\mu\nu}A_{\rho\sigma}}{4p\cdot q}
:\sum_{i}(p^{\mu}q^{d}-p\cdot q\eta^{\mu d})(q^{\rho}p^{\nu}-p\cdot q \eta^{\nu\rho}) \partial X_{d}
\bar{\partial}X^{\sigma}\exp\left(ip\cdot X(w)\right ):\nonumber\\
=&\dfrac{\pi\alpha'^2A_{\rho\sigma}}{4p\cdot q}
:\sum_{i}(p^2q^{d}q^{\rho}-p^{\rho}q^{d}p\cdot q-p^{d}q^{\rho}p\cdot q+(p\cdot q)^2\eta^{d\rho}) \partial X_{d}
\bar{\partial}X^{\sigma}\exp\left(ip\cdot X(w)\right ):\nonumber\\
=&\dfrac{\pi\alpha'^2A_{\rho\sigma}}{4p\cdot q}
:\sum_{i}(p^{d}q^{\rho}p\cdot q+(p\cdot q)^2\eta^{d\rho}) \partial X_{d}
\bar{\partial}X^{\sigma}\exp\left(ip\cdot X(w)\right ):\nonumber\\
=&\dfrac{\pi\alpha'^2}{4}
:\sum_{i}(A_{\rho\sigma}q^{\rho}\bar{\partial}X^{\sigma}\partial +A_{\rho\sigma}p\cdot q\partial X^{\rho}\bar{\partial}X^{\sigma}) 
\exp\left(ip\cdot X(w)\right ):.\label{eq.alpha term in soft dilaton}
\end{align}
Here we have used $ p^2=0 $ and $ A_{\rho\sigma}p^{\rho}=0 $.
We can drop the first term in eq.(\ref{eq.alpha term in soft dilaton}) because it is total derivative. The second term becomes proportional to the original vertex operator, and vanishes by the momentum conservation when we consider all the hard vertex operators:
\begin{align}
&\int d^2z \langle:V_{s}(q;z)::V_{1}(p_{1};w_{1}):\cdots:V_{i}(p_{i};w_{i}):\cdots:V_{n}(p_{n;}w_{n})\rangle\nonumber\\
&\rightarrow \dfrac{\pi\alpha'^2}{4}
\sum_{i}p_{i}\cdot q \langle:V_{1}(p_{1};w_{1}):\cdots:V_{i}(p_{i};w_{i}):\cdots:V_{n}(p_{n;}w_{n})\rangle\nonumber\\
&=0,
\end{align}
 where the arrow means that we focus only on the contribution from $ C_{1}(w) $. The same is true for $ C_{2}(w) $.
 
 Therefore, the higher order terms in $ \alpha' $ for the soft dilaton and hard massless vertices become 0. The explicit form of the subsubleading soft dilaton theorem is
 \begin{align}
 &A_{\rho\sigma}\left [\left (-2q_{a}p_{\nu}\dfrac{\partial}{\partial p_{\nu}}\dfrac{\partial}{\partial p_{a}}+p\cdot q\dfrac{\partial}{\partial p_{a}}\dfrac{\partial}{\partial p_{a}}\right)\partial X^{\rho}\bar{\partial}X^{\sigma}\right .\nonumber\\
 +&\dfrac{2}{p\cdot q}\left( \eta^{\rho\sigma}q_{\alpha}q_{\beta}\partial X^{\alpha}\bar{\partial}X^{\beta}-q_{\alpha}q^{\sigma}\partial X^{\alpha}\bar{\partial}X^{\rho}-q_{\sigma}q_{\beta}\partial X^{\sigma}\bar{\partial}X^{\beta}+q^{\rho}q^{\sigma}\partial X^{\alpha}\bar{\partial}X_{\alpha}
 \right)\nonumber\\
 +&\left .2\left(
 - \dfrac{\partial}{\partial p_{\rho}}\partial X_{\beta}\bar{\partial}X^{\sigma}
 + q^{\rho}\dfrac{\partial}{\partial p_{\beta}}\partial X^{\beta}\bar{\partial}X^{\sigma}
 \right)\right ]\exp(ip\cdot X(w)).
 \end{align} 
 This is the same result as in \cite{DiVecchia:2016amo}. 

\subsubsection{Example : soft graviton/dilaton theorem for hard massive particles}\quad

\quad A physical state at the next level is expressed by the second-order traceless transverse symmetric tensor, when we add the appropriate spurious state. Then we can assume that the hard vertex is 
\begin{align}
 V_{hard}(w)=A_{\rho_{1}\rho_{2}\sigma_{1}\sigma_{2}}\partial X^{\rho_{1}}\partial X^{\rho_{2}}\bar{\partial}X^{\sigma_{1}}\bar{\partial}X^{\sigma_{2}}\exp(ip\cdot X(w)).
 \label{eq.massive hard vertex} 
\end{align}
\begin{align}
A_{\rho_{1}\rho_{2}\sigma_{1}\sigma_{2}}p^{\rho_{1}}=0,\quad A^{\alpha}_{\ \alpha\sigma_{1}\sigma_{2}}=0,\qquad
A_{\rho_{1}\rho_{2}\sigma_{1}\sigma_{2}}p^{\sigma_{1}}=0,\quad A_{\rho_{1}\rho_{2}\alpha}^{ \ \qquad \alpha}=0.
\label{eq.massive hard polarization}
\end{align}
We apply the formula eq.(\ref{graviton leading}), eq.(\ref{eq.subleading graviton theorem}) and eq.(\ref{eq.graviton subsubleading}) for this case, and obtain
\begin{align}
:V_{s}(z)::V_{hard}(w)=:&\dfrac{\pi\alpha' h_{\mu\nu}p^{\mu}p^{\nu}A_{\rho_{1}\rho_{2}\sigma_{1}\sigma_{2}}}{2p\cdot q} :\partial X^{\rho_{1}}\partial X^{\rho_{2}} \bar{\partial}X^{\sigma_{1}}\bar{\partial}X^{\sigma_{2}} \exp(ip\cdot X(w)):\nonumber\\
+&\dfrac{\alpha'h_{\mu\nu}\pi p^{\mu}q_{a}A_{\rho_{1}\rho_{2}\sigma_{1}\sigma_{2}}}{2p\cdot q}
\left (:J^{\nu a}\partial X^{\rho_{1}}\partial X^{\rho_{2}}\bar{\partial}X^{\sigma_{1}}\bar{\partial}X^{\sigma_{2}}\exp\left(ip\cdot X(w)\right ):\right )^{\rho_{1}\rho_{2}\sigma_{1}\sigma_{2}}\nonumber\\
+&\dfrac{\pi\alpha'h_{\mu\nu}A_{\rho_{1}\rho_{2}\sigma_{1}\sigma_{2}}}{4p\cdot q}
\left[q_{a}q_{b}\left (L^{\mu a}L^{\nu b}\eta^{\rho_{1}\alpha}\eta^{\rho_{2}\beta}\eta^{\sigma_{1}\gamma}\eta^{\sigma_{2}\delta}
+2(S^{\mu a})^{\rho_{1}\alpha}(\bar{S}^{\nu b})^{\sigma_{1}\gamma}\eta^{\rho_{2}\beta}\eta^{\sigma_{2}\beta}\right )\right .\nonumber\\
&\left .+2q_{a}L^{\mu a}\left(q_{b}(S^{b\nu})^{\rho_{1}\alpha}\eta^{\rho_{2}\beta}\eta^{\sigma_{1}\gamma}\eta^{\sigma_{2}\delta}+q_{b}(\bar{S}^{b\nu})^{\sigma_{1}\gamma}\eta^{\rho_{1}\alpha}\eta^{\rho_{2}\beta}\eta^{\sigma_{2}\delta}  \right)
\right]
\partial X_{\alpha}\partial X_{\beta}\bar{\partial}X_{\gamma}\bar{\partial}X_{\delta}\exp\left(ip\cdot X(w)\right )\nonumber\\
+&\dfrac{\pi\alpha'^2h_{\mu\nu}A_{\rho_{1}\rho_{2}\sigma_{1}\sigma_{2}}}{2p\cdot q}
(S^{\mu a})^{ed}(S^{b\nu})^{\rho_{1}f}q_{a}q_{b}\partial X_{d}p_{e}p_{f} 
\partial X^{\rho_{2}}\bar{\partial}X^{\sigma_{1}}\bar{\partial}X^{\sigma_{2}}\exp\left(ip\cdot X(w)\right ):\nonumber\\
+&\dfrac{\pi\alpha'^2h_{\mu\nu}A_{\rho_{1}\rho_{2}\sigma_{1}\sigma_{2}}}{2p\cdot q}(S^{\mu a})^{ed}(S^{b\nu})^{\sigma_{1}f}
q_{a}q_{b}\bar{\partial}X_{d}p_{e}p_{f} 
:\partial X^{\rho_{1}}\partial X^{\rho_{2}}\bar{\partial}X^{\sigma_{2}}\exp\left(ip\cdot X(w)\right ):\nonumber\\
+&\dfrac{i\pi\alpha'^2h_{\mu\nu}A_{\rho_{1}\rho_{2}\sigma_{1}\sigma_{2}}}{4p\cdot q}
\left[
q_{a}p_{d}(S^{\mu a})^{\sigma_{1}d}q_{b}\bar{\partial}^{2}X_{h}(S^{\nu b})^{\sigma_{2}h}\right]
:\partial X^{\rho_{1}}\partial X^{\rho_{2}}\exp\left(ip\cdot X(w)\right ):\nonumber\\
+&\dfrac{i\pi\alpha'^2h_{\mu\nu}A_{\rho_{1}\rho_{2}\sigma_{1}\sigma_{2}}}{4p\cdot q}\left [
q_{a}p_{d}(S^{\mu a})^{\rho_{1}d}q_{b}\partial^{2}X_{h}(S^{\nu b})^{\rho_{2}h}
\right]
\bar{\partial}X^{\sigma_{1}}\bar{\partial}X^{\sigma_{2}}\exp\left(ip\cdot X(w)\right ):
+\mathcal{O}(q^2).
\label{eq.soft graviton theorem for massive}
\end{align}
In this case, all terms are the same order because $ p^2=-\dfrac{\alpha'}{4} $. 
We can rewrite the above equation by using the total angular momentum as
\footnote{
	As we have seen for the massless particles, the contribution vanishes when the spin angular momentum operators act on the holomorphic part twice as in eq.(\ref{eq.spinspin vanish}).  On the other hand, for the massive particles, $S^{2}(\bar{S}^{2}) $ need not be 0 when each of the $ S's(\bar{S}'s) $ acts on the different indexes, such as 
	\begin{align}
	h_{\mu\nu}q_{a}q_{b}(S^{\mu a})^{\rho_{1}}_{\alpha}(S^{\nu b})^{\rho_{2}}_{\beta}\partial X^{\alpha}\partial X^{\beta}\neq 0.
	\end{align}
}
\begin{align}
&\dfrac{\pi\alpha' h_{\mu\nu}p^{\mu}p^{\nu}A_{\rho_{1}\rho_{2}\sigma_{1}\sigma_{2}}}{2p\cdot q} :\partial X^{\rho_{1}}\partial X^{\rho_{2}} \bar{\partial}X^{\sigma_{1}}\bar{\partial}X^{\sigma_{2}} \exp(ip\cdot X(w)):\nonumber\\
+&\dfrac{\alpha'h_{\mu\nu}\pi p^{\mu}q_{a}A_{\rho_{1}\rho_{2}\sigma_{1}\sigma_{2}}}{2p\cdot q}
\left (:J^{\nu a}\partial X^{\rho_{1}}\partial X^{\rho_{2}}\bar{\partial}X^{\sigma_{1}}\bar{\partial}X^{\sigma_{2}}\exp\left(ip\cdot X(w)\right )\right )^{\rho_{1}\rho_{2}\sigma_{1}\sigma_{2}}\nonumber\\
+& \dfrac{\pi\alpha'h_{\mu\nu}A_{\rho_{1}\rho_{2}\sigma_{1}\sigma_{2}}q_{a}q_{b}}{4p\cdot q}
\left[\left (J^{\mu a}J^{\nu b}-S^{\mu a}S^{\nu b}-\bar{S}^{\mu a}\bar{S}^{\nu b}\right )\partial X \partial X \bar{\partial}X \bar{\partial}X \exp(ip\cdot X(w))\right]^{\rho_{1}\rho_{2}\sigma_{1}\sigma_{2}} \nonumber\\
+&\dfrac{\pi\alpha'^2h_{\mu\nu}A_{\rho_{1}\rho_{2}\sigma_{1}\sigma_{2}}q_{a}q_{b}}{2p\cdot q}
(S^{\mu a}p)^{\rho_{1}}p_{c}(S^{b\nu}\partial X)^{c} 
\partial X^{\rho_{2}}\bar{\partial}X^{\sigma_{1}}\bar{\partial}X^{\sigma_{2}}\exp\left(ip\cdot X(w)\right ):\nonumber\\
+&\dfrac{\pi\alpha'^2h_{\mu\nu}A_{\rho_{1}\rho_{2}\sigma_{1}\sigma_{2}}q_{a}q_{b}}{2p\cdot q}(S^{\mu a}p)^{\rho_{1}}p_{c}(S^{b\nu}\bar{\partial}X)^{c} 
\partial X^{\rho_{1}}\partial X^{\rho_{2}}\bar{\partial}X^{\sigma_{2}}\exp\left(ip\cdot X(w)\right ):\nonumber\\
+&\dfrac{i\pi\alpha'^2h_{\mu\nu}A_{\rho_{1}\rho_{2}\sigma_{1}\sigma_{2}}q_{a}q_{b}}{4p\cdot q}
\left[
(S^{\mu a}p)^{\sigma_{1}}(S^{\nu b}\bar{\partial}^{2}X)^{\sigma_{2}}\right]
\partial X^{\rho_{1}}\partial X^{\rho_{2}}\exp\left(ip\cdot X(w)\right )\nonumber\\
+&\dfrac{i\pi\alpha'^2h_{\mu\nu}A_{\rho_{1}\rho_{2}\sigma_{1}\sigma_{2}}q_{a}q_{b}}{4p\cdot q}\left [
(S^{\mu a}p)^{\rho_{1}}(S^{\nu b}\partial^{2}X)^{\rho_{2}}
\right]
\bar{\partial}X^{\sigma_{1}}\bar{\partial}X^{\sigma_{2}}\exp\left(ip\cdot X(w)\right ).
\label{eq.soft graviton theorem for massive 2}
\end{align}
The first, second and rest lines represent the leading, subleading and subsubleading soft theorems. This is also derived by using Ward identity in Appendix \ref{section;from Ward identity}.

We can see that the resulting vertex operator in the subsubleading soft theorem satisfy the physical condition as follows.
It is obvious that the first and second lines are physical.
The term including the total angular momentum, $ J^{\mu a}J^{\nu b} $, is physical.
The remaining terms are divided into the two parts, each of which involves the transformations of only the holomorphic or antiholomorphic prefactors.
We focus on the holomorphic part and omit the polarization tensors in the antiholomorphic part.

From eq.(\ref{eq.soft graviton theorem for massive 2}), we find that the coefficients $ \zeta_{\alpha\beta} $ of $ \partial X^{\alpha}\partial X^{\beta}\bar{\partial}X^{\sigma_{1}}\bar{\partial}X^{\sigma_{2}}\exp(ip\cdot X) $ is
\begin{align}
\zeta^{\alpha\beta}=&\dfrac{\pi\alpha'h_{\mu\nu}A_{\rho_{1}\rho_{2}\sigma_{1}\sigma_{2}}q_{a}q_{b}}{2p\cdot q}
\left[
(S^{\mu a})^{\rho_{1}\alpha}(S^{\nu b})^{\rho_{2}\beta}
+\alpha'(S^{\mu a})^{\rho_{1}d}p_{d}p_{c}(S^{\nu b})^{c\alpha}\eta^{\beta\rho_{2}}
\right],
\end{align}
and that the coefficient $ \xi_{\alpha} $ of $ \partial^{2} X^{\alpha}\bar{\partial}X^{\sigma_{1}}\bar{\partial}X^{\sigma_{2}} $ is 
\begin{align}
\xi_{\alpha}=&\dfrac{i\pi\alpha'^2h_{\mu\nu}A_{\rho_{1}\rho_{2}\sigma_{1}\sigma_{2}}q_{a}q_{b}}{4p\cdot q}\left [
(S^{\mu a})^{\rho_{1}c}p_{c}(S^{\nu b})^{\rho_{2}}_{\alpha}
\right].
\end{align}
For convenience, the polarization tensors can be replaced by the following form:
\begin{align}
&h_{\mu\nu}=h_{\mu}h_{\nu},
\quad  h\cdot q=0,\\
&A_{\rho_{1}\rho_{2}\sigma_{1}\sigma_{2}}=\zeta_{\rho_{1}}\zeta_{\rho_{2}}\bar{\zeta}_{\sigma_{1}}\bar{\zeta}_{\sigma_{2}},
\quad p\cdot \zeta=0, \quad p\cdot \bar{\zeta}=0.
\end{align}
We can check that these polarizations satisfy the following equations:
\begin{align}
\xi_{\alpha}&=\dfrac{i\alpha'}{2}\zeta_{\alpha\beta}p^{\beta},\\
\zeta^{\alpha}_{\alpha}&=\alpha'p_{\alpha}p_{\beta}\zeta^{\alpha\beta}.
\end{align}
This is nothing but the physical condition. Thus, the subsubleading term is physical.

Next, by adding spurious states, we can transform the resulting vertex, $ \zeta_{\alpha\beta}\partial X^{\alpha}\partial X^{\beta}+\xi_{\alpha}\partial^{2}X^{\alpha} $, to the original form of eq.(\ref{eq.massive hard vertex}) and (\ref{eq.massive hard polarization}):
\begin{align}
&V_{phys}(p;w)=\psi_{\mu\nu}\partial X^{\mu}\partial X^{\nu}\mathrm{e}^{ip\cdot X}(w),\\
&\psi^{\mu}_{\mu}(p)=0,p_{\mu}\psi^{\mu\nu}=0.
\end{align}
In fact, the states generated by Virasoro generators are
\begin{align}
V_{spurious}(p;w)=-(\sqrt{\dfrac{1}{2\alpha'}}(g_{\mu}p_{\nu}+g_{\nu}p_{\mu})+\dfrac{2}{\alpha'}c\eta_{\mu\nu})\partial X^{\mu}\partial X^{\nu}
+i(\sqrt{\dfrac{2}{\alpha'}}g_{\mu}+cp_{\mu})\partial^2 X^{\mu},
\end{align}
where $ g_{\mu} $ and $ c $ are an arbitrary vector and constant, respectively. 
Assuming that the polarizations $ \zeta_{\mu\nu} $ and $ \xi_{\mu} $ are written by the linear combinations of these vertices,
\begin{align}
\xi_{\mu}=&i\sqrt{\dfrac{2}{\alpha'}}g_{\mu}+icp_{\mu},\\
\zeta_{\mu\nu}=&\psi_{\mu\nu}-(\sqrt{\dfrac{1}{2\alpha'}}(g_{\mu}p_{\nu}+g_{\nu}p_{\mu})-\dfrac{2}{\alpha'}c\eta_{\mu\nu})\nonumber\\
=&\psi_{\mu\nu}+\dfrac{i}{2}(\xi_{\mu}p_{\nu}+p_{\mu}\xi_{\nu})+cp_{\mu}p_{\nu}-\dfrac{c}{\alpha'}\eta_{\mu\nu}\label{eq.polari eq},
\end{align}
we can fix $ c $ by $ \psi^{\mu}_{\mu}=0 $ and $ \psi_{\mu\nu}p^{\nu}=0 $:
\begin{align}
c=-\dfrac{\pi\alpha'^3}{40p\cdot q}
\left[
(h\cdot \zeta)(p\cdot q)-(q\cdot \zeta)(h\cdot p)
\right]^2.
\end{align}
Therefore resulting vertex in the subsubleading soft theorem is given by
\begin{align}
\psi_{\alpha\beta}&=
\dfrac{\pi\alpha'}{4p\cdot q}
\left[
2c_{\alpha}c_{\beta}-\alpha'(c\cdot p)(a_{\alpha}\zeta_{\beta}+\zeta_{\alpha}a_{\beta})
+\dfrac{\alpha'}{2}(c\cdot p)(c_{\alpha}p_{\beta}+p_{\alpha}c_{\beta})
+\dfrac{\alpha'^2}{10}(c\cdot p)^2(p_{\alpha}p_{\beta}-\dfrac{1}{\alpha'}\eta_{\alpha\beta})
\right],\\
c_{\alpha}&=(h\cdot \zeta)q_{\alpha}-(q\cdot \zeta)h_{\alpha},\\
a_{\alpha}&=(h\cdot p)q_{\alpha}-(p\cdot q)h_{\alpha}.
\end{align}

\subsection{Compact formula for soft graviton theorem }\label{sec;compact form}
We point out that the derivation of soft graviton theorem can be simplified as follows.
First, we consider the contractions of the soft and hard vertices except for the hard prefactors $ O(w) $.
\begin{align}
:V_{s}(z)::V_{hard}(w):
=&:h_{\mu}h_{\nu}X^{\mu}X^{\nu}\partial_{z}\bar{\partial_{z}}\mathrm{e}^{iq\cdot X}(z):
:O \mathrm{e}^{ip\cdot X}(w):\nonumber\\
=&:h_{\mu}h_{\nu}
\left(
X^{\mu}(z)-i\alpha'p^{\mu}\ln{|z-w|}
\right)
\left(
X^{\nu}(z)-i\alpha'p^{\nu}\ln{|z-w|}
\right)
\partial_{z}\bar{\partial_{z}}
\left(
\mathrm{e}^{iq\cdot X(z)+ip\cdot X(w)}|z-w|^{\alpha'p\cdot q}
\right):
:O(w):\nonumber\\
=&:\left [(h\cdot X(z))(h\cdot X(z))-2i\alpha'(h\cdot p)(h\cdot X(z))\ln|z-w|-\alpha'^2(h\cdot p)(h\cdot p)\left(\ln|z-w|\right)^2 \right ]\nonumber\\
&\times(\alpha'p\cdot q)^2|z-w|^{\alpha'p\cdot q-2}\left(1+\dfrac{z-w}{\alpha'p\cdot q}\partial_{z}\right)\left(1+\dfrac{\bar{z}-\bar{w}}{\alpha'p\cdot q}\bar{\partial_{z}}\right)
\mathrm{e}^{iq\cdot X(z)+ip\cdot X(w)}::O(w):.
\label{eq.new folmula}
\end{align}
Here we assume that the contractions between $ X(w) $ and $ O(w) $ should not be taken in the last line.
We can easily check the following equations.
\begin{align}
&\int_{|z|<\epsilon}d^2z |z-w|^{\alpha'p\cdot q-2}=\dfrac{2\pi}{\alpha'p\cdot q}+\mathcal{O}(1)\nonumber\\
&\int_{|z|<\epsilon}d^2z |z-w|^{\alpha'p\cdot q-2}\ln|z-w|=-\dfrac{2\pi}{(\alpha'p\cdot q)^2}+\mathcal{O}(1)\nonumber\\
&\int_{|z|<\epsilon}d^2z |z-w|^{\alpha'p\cdot q-2}\left(\ln|z-w|\right)^2=\dfrac{4\pi}{(\alpha'p\cdot q)^3}+\mathcal{O}(1)\nonumber\\
&\int_{|z|<\epsilon}d^2z |z-w|^{\alpha'p\cdot q-m}\left(\ln|z-w|\right)^n=\mathcal{O}(1)
\qquad(m\neq2, n=0,1,2)
\label{eq.integral formula}
\end{align}
Therefore, when we take contractions and expansions with respect to $ z-w $ in eq.(\ref{eq.new folmula}), only the terms that contain $ |z-w|^{\alpha'p\cdot q-2} $ are effective through subsubleading order in eq.(\ref{eq.new folmula}).
By using eq.(\ref{eq.integral formula}) eq.(\ref{eq.new folmula}) becomes
\begin{align}
\int_{|z|<\epsilon}d^2z:V_{s}(z)::V_{hard}(w):
=&\dfrac{4\pi\alpha'}{p\cdot q}
:\left[ -(h\cdot p)^2+i(h\cdot p)(h\cdot X(z))(p\cdot q)+\dfrac{(h\cdot X(z))^2(p\cdot q)^2}{2}
\right]\nonumber\\
&\times\left .\left(1+\dfrac{z-w}{\alpha'p\cdot q}iq\cdot \partial X(z)\right)
\left(1+\dfrac{\bar{z}-\bar{w}}{\alpha'p\cdot q}iq\cdot \bar{\partial} X(\bar{z})\right)
\mathrm{e}^{iq\cdot X(z)+ip\cdot X(w)}:
:O(w):
\right |_{|z-w|^{0}}\nonumber\\
&+\mathcal{O}(q^2)\nonumber\\
=&-\dfrac{4\pi\alpha'(h\cdot p)^2}{p\cdot q}
:\exp\left({i\dfrac{p\cdot q}{h\cdot p}h\cdot X(z)}+iq\cdot X(z)+ip\cdot X(w)\right)\nonumber\\
&\left .\times\left(1+\dfrac{z-w}{\alpha'p\cdot q}iq\cdot \partial X(z)\right)
\left(1+\dfrac{\bar{z}-\bar{w}}{\alpha'p\cdot q}iq\cdot \bar{\partial} X(\bar{z})\right):
:O(w):\right |_{|z-w|^{0}}+\mathcal{O}(q^2)\nonumber\\
=&\left .-\dfrac{4\pi\alpha'(h\cdot p)^2}{p\cdot q}
:\mathrm{e}^{ia\cdot X(z)+ip\cdot X(w)}
\left(1+(z-w)b\cdot \partial X(z)\right)
\left(1+(\bar{z}-\bar{w})b\cdot \bar{\partial} X(\bar{z})\right)::O(w):\right |_{|z-w|^{0}}\nonumber\\
&+\mathcal{O}(q^2),
\label{eq.new formula 2}
\end{align}
where
\begin{align}
&a^{\mu}=q^{\mu}-\dfrac{p\cdot q}{h\cdot p}h^{\mu},\\
&b^{\mu}=\dfrac{iq^{\mu}}{\alpha'p\cdot q}.
\end{align}
In the following we take $ O(w)=(\zeta_{a}\cdot \partial^{a}X(w))(\zeta_{b}\cdot\partial^{b}X(w))\cdots
(\bar{\zeta}_{c}\cdot\bar{\partial}^{c}X(\bar{w}))(\bar{\zeta}_{d}\cdot\bar{\partial}^{d}X(\bar{w}))\cdots $.
The first, second and third order terms in $ a^{\mu} $ correspond to the leading, subleading and subsubleading soft theorem, respectively.
\begin{enumerate}
	\item leading order
	
	\begin{align}
	\left .\int_{|z|<\epsilon}d^2z:V_{s}(z)::V_{hard}(w):\right |_{\text{leading}}
	=-\dfrac{4\pi\alpha'(h\cdot p)^2}{p\cdot q}
	:\mathrm{e}^{ip\cdot X(w)}
	O(w):
	\end{align}
	This is the leading soft graviton theorem.
	\item subleading order

	\begin{align}
	&\left .-\dfrac{4\pi\alpha'(h\cdot p)^2}{p\cdot q}
	:\mathrm{e}^{ip\cdot X(w)}ia\cdot X(z)
	\left(1+(z-w)b\cdot \partial X(z)\right)
	\left(1+(\bar{z}-\bar{w})b\cdot \bar{\partial} X(\bar{z})\right):
	:((\zeta_{a}\cdot \partial^{a}X(w))\cdots
	(\bar{\zeta}_{c}\cdot\bar{\partial}^{c}X(\bar{w}))\cdots:\right |_{|z-w|^{0}}\nonumber\\
	=&
	-\dfrac{4i\pi\alpha'(h\cdot p)^2}{p\cdot q}
	\left[ a\cdot X(w) 
	+\dfrac{\alpha'}{2}(a\cdot \zeta_{a})(b\cdot \partial^{a}X(w))
	-\dfrac{\alpha'}{2}(b\cdot\zeta_{a})(a\cdot \partial^{a}X(w))
	+\dfrac{\alpha'}{2}(a\cdot \bar{\zeta}_{c})(b\cdot \bar{\partial}^{c}X(w))
	-\dfrac{\alpha'}{2}(b\cdot\bar{\zeta}_{c})(a\cdot \bar{\partial}^{c}X(w))
	\right].
	\end{align}
	Here we omit the unchanged operators. This is subleading soft graviton theorem.
	In fact, the first term in square bracket represents the orbital angular momentum operator. The second and third terms (the fourth and fifth terms) are combined to the spin angular momentum operator for the holomorphic part.
	
	\item subsubleading order
	\begin{align}
	&\dfrac{2\pi\alpha'(h\cdot p)^2}{p\cdot q}
	:\mathrm{e}^{ip\cdot X(w)}(a\cdot X(z))^2
	\left(1+(z-w)b\cdot \partial X(z)\right)
	\left(1+(\bar{z}-\bar{w})b\cdot \bar{\partial} X(\bar{z})\right):\nonumber\\
	&\left .:(\zeta_{a}\cdot \partial^{a}X(w))(\zeta_{b}\cdot\partial^{b}X(w))\cdots
	(\bar{\zeta}_{c}\cdot\bar{\partial}^{c}X(\bar{w}))(\bar{\zeta}_{d}\cdot\bar{\partial}^{d}X(\bar{w}))\cdots:\right |_{|z-w|^{0}}\nonumber\\
	=&\dfrac{2\pi\alpha'(h\cdot p)^2}{p\cdot q}
	\left[
	(a\cdot X)^2
	+\left(-\dfrac{\alpha'}{2}\sum_{k=0}^{a}\dfrac{a!}{k!(a-k)!}(b\cdot\zeta_{a})(a\cdot\partial^{k}X)(a\cdot\partial^{a-k}X)
	+(a\rightarrow b)\right)\right.\nonumber\\
	+&\left(-\dfrac{\alpha'}{2}\sum_{k=0}^{c}\dfrac{c!}{k!(c-k)!}(b\cdot\bar{\zeta}_{c})(a\cdot\bar{\partial}^{k}X)(a\cdot\bar{\partial}^{c-k}X)
	+(c\rightarrow d)\right)\nonumber\\
	&\left .+\left(
	\dfrac{\alpha'^2}{2}(b\cdot\zeta_{a})(b\cdot\bar{\zeta}_{c})(a\cdot\partial^{a}X)(a\cdot\bar{\partial}^{n}X)
	+(a\rightarrow b)+(c\rightarrow d)+(a\rightarrow b, c\rightarrow d)
	\right)\right .\nonumber\\
	+&\left [\alpha'(a\cdot\zeta_{a})
	\left\{
	\dfrac{a\cdot\partial^{a}X}{a}
	+\sum_{k=0}^{a-1}\dfrac{(a-1)!(a\cdot\partial^{k}X)(b\cdot\partial^{a-k}X)}{k!(m-k-1)!}
	-\dfrac{\alpha'(a-1)!b!(b\cdot\zeta_{b})(a\cdot\partial^{a+b}X)}{2(a+b)!}
	-\dfrac{\alpha'}{2}(b\cdot\bar{\zeta_{c}})(b\cdot\partial^{a}X)(a\cdot\bar{\partial}^{n}X)
	+(c\rightarrow d)
	\right\}\right .\nonumber\\
	&+\left .(a\rightarrow b)\right ]\nonumber\\
	+&\left [\alpha'(a\cdot\bar{\zeta}_{c})
	\left\{
	\dfrac{a\cdot\bar{\partial}^{c}X}{c}
	+\sum_{k=0}^{c-1}\dfrac{(c-1)!(a\cdot\bar{\partial}^{k}X)(b\cdot\bar{\partial}^{c-k}X)}{k!(c-k-1)!}
	-\dfrac{\alpha'(c-1)!d!(b\cdot\bar{\zeta}_{d})(a\cdot\bar{\partial}^{c+d}X)}{2(c+d)!}
	-\dfrac{\alpha'}{2}(b\cdot\zeta_{a})(b\cdot\bar{\partial}^{c}X)(a\cdot\partial^{a}X)
	+(a\rightarrow b)\right\}\right .\nonumber\\
	&+\left .\left (c\rightarrow d\right )\right ] 
	\nonumber\\
	&\left .+\dfrac{\alpha'^2}{2}\dfrac{(a-1)!(b-1)!}{(a+b-1)!}(a\cdot\zeta_{a})(a\cdot\zeta_{b})(b\cdot\partial^{a+b}X)
	+\dfrac{\alpha'^2}{2}\dfrac{(c-1)!(d-1)!}{(c+d-1)!}(a\cdot\bar{\zeta}_{c})(a\cdot\bar{\zeta}_{d})(b\cdot\bar{\partial}^{c+d}X)
	+\dfrac{\alpha'^2}{2}(a\cdot\zeta_{a})(a\cdot\bar{\zeta}_{c})(b\cdot\partial^{a}X)(b\cdot\bar{\partial}^{c}X)
	\right]\nonumber\\
	=&\dfrac{2\pi\alpha'(h\cdot p)^2}{p\cdot q}
	\left[
	(a\cdot X)^2
	-\alpha'(a\cdot X)\left (\left\{
	(b\cdot\zeta_{a})(a\cdot\partial^{a}X)-(a\cdot\zeta_{a})(b\cdot\partial^{a}X)
	\right\}+(a\rightarrow b)\right )\right .\nonumber\\
	&-\alpha'(a\cdot X)\left (\left\{
	(b\cdot\bar{\zeta}_{c})(a\cdot\bar{\partial}^{c}X)-(a\cdot\bar{\zeta}_{c})(b\cdot\bar{\partial}^{c}X)
	\right\}+(c\rightarrow d)\right )
	\nonumber\\
	&+\dfrac{\alpha'^2}{2}\left [
	\left\{(b\cdot\zeta_{a})(a\cdot\partial^{a}X)-(a\cdot\zeta_{a})(b\cdot\partial^{a}X)
	\right\}
	\left\{(b\cdot\bar{\zeta}_{c})(a\cdot\bar{\partial}^{c}X)-(a\cdot\bar{\zeta}_{c})(b\cdot\bar{\partial}^{c}X)
	\right\}+(a\rightarrow b)+(c\rightarrow d)+(a\rightarrow b, c\rightarrow d)\right ]\nonumber\\
	&+\left(\dfrac{\alpha'(a\cdot\zeta_{a})(a\cdot\partial^{a}X)}{a}+(a\rightarrow b) \right)
	+\left(\dfrac{\alpha'(a\cdot\bar{\zeta}_{c})(a\cdot\bar{\partial}^{c}X)}{c}+(c\rightarrow d) \right)\nonumber\\
	&-\dfrac{\alpha'}{2}\left (\sum_{k=1}^{a-1}\dfrac{a!}{k!(a-k)!}(a\cdot\partial^{k}X)
	\left((b\cdot\zeta_{a})(a\cdot\partial^{a-k}X)-(a\cdot\zeta_{a})(b\cdot\partial^{a-k}X) \right)+(a\rightarrow b)
	\right )\nonumber\\
	&-\dfrac{\alpha'}{2}\left (\sum_{k=1}^{c-1}\dfrac{c!}{k!(c-k)!}(a\cdot\bar{\partial}^{k}X)
	\left((b\cdot\bar{\zeta}_{c})(a\cdot\bar{\partial}^{c-k}X)-(a\cdot\bar{\zeta}_{c})(b\cdot\bar{\partial}^{c-k}X) \right)+(c\rightarrow d)
	\right )\nonumber\\
	&\left .+\dfrac{\alpha'^2(a-1)!(b-1)!}{2(a+b)!}
	\left(
	a(a\cdot\zeta_{b})\left\{
	(a\cdot\zeta_{a})(b\cdot\partial^{a+b}X)-(b\cdot\zeta_{a})(a\cdot\partial^{a+b}X)
	\right\}
	+b(a\cdot\zeta_{a})\left\{
	(a\cdot\zeta_{b})(b\cdot\partial^{a+b}X)-(b\cdot\zeta_{b})(a\cdot\partial^{a+b}X)
	\right\}
	\right)
	\right]
	\label{new formula subsubleading}
	\end{align}
	When we substitute $ a^{\mu}=q^{\mu}-\dfrac{p\cdot q}{h\cdot p}h^{\mu}$ and $b^{\mu}=\dfrac{iq^{\mu}}{\alpha'p\cdot q} $ into this equation, we find that eq.(\ref{new formula subsubleading}) agrees with eq.(\ref{eq.graviton subsubleading}).
	
	Furthermore, we can rewrite this result in the operator formalism and obtain the compact form as follows:
	\begin{align}
	&\dfrac{2\pi\alpha'(h\cdot p)^2}{p\cdot q}
	:\mathrm{e}^{ip\cdot X(w)}(a\cdot X(z))^2
	\left(1+(z-w)b\cdot \partial X(z)\right)
	\left(1+(\bar{z}-\bar{w})b\cdot \bar{\partial} X(\bar{z})\right):\nonumber\\
	&\left .:(\zeta_{a}\cdot \partial^{a}X(w))(\zeta_{b}\cdot\partial^{b}X(w))\cdots
	(\bar{\zeta}_{c}\cdot\bar{\partial}^{c}X(\bar{w}))(\bar{\zeta}_{d}\cdot\bar{\partial}^{d}X(\bar{w}))
	:\right |_{|z-w|^{0}}\nonumber\\
	\rightarrow&\dfrac{2\pi\alpha'}{p\cdot q}
	\left[
	q_{a}q_{b}\left(L^{\mu a}L^{\nu b}+S^{\mu a}\bar{S^{\nu b}}
	\right)
	+Q+\bar{Q}
	\right ]
	\left|\text{hard massive state};k\right> ,
	\end{align}
	where
	\begin{align}
	&Q\equiv \left .\left(\sqrt{\dfrac{\alpha'}{2}}\sum_{k\neq 0}\dfrac{(a\cdot \alpha_{k})}{kz^{k}}\right)^2
	\left (1+\sqrt{\dfrac{\alpha'}{2}}\sum_{k\neq 0}\dfrac{b\cdot\alpha_{k}}{iz^{k}} \right )\right |_{z^0},\nonumber\\
	&\bar{Q}\equiv \left .\left(\sqrt{\dfrac{\alpha'}{2}}\sum_{k\neq 0}\dfrac{(a\cdot \alpha_{k})}{k\bar{z}^{k}}\right)^2
	\left (1+\sqrt{\dfrac{\alpha'}{2}}\sum_{k\neq 0}\dfrac{b\cdot\alpha_{k}}{i\bar{z}^{k}} \right )\right |_{\bar{z}^0},\nonumber\\
	&\left|\text{hard massive state};p\right>
	=:(i\sqrt{\dfrac{\alpha'}{2}}(a-1)!\zeta_{a}\cdot \alpha_{-a})\cdots
	(i\sqrt{\dfrac{\alpha'}{2}}(c-1)!\bar{\zeta}_{c}\cdot\bar{\alpha}_{-c})\cdots:
	\left|\text{vacuum},p\right>.
	\end{align}
\end{enumerate}

\section{Soft B field theorem}\label{sec;bfield}
\quad The soft B field theorem has been examined in \cite{DiVecchia:2017gfi}, and its universal behavior is determined through subleading order. This can also be seen in our formulation. We can rewrite the B field vertex operator as follows:
\begin{align}
&:\partial X^{\mu}(z)\bar{\partial}X^{\nu}\exp\left(iq\cdot X(z,\bar{z})\right):\nonumber\\
=&\partial:X^{\mu}(z,\bar{z})\bar{\partial}X^{\nu}(\bar{z})\exp\left(iq\cdot X(z,\bar{z})\right):
-:X^{\mu}(z,\bar{z})\bar{\partial}X^{\nu}(z,\bar{z})\partial_{z} \exp\left(iq\cdot X(z,\bar{z})\right):.
\label{eq.soft B field vertex}
\end{align}
For the B field we cannot further rewrite it by using partial integration. 
We find that the contribution from the bulk is $ \mathcal{O}(q) $. Therefore the soft theorem holds universally through subleading order, and is obtained in a similar manner to the previous subsection.
First, we rewrite the soft vertex as
\begin{align}
-h_{\mu\nu}X^{\mu}(z,\bar{z})\bar{\partial}X^{\nu}(\bar{z})\partial \exp(iq\cdot X(z,\bar{z}))
=\lim_{\xi,\omega \to 0}\lim_{z'\to z}\dfrac{\partial}{\partial \xi_{\mu}}\dfrac{\partial}{\partial \omega_{\nu}}h_{\mu\nu}\partial_{z}\bar{\partial}_{\bar{z}}\exp(iq\cdot X(z,\bar{z})+i\xi\cdot X(z',\bar{z'})+i\omega\cdot \bar{\partial}X(\bar{z}')).
\end{align}
The details are given in Appendix \ref{sec;Bfield} and the result is
\begin{align}
&-\int_{|z|<\epsilon}h_{\mu\nu}:X^{\mu}(z,\bar{z})\bar{\partial}X^{\nu}(z,\bar{z})\partial_{z} \exp\left(iq\cdot X(z,\bar{z})\right)::V(w):\nonumber\\
=&\dfrac{-i\pi\alpha'h_{\mu\nu}}{2}\left(\dfrac{p^{\nu}q_{a}(S^{\mu a}-\bar{S}^{\mu a})-(\mu\leftrightarrow\nu)}{p\cdot q}+\dfrac{S^{\mu\nu}-\bar{S}^{\mu\nu}}{2} \right):V(w):+\mathcal{O}(q^2),
\end{align}
where $ V(w) $ is any hard vertex operator.

\section{Soft theorem in other string theories}\label{superstring}
\subsection{superstring}
\quad The soft theorems in superstring theory have been examined in various contexts \cite{Bianchi:2014gla}\cite{DiVecchia:2016szw}\cite{Sen:2017nim}. They can be seen also from our formulation. 
 
The vertex operator of the graviton or dilaton in (0,0) picture can be written as 
\begin{align}
h_{\mu\nu}:\left(i\partial X^{\mu}(z)-\dfrac{\alpha'}{2}q_{\alpha}\psi^{\mu}(z)\psi^{\alpha}(z)\right)
\left(i\bar{\partial} X^{\nu}(\bar{z})-\dfrac{\alpha'}{2}q_{\beta}\tilde{\psi}^{\nu}\tilde{\psi}^{\beta}\right)
\exp\left(iq\cdot X(z,\bar{z})\right):.\label{eq.supergraviton}
\end{align}
Note that $ \psi^{\mu}(z) \psi^{\alpha}(z) $ can be regarded as the spin angular momentum operator. We classify the terms in eq.(\ref{eq.supergraviton}) as follows:\\
\begin{enumerate}
 \item The product of the bosonic parts is the same as in the previous section. It gives the momentum and the angular momentum as well as the other terms in eq.(\ref{eq.soft graviton theorem for massive}).\\
 \item The product of the fermionic parts is quadratic in q and contribute only to the subsubleading soft theorem.
   This gives the product of the spin angular momenta of the holomorphic and antiholomorphic parts of the hard vertices.\\
 \item For the products of the bosonic and fermionic part, we perform partial integration in the bosonic part as in the previous section.
 \begin{align}
 \partial X^{\mu}(z)\exp\left(iq\cdot X(z,\bar{z})\right)=\partial\left(X^{\mu}\exp\left(iq\cdot X(z,\bar{z})\right)\right)-X^{\mu}(z,\bar{z})\partial \exp\left(iq\cdot X(z,\bar{z})\right).
 \end{align}
 Because the fermionic part is multiplied by q, the contribution from the bulk is at least $ \mathcal{O}(q^2) $.
 Therefore, these terms contribute to the subleading and the subsubleading soft theorem.
\end{enumerate}

As an example, let's consider the contraction of the soft graviton with a hard dilatino or gravitino in NS-R sector.
The vertex operator in $ (-1,-\dfrac{1}{2}) $ picture in the NS - R sector is given by
\begin{align}
V(w,\bar{w})=u^{\alpha}_{\rho}\psi^{\rho}(w)\exp(-\phi(w))\tilde{S_{\alpha}}(\bar{w})\exp\left (-\dfrac{\tilde{\phi}(\bar{w})}{2}\right )\exp\left(ip\cdot X(w,\bar{w})\right),
\end{align}
where $ u^{\alpha}_{\mu} $ is a polarization, $ \phi,\tilde{\phi} $ are superghosts, $ \tilde{S_{\alpha}} $ generates the ground state in R sector. In this case the prefactors are purely fermionic, so the spin angular momentum comes from the fermionic part and the orbital angular momentum from the bosonic part separately.

The contraction rules are as follows:
\begin{align}
\psi^{\mu}(z)\psi^{\nu}(w)\sim \dfrac{\eta^{\mu\nu}}{z-w},\\
\psi^{\mu}(z)S_{\alpha}(w)\sim \dfrac{1}{\sqrt{2(z-w)}}(\Gamma^{\mu})_{\alpha}^{ \ \beta}S_{\beta}(w),\\
\phi(z)\phi(w)\sim -\ln(z-w).
\end{align}
The contributions from the above 1. $ \sim $ 3. are as follows:
\begin{enumerate}
	\item This is essentially the same as in the hard tachyon in bosonic string theory:
	\begin{align}
	-\dfrac{\pi\alpha'h_{\mu\nu}}{4p\cdot q}
	\left( 2p^{\mu}p^{\nu}+2q_{a}p^{\mu}L^{\nu a}+q_{a}q_{b}L^{\mu a}L^{\nu b}\right )u^{\alpha}_{\mu}\psi^{\mu}(w)\exp(-\phi(w))\tilde{S_{\alpha}}(\bar{w})\exp\left (-\dfrac{\tilde{\phi}(\bar{w})}{2}\right )
	\exp \left( ip\cdot X(w,\bar{w})\right).
	\end{align}
	\item The holomorphic part gives
	\begin{align}
	-\dfrac{\alpha'}{2}q_{a}\psi^{\mu}(z)\psi^{a}(z)\psi^{\rho}(w)
	\sim -\dfrac{\alpha'(\eta^{\mu\rho}\psi^{a}-\eta^{a\rho}\psi^{\mu})}{2(z-w)}
	=-\dfrac{\alpha'q_{a}(S^{\mu a})^{\rho}_{ \ m}\psi^{m}(w)}{2(z-w)},\label{eq.super holomorphic spin}
	\end{align}
	while the antiholomorphic part gives
	\begin{align}
	-\dfrac{\alpha'}{2}q_{b}\tilde{\psi}^{\mu}\tilde{\psi}^{b}\tilde{S_{\alpha}}(\bar{w})
	\sim -\dfrac{\alpha'q_{b}(\Gamma^{b})_{\beta}^{ \ \gamma}(\Gamma^{\nu})_{\gamma}^{ \ \delta}\tilde{S_{\delta}}(\bar{w})}{4(z-w)}
	=-\dfrac{\alpha'q_{b}(S^{\nu b})_{\beta}^{ \ \delta}\tilde{S_{\delta}}}{2(\bar{z}-\bar{w})}.\label{eq.super antiholomorphic spin}
	\end{align}
	Taking the products of these results and integrating over z, we obtain
	\begin{align}
	\dfrac{\pi\alpha'q_{a}q_{b}S^{\mu a}S^{\nu b}}{2p\cdot q}
	u^{\alpha}_{\rho}\psi^{\rho}(w)\exp(-\phi(w))\tilde{S_{\alpha}}(\bar{w})\exp\left (-\dfrac{\tilde{\phi}(\bar{w})}{2}\right )\exp\left(ip\cdot X(w,\bar{w})\right).
	\end{align}
	
	\item For the bosonic part we have
	\begin{align}
	& :iX^{\mu}(z,\bar{z})\partial \exp\left(iq\cdot X(z,\bar{z})\right)::\exp\left(ip\cdot X(w,\bar{w})\right):\nonumber\\
	=&i\lim_{\xi \to 0}\lim_{z'\to z}\dfrac{\partial}{\partial \xi_{\mu}}:\exp\left(iq\cdot X(z,\bar{z})+i\xi\cdot X(z',\bar{z'})\right)::\exp\left(ip\cdot X(w,\bar{w})\right):\nonumber\\
	=&i\dfrac{\alpha'p\cdot q}{2(z-w)}\dfrac{\partial}{\partial q_{\mu}}
	:\left(\dfrac{|z-w|^{\alpha'p\cdot q}\exp\left(iq\cdot X(z,\bar{z})\right)}{\alpha'p\cdot q}  \right)\exp\left(ip\cdot X(w,\bar{w})\right):.
	\end{align}
	After the z integration this gives the momentum and the orbital angular momentum.
	
	For the fermionic part, by the same calculation as eq.(\ref{eq.super holomorphic spin}) and eq.(\ref{eq.super antiholomorphic spin}), we get the spin angular momentum for each of the holomorphic and antiholomorphic parts. \\
	Combining these results , we obtain
	\begin{align}
	-\dfrac{\pi \alpha'h_{\mu\nu} (ip^{\mu}+iq_{a}L^{\mu a})q_{b}S^{\mu b}}{2p\cdot q}
	u^{\alpha}_{\rho}\psi^{\rho}(w)\exp(-\phi(w))\tilde{S_{\alpha}}(\bar{w})\exp\left (-\dfrac{\tilde{\phi}(\bar{w})}{2}\right )\exp\left(ip\cdot X(w,\bar{w})\right).
	\end{align}
\end{enumerate}
Summing up these contributions, we can express the soft theorem as follows:
\begin{align}
&-\dfrac{\pi\alpha'h_{\mu\nu}}{4p\cdot q}
\left( 2p^{\mu}p^{\nu}+2q_{a}p^{\mu}J^{\nu a}+q_{a}q_{b}J^{\mu a}J^{\nu b}-q_{a}q_{b}S^{\mu a}S^{\nu b}-q_{a}q_{b}\bar{S}^{\mu a}\bar{S}^{\nu b}\right )\nonumber\\
&\times u^{\alpha}_{\mu}\psi^{\mu}(w)\exp(-\phi(w))\tilde{S_{\alpha}}(\bar{w})\exp\left (-\dfrac{\tilde{\phi}(\bar{w})}{2}\right )
\exp \left( ip\cdot X(w,\bar{w})\right).\label{eq.superstring soft theorem}
\end{align}
As in eq.(\ref{eq.spinspin vanish}) the same combination of the spin operators vanishes when it acts on a single spinor :

\begin{align}
h_{\mu\nu}q_{a}q_{b}(S^{\mu a})^{i}_{j}(S^{\nu b})^{j}_{k}=0,\label{eq.spinspin vanish for spinor}
\end{align}
where $ i,j,k $ are spinor indexes\footnote{Eq.(\ref{eq.spinspin vanish for spinor}) holds for dilaton as well as graviton contrary to the case of a vector index in eq.(\ref{eq.spinspin vanish})}. Therefore the fourth and fifth terms are 0 and eq.(\ref{eq.superstring soft theorem}) is written in terms of the total angular momentum for soft graviton.

In superstring theory, if there is no bosonic prefactor $ \partial^{m} X$ or $\bar{\partial}^{m}X $, there appears neither the higher order correction in $ \alpha' $ nor a mixing with different vertices, because the spin angular momentum operator comes only from the fermionic part.

\subsection{heterotic string}
\quad In this section we discuss the soft theorems for gauge bosons in heterotic string theory. The vertex operator of the gauge boson is given by
\begin{align}
V_{gauge}(z)=\zeta_{a\mu}j^{a}(z)\bar{\partial}X^{\mu}(\bar{z})\exp\left(ik\cdot X(z,\bar{z}) \right).
\end{align}
Here $ j^{a} $ is a holomorphic (1,0) operator that satisfies the current algebra:
\begin{align}
j^{a}(z)j^{b}(w)\sim \dfrac{k^{ab}}{(z-w)^2}+\dfrac{ic^{ab}_{c}}{z-w}j^{c}(w),
\end{align}
where $ k^{ab},c^{ab}_{c} $ are constants.

As in the previous sections, we rewrite $ V_{gauge} $ as 
\begin{align}
V_{gauge}(z)=\bar{\partial}\left(\zeta_{a\mu}j^{a}(z)\exp\left(ik\cdot X(z,\bar{z}) \right) \right)
-\zeta_{a\mu}j^{a}(z)\bar{\partial}\exp\left(ik\cdot X(z,\bar{z}) \right).
\end{align}
We can drop the first term because it is a total derivative. The contribution from the bulk of the second term is $ \mathcal{O}(q) $. Therefore the soft theorem holds through the 0-th order in q.

When we evaluate the OPE with a hard vertex, the holomorphic part gives the generator of the gauge symmetry with a pole $ \dfrac{1}{z-w} $.
The antiholomorphic part has exactly the same form as in the case of superstring, and it gives the momentum, the orbital angular momentum and the spin angular momentum.

This is consistent with the results \cite{low}\cite{Casali:2014xpa}\cite{Bern:2014vva} that the soft gauge boson theorem at the tree level is universal through subleading order.

\section{Conclusion}
\quad We have discussed the soft theorem in string theory in terms of the OPE of the soft and hard vertices.
When the soft vertex is expanded with the soft momentum after some partial integration, it turns out to be a superlocal operator through a certain order in q. As a result, we find that the scattering amplitude is evaluated only by the local integral around the hard vertices through that order of the soft momentum, which leads to the universal soft behavior. \\

We have confirmed that the soft behavior of massless particles of bosonic closed string, closed superstring and heterotic string can actually be reproduced by that method. When the hard vertex represents a massive particle, we find that the subsubleading soft theorem can no longer be written in terms of the total angular momentum. \\

It is known that the universal behavior of the soft theorem breaks down when loop effects are taken into account \cite{Sen:2017nim}\cite{Bern:2014oka}\cite{He:2014bga}. It is interesting to see whether loop effects can be evaluated with the superlocal operator. Because the loop effect can be examined by factorizing diagrams in field theory, we expect that the loop effects appear as the pinches of the world-sheet in string theory.\\

In this paper we have considered scattering amplitudes with one soft particle. It should be able to consider the scattering amplitudes with more than one soft particle in this formulation. \\

Although we have obtained a simple picture of soft theorems based on the superlocal operator, we still do not have a brief explanation why the total angular momentum operator emerges.

\section*{Acknowledgements}
We would like to thank Yu-tin Huang, Takeshi Morita, Sotaro Sugishita and Yuta Hamada for valuable discussions. 
SH thanks to the organizers and participants of the workshop, "Infrared physics of gauge theories and quantum dynamics of inflation".

\appendix
\section{Analogy with field theory}\label{simple}
\quad In this appendix we show that the soft theorem has the same structure as field theory, if we ignore the higher order corrections of $\alpha'$ and the mixing with different vertex operators.
Note that here we consider the soft vertex before partial integration. 

Let's calculate the OPE between the soft graviton vertex $ V_{soft}(z,\bar{z})=h_{\mu\nu}\partial X^{\mu}(z)\bar{\partial}X^{\nu}(\bar{z})\exp(iq\cdot X(z,\bar{z})) $ and a hard vertex. For simplicity we take a graviton as the hard vertex operator: $ V_{hard}(w,\bar{w})=A_{\rho\sigma}\partial X^{\rho}(w)\bar{\partial}X^{\sigma}(\bar{w})\exp(ip\cdot X(w,\bar{w}))$.
In the following we omit polarization tensors. \\

First, we consider the contraction between the exponential functions in the soft and the hard vertices.

\begin{align}
&\int d^2z\partial X^{\mu}(z) \overline{\partial}X^{\nu}(\bar{z}) \exp\left( iq\cdot X\left( z,\bar{z}\right) \right) \partial X^{\rho}(w) \overline{\partial}X^{\sigma}(\bar{w}) \exp\left( ip\cdot X_{G}\left(w,\bar{w}\right) \right)\nonumber \\
=&\int d^2z|z-w|^{\alpha' q\cdot p} \partial X^{\mu}(z) \overline{\partial}X^{\nu}(\bar{z}) \partial X^{\rho}(w) \overline{\partial}X^{\sigma}(\bar{w}):\exp{\left( iq\cdot X\left( z,\bar{z}\right) +ip\cdot X\left(w,\bar{w}\right)\right)}:.\label{eq.field theory}
\end{align}
Here we assume that we do not take any contraction among the operators defined on the same points, z or w, even if the symbol of the normal ordering is not explicitly written.
Then we expand the exponential functions with the power of q. In order to indicate that the expanded operators should not be contracted with $\exp{\left(ip \cdot X \left(w,\bar{w} \right) \right)} $ anymore, we denote them by the symbol $ X'(z, \bar{z}) $. We classify the contributions to eq.(\ref{eq.field theory}) into the following five cases by the order of q and the integration regions.

\begin{enumerate}
\item \uline{The contribution of the 0-th order in q from the disk around the hard vertex}\label{0v}\\
It is given by
\begin{align}
&\int_{|z|<\epsilon} d^2z\partial X^{\mu}(z)\bar{\partial}X^{\nu}(\bar{z})\partial X^{\rho}(w)\bar{\partial}X^{\sigma}(\bar{w})\exp(ip\cdot X(w,\bar{w}))\nonumber\\
\sim &\dfrac{-\pi\alpha' \epsilon^{\alpha'p\cdot q}p^{\mu}p^{\nu}}{2p\cdot q}\partial X^{\rho}(w)\bar{\partial}X^{\sigma}(\bar{w})\exp(ip\cdot X(w,\bar{w}))\nonumber\\
\sim &\dfrac{-\pi\alpha' p^{\mu}p^{\nu}}{2p\cdot q}\partial X^{\rho}(w)\bar{\partial}X^{\sigma}(\bar{w})\exp(ip\cdot X(w,\bar{w})),
\end{align}
where we have ignored the higher order terms in $ \alpha' $. This gives the leading soft theorem.

\item \uline{The contribution of the 0-th order in q from the bulk} \label{0b}
\\
Because the singularity in q is not yielded from the bulk, this contribution gives a part of the subleading soft theorem. In fact, the soft vertex operator at this order is a total derivative with respect to z, and we can rewrite it to the contour integral around the hard vertex as follows:
\begin{align}
\int_{bulk} d^2z \partial X^{\mu}\bar{\partial}X^{\nu}(\bar{z})V(w_{1},\bar{w_{1}})\cdots V(w_{n},\bar{w_{n}})
=-\dfrac{i}{2}\sum_{i=1}^{n}\oint_{|z-w_{i}|=\epsilon}d\bar{z}X^{\mu}(z,\bar{z})\bar{\partial}X^{\nu}(\bar{z})V(w_{1},\bar{w_{1}})\cdots V(w_{n},\bar{w_{n}}).
\end{align}
Then by looking at the pole in the antiholomorphic part of the OPE, we have
\begin{align}
&-\dfrac{i}{2}\oint_{|z-w|=\epsilon}d\bar{z}X^{\mu}(z,\bar{z})\bar{\partial}X^{\nu}(\bar{z})\partial X^{\rho}(w)\bar{\partial}X^{\sigma}(\bar{w})\exp(ip\cdot X(w,\bar{w}))\nonumber\\
\sim&\dfrac{i\pi\alpha'p^{\nu}}{2}\dfrac{\partial}{i\partial p_{\mu}}
\left(\partial X^{\rho}(w)\bar{\partial}X^{\sigma}(\bar{w})\exp(ip\cdot X(w,\bar{w}))\right).
\end{align}
This is a part of the orbital angular momentum of the subleading soft theorem. The fact that a part of the angular momentum comes out from the bulk is similar to the structure in field theory.

\item \uline{The contribution of the first order in q from the disk around the hard vertex} \label{1v}\\
It is given by
\begin{align}
&iq_{\alpha}\int_{|z|<\epsilon} d^2z\partial X^{\mu}(z)\bar{\partial}X^{\nu}(\bar{z})X'^{\alpha}(z,\bar{z})
\partial X^{\rho}(w)\bar{\partial}X^{\sigma}(\bar{w})\exp(ip\cdot X(w,\bar{w})).
\end{align}
As in the above cases, we take the OPE, and then ignore the higher order terms in $ \alpha' $ and the mixing with different vertex operators. We have
\begin{align}
\sim -\dfrac{\pi\alpha'}{2 q \cdot p}\exp(ip\cdot X(w,\bar{w}))
&\left[
iq_{\alpha}p^{\mu}p^{\nu}\dfrac{\partial}{i\partial p_{\alpha}}
\partial X^{\rho}(w)\bar{\partial}X^{\sigma}(\bar{w})
+
q_{\alpha}p^{\mu}_{k}\eta^{\nu \sigma}
\partial X^{\rho}(w)\bar{\partial}X^{\alpha}(\bar{w})
-
p^{\mu}_{k}q^{\sigma}\partial X^{\rho}(w)\bar{\partial}X^{\nu}(\bar{w})\right. \nonumber\\
&\left .+
q_{\alpha}p^{\nu}_{k}\eta^{\mu \rho}\partial X^{\alpha}(w)\bar{\partial}X^{\sigma}(\bar{w})
-
p^{\nu}_{k}q^{\rho}\partial X^{\mu}(w)\bar{\partial}X^{\sigma}(\bar{w})
\right] .
\end{align}

\item \uline{The contribution of the first order in q from the bulk}\label{1b}

By the symmetry of the polarization tensor, we can rewrite the soft vertex operator as follows:
\begin{align}
iq_{\alpha}\partial X^{\mu}(z)\bar{\partial}X^{\nu}(\bar{z})X'^{\alpha}(z,\bar{z})
=iq_{\alpha}\partial \left(X^{\mu}(z,\bar{z})\bar{\partial}X^{\nu}X'^{\alpha}(z,\bar{z} )\right)
-\dfrac{iq_{\alpha}}{2}\bar{\partial}\left(X^{\mu}(z,\bar{z})X^{\nu}(z,\bar{z})\partial X'^{\alpha}(z)\right).
\end{align}
As in \ref{0b}., the z integration can be replaced by contour integrals around the hard vertices:
\begin{align}
&\int_{bulk} d^2z \partial X^{\mu}\bar{\partial}X^{\nu}(\bar{z})iq_{\alpha}X'^{\alpha}(z,\bar{z})
V(w_{1},\bar{w_{1}})\cdots V(w_{n},\bar{w_{n}})\\
&=\dfrac{q_{\alpha}}{2}\sum_{i=1}^{n}\oint_{|z-w_{i}|=\epsilon}d\bar{z}X^{\mu}(z,\bar{z})\bar{\partial}X^{\nu}(\bar{z})X'^{\alpha}(z,\bar{z})
V(w_{1},\bar{w_{1}})\cdots V(w_{n},\bar{w_{n}})\nonumber\\
&+\dfrac{q_{\alpha}}{4}\sum_{i=1}^{n}\oint_{|z-w_{i}|=\epsilon}dzX^{\mu}(z,\bar{z})X^{\nu}(z,\bar{z})\partial X'^{\alpha}(z)V(w_{1},\bar{w_{1}})\cdots V(w_{n},\bar{w_{n}})
\end{align}
When a hard vertex is a graviton, we get
\begin{align}
&\dfrac{q_{\alpha}}{2}\oint_{|z-w|=\epsilon}d\bar{z}X^{\mu}(z,\bar{z})\bar{\partial}X^{\nu}(\bar{z})X^{\alpha}(z,\bar{z})
\partial X^{\rho}(w)\bar{\partial}X^{\sigma}(\bar{w})\exp(ip\cdot X(w,\bar{w}))\nonumber\\
&+\dfrac{q_{\alpha}}{4}\sum_{i=1}^{n}\oint_{|z-w_{i}|=\epsilon}dzX^{\mu}(z,\bar{z})X^{\nu}(z,\bar{z})\partial X^{\alpha}(z)
\partial X^{\rho}(w)\bar{\partial}X^{\sigma}(\bar{w})\exp(ip\cdot X(w,\bar{w}))\nonumber\\
&=-\dfrac{i\pi\alpha'q_{\alpha}}{2}
\left[
-ip^{\nu}\dfrac{\partial}{i\partial p_{\mu}}\dfrac{\partial}{i\partial p_{\alpha}}\partial X^{\rho}(w)\bar{\partial}X^{\sigma}(\bar{w})
-\eta^{\nu \sigma}\dfrac{\partial}{i\partial p_{\mu}}\partial X^{\rho}(w)\bar{\partial}X^{\alpha}(\bar{w})\right .\nonumber\\
&+\eta^{\alpha \sigma}\dfrac{\partial}{i\partial p_{\mu}}\partial X^{\rho}(w)\bar{\partial}X^{\nu}(\bar{w})
+\dfrac{1}{2}ip^{\alpha}\dfrac{\partial}{i\partial p_{\mu}}\dfrac{\partial}{i\partial p_{\nu}}\partial X^{\rho}(w)\bar{\partial}X^{\sigma}(\bar{w})\nonumber\\
&\left .+\eta^{\alpha \rho}\dfrac{\partial}{i\partial p_{\mu}}\partial X^{\nu}(w)\bar{\partial}X^{\sigma}(\bar{w})
-\eta^{\mu \rho}\dfrac{\partial}{i\partial p_{\nu}}\partial X^{\alpha}(w)\bar{\partial}X^{\sigma}(\bar{w})
\right]\exp(ip\cdot X(w,\bar{w}))
\end{align}
This contributes to subsubleading soft theorem.

\item \uline{The contribution of the second order in q from the disk around the hard vertex operators} \label{[2v]}\\
It is given by
\begin{align}
&\dfrac{iq_{\alpha}iq_{\beta}}{2}\int_{|z|<\epsilon} d^2z\partial X^{\mu}(z)\bar{\partial}X^{\nu}(\bar{z})X'^{\alpha}(z,\bar{z})X'^{\beta}(z,\bar{z})
\partial X^{\rho}(w)\bar{\partial}X^{\sigma}(\bar{w})\exp(ip\cdot X(w,\bar{w}))\nonumber\\
\sim& -\dfrac{\pi\alpha'q_{\alpha}q_{\beta}}{4p\cdot q}
\left[
-p^{\mu}p^{\nu} \dfrac{\partial}{i\partial p_{\alpha}}\dfrac{\partial}{i\partial p_{\beta}}
\partial X^{\rho}(w)\bar{\partial}X^{\sigma}(\bar{w}) +
2ip^{\mu}\eta^{\nu \sigma}
\dfrac{\partial}{i\partial p_{\alpha}}\partial X^{\rho}(w)\bar{\partial}X^{\beta}(\bar{w}) \right.\nonumber\\
&\qquad\qquad\qquad
-2ip^{\mu} \eta^{\alpha \sigma}
\dfrac{\partial}{i\partial p_{\beta}}\partial X^{\rho}(w)\bar{\partial}X^{\nu}(\bar{w}) 
+
2i\eta^{\mu \rho} p^{\nu} \dfrac{\partial}{i\partial p_{\alpha}} \partial X^{\beta}(w)\bar{\partial}X^{\sigma}(\bar{w}) \nonumber\\
&\qquad\qquad\qquad+
2\eta^{\mu \rho} \eta^{\nu \sigma}\partial X^{\alpha}(w)\bar{\partial}X^{\beta}(\bar{w})
-2\eta^{\mu \rho}\eta^{\alpha \sigma}  \partial X^{\beta}(w)\bar{\partial}X^{\nu}(\bar{w})\nonumber\\
&\qquad\qquad\qquad-
2i\eta^{\alpha \rho}p^{\nu}_{k} \dfrac{\partial}{i\partial p_{k\beta}}\partial X^{\mu}(w)\bar{\partial}X^{\sigma}(\bar{w})
-
2 \eta^{\alpha \rho}\eta^{\nu \sigma}  \partial X^{\mu}(w)\bar{\partial}X^{\beta}(\bar{w})\nonumber\\
&\qquad\qquad\qquad\left. +
2\eta^{\alpha \rho}\eta^{\beta \sigma} \partial X^{\mu}(w)\bar{\partial}X^{\nu}(\bar{w})
\right]
\exp(ip\cdot X(w,\bar{w})),
\end{align}
where we have ignored the higher order terms in $ \alpha' $ and the mixing with different vertex operators.
The sum of the contribution of 4. and 5. gives the subsubleading soft theorem.
\end{enumerate}

\section{Subsubleading soft theorem for graviton and dilaton}\label{sec;formula}
\quad We focus on the second order terms in $ (q+\xi) $ in eq.(\ref*{eq.before}).

	For convenience we classify the terms by three types of underlines, $ \underline{ \qquad}$, $ \uuline{\qquad}$ and $ \uwave{\qquad} $.
	The simple and double lines do not change the number of prefactors, but the wavy lines do.
	The simple line represents the terms without $ \alpha' $ correction, while the double line stands for the higher order terms in $ \alpha' $.
	\begin{enumerate}
		\item \uline{(0,2;0) terms}
		\begin{align}
		&-\lim_{\xi \to 0}\dfrac{1}{2}h_{\mu\nu}\partial_{\xi}^{\mu}\partial_{\xi}^{\nu}
		|z-w|^{\alpha'p\cdot (q+\xi)}
		\exp\left(ip\cdot X(w)+i\sum_{i}\zeta_{i}\cdot
		\partial_{w}^{m_{i}}X(w)+i\sum_{i}\lambda_{i}\cdot \bar{\partial}^{n_{i}}X(\bar{w_{i}})\right)\nonumber\\
		\times&\dfrac{1}{2}
		\left( -\dfrac{\alpha'}{2}\sum_{i}\dfrac{(n_{i}-1)!(q+\xi)\cdot \lambda_{i}}{(\bar{z}-\bar{w})^{n_{i}}} \right)
		\left( -\dfrac{\alpha'}{2}\sum_{j}\dfrac{(n_{j}-1)!(q+\xi)\cdot \lambda_{j}}{(\bar{z}-\bar{w})^{n_{j}}} \right) \nonumber\\
		\times&\left .\left(\dfrac{\alpha'p\cdot q}{2(z-w)}+\sum_{i}\dfrac{\alpha'm_{i}!q\cdot \zeta_{i}}{2(z-w)^{m_{i}+1}}+iq\cdot \partial X(z)  \right)\left(\dfrac{\alpha'p\cdot q}{2(\bar{z}-\bar{w})}+\sum_{i}\dfrac{\alpha'n_{i}!q\cdot \lambda_{i}}{2(\bar{z}-\bar{w})^{n_{i}+1}}+iq\cdot \bar{\partial} X(\bar{z})  \right)
		\right|_{\text{multilinear in }  i\zeta_{i},i\lambda_{i} }.
		\end{align}
		In the last line only the first term in the first bracket gives the pole $ \dfrac{1}{z-w} $, and only the third term in the second bracket gives $ \dfrac{1}{\bar{z}-\bar{w}} $. 
		\begin{align}
		&-\lim_{\xi \to 0}\dfrac{1}{2}h_{\mu\nu}\partial_{\xi}^{\mu}\partial_{\xi}^{\nu}
		|z-w|^{\alpha'p\cdot (q+\xi)}\exp\left(ip\cdot X(w)+i\sum_{i}\zeta_{i}\cdot
		\partial_{w}^{m_{i}}X(w)+i\sum_{i}\lambda_{i}\cdot \bar{\partial}^{n_{i}}X(\bar{w_{i}})\right)\nonumber\\
		&\times \left .
		\dfrac{1}{2}\left( -\dfrac{\alpha'}{2}\sum_{i}\dfrac{(n_{i}-1)!(q+\xi)\cdot \lambda_{i}}{(\bar{z}-\bar{w})^{n_{i}}} \right)
		\left( -\dfrac{\alpha'}{2}\sum_{j}\dfrac{(n_{j}-1)!(q+\xi)\cdot \lambda_{j}}{(\bar{z}-\bar{w})^{n_{j}}} \right) 
		\left(\dfrac{\alpha'p\cdot q}{2(z-w)} \right)\left(iq\cdot \bar{\partial} X(\bar{z})  \right)
		\right|_{\text{multilinear in }  i\zeta_{i},i\lambda_{i} }\nonumber\\
		=&-\lim_{\xi \to 0}\dfrac{1}{2}h_{\mu\nu}\partial_{\xi}^{\mu}\partial_{\xi}^{\nu}
		\dfrac{2\pi}{\alpha'p\cdot(q+\xi)}
		\exp\left(ip\cdot X(w)+i\sum_{i}\zeta_{i}\cdot
		\partial_{w}^{m_{i}}X(w)+i\sum_{i}\lambda_{i}\cdot \bar{\partial}^{n_{i}}X(\bar{w_{i}})\right)\nonumber\\
		&\left .\times\uwave{ \dfrac{1}{2}\left( -\dfrac{\alpha'}{2}\sum_{i}(n_{i}-1)!(q+\xi)\cdot \lambda_{i} \right)\left( -\dfrac{\alpha'}{2}\sum_{j}(n_{j}-1)!(q+\xi)\cdot \lambda_{j} \right)
			\dfrac{\alpha'p\cdot q}{2}\dfrac{iq\cdot \bar{\partial}^{n_{i}+n_{j}}X}{(n_{i}+n_{j}-1)!}}
		\right|_{\text{multilinear in }  i\zeta_{i},i\lambda_{i} }.\label{eq.(0,2;0)}
		\end{align}
		
		\item \uline{(2,0;0) terms}\\
		By replacing $ \lambda_{i}, n_{i}, \bar{X} $ with $ \zeta_{i}, m_{i}, X $ in the above expression we get the result for the holomorphic part.
		\begin{align}
		&-\lim_{\xi \to 0}\dfrac{1}{2}h_{\mu\nu}\partial_{\xi}^{\mu}\partial_{\xi}^{\nu}
		\dfrac{2\pi}{\alpha'p\cdot(q+\xi)}
		\exp\left(ip\cdot X(w)+i\sum_{i}\zeta_{i}\cdot
		\partial_{w}^{m_{i}}X(w)+i\sum_{i}\lambda_{i}\cdot \bar{\partial}^{n_{i}}X(\bar{w_{i}})\right)
		\nonumber\\
		&\left .\times \uwave{\dfrac{1}{2}\left( -\dfrac{\alpha'}{2}\sum_{i}(m_{i}-1)!(q+\xi)\cdot \zeta_{i} \right)\left( -\dfrac{\alpha'}{2}\sum_{j}(m_{j}-1)!(q+\xi)\cdot \zeta_{j} \right)
			\dfrac{\alpha'p\cdot q}{2}\dfrac{iq\cdot \partial^{m_{i}+m_{j}}X}{(m_{i}+m_{j}-1)!}}
		\right|_{\text{multilinear in }  i\zeta_{i},i\lambda_{i} }.
		\label{eq.(2,0;0)}
		\end{align}
		
		\item \uline{(1,1;0) terms}
		
		\begin{align}
		&-\lim_{\xi \to 0}\dfrac{1}{2}h_{\mu\nu}\partial_{\xi}^{\mu}\partial_{\xi}^{\nu}
		|z-w|^{\alpha'p\cdot (q+\xi)}\exp\left(ip\cdot X(w)+i\sum_{i}\zeta_{i}\cdot
		\partial_{w}^{m_{i}}X(w)+i\sum_{i}\lambda_{i}\cdot \bar{\partial}^{n_{i}}X(\bar{w_{i}})\right)
		\nonumber\\
		\times&\left(-\dfrac{\alpha'}{2}\sum_{i}\dfrac{(m_{i}-1)!(q+\xi)\cdot \zeta_{i}}{(z-w)^{m_{i}}}\right ) \left (-\dfrac{\alpha'}{2}\sum_{i}\dfrac{(n_{i}-1)!(q+\xi)\cdot \lambda_{i}}{(\bar{z}-\bar{w})^{n_{i}}} \right) \nonumber\\
		\times&\left .\left(\dfrac{\alpha'p\cdot q}{2(z-w)}+\sum_{i}\dfrac{\alpha'm_{i}!q\cdot \zeta_{i}}{2(z-w)^{m_{i}+1}}+iq\cdot \partial X(z)  \right)\left(\dfrac{\alpha'p\cdot q}{2(\bar{z}-\bar{w})}+\sum_{i}\dfrac{\alpha'n_{i}!q\cdot \lambda_{i}}{2(\bar{z}-\bar{w})^{n_{i}+1}}+iq\cdot \bar{\partial} X(\bar{z})  \right)
		\right|_{\text{multilinear in }  i\zeta_{i},i\lambda_{i} }.
		\end{align}
		In the last line only the product of the third terms in each bracket, $ iq\cdot \partial X\times iq\cdot \bar{\partial}X $ gives the factor $ |z-w|^{-2} $.
		\begin{align}
		&-\lim_{\xi \to 0}\dfrac{1}{2}h_{\mu\nu}\partial_{\xi}^{\mu}\partial_{\xi}^{\nu}
		|z-w|^{\alpha'p\cdot (q+\xi)}\exp\left(ip\cdot X(w)+i\sum_{i}\zeta_{i}\cdot
		\partial_{w}^{m_{i}}X(w)+i\sum_{i}\lambda_{i}\cdot \bar{\partial}^{n_{i}}X(\bar{w_{i}})\right)\nonumber\\
		\times&\left .\left(-\dfrac{\alpha'}{2}\sum_{i}\dfrac{(m_{i}-1)!(q+\xi)\cdot \zeta_{i}}{(z-w)^{m_{i}}}\right ) \left (-\dfrac{\alpha'}{2}\sum_{i}\dfrac{(n_{i}-1)!(q+\xi)\cdot \lambda_{i}}{(\bar{z}-\bar{w})^{n_{i}}} \right) \left(iq\cdot \partial X(z)  \right)\left(iq\cdot \bar{\partial} X(\bar{z})  \right)
		\right|_{\text{multilinear in }  i\zeta_{i},i\lambda_{i} }\nonumber\\
		=&-\lim_{\xi \to 0}\dfrac{1}{2}h_{\mu\nu}\dfrac{2\pi}{\alpha'p\cdot (q+\xi)}\uline{\left(-\dfrac{\alpha'}{2}\sum_{i}(q+\xi)\cdot \zeta_{i}\right ) \left (-\dfrac{\alpha'}{2}\sum_{j}(q+\xi)\cdot \lambda_{j} \right)} \nonumber\\
		\times&\left .\underline{iq\cdot \partial^{m_{i}}X(w) iq\cdot \bar{\partial}^{n_{i}}X(\bar{w})}\exp\left(ip\cdot X(w)+i\sum_{i}\zeta_{i}\cdot
		\partial_{w}^{m_{i}}X(w)+i\sum_{i}\lambda_{i}\cdot \bar{\partial}^{n_{i}}X(\bar{w_{i}})\right)
		\right|_{\text{multilinear in }  i\zeta_{i},i\lambda_{i} }.
		\label{eq.(1,1;0)}
		\end{align}
		
		\item \uline{(0,1;1) terms}
		\begin{align}
		&-\lim_{\xi \to 0}\dfrac{1}{2}h_{\mu\nu}\partial_{\xi}^{\mu}\partial_{\xi}^{\nu}
		|z-w|^{\alpha'p\cdot (q+\xi)}\left( -\dfrac{\alpha'}{2}\sum_{i}\dfrac{(n_{i}-1)!(q+\xi)\cdot \lambda_{i}}{(\bar{z}-\bar{w})^{n_{i}}} \right) \nonumber\\
		\times&i(q+\xi)\cdot X(z):\exp\left(ip\cdot X(w)+i\sum_{i}\zeta_{i}\cdot
		\partial_{w}^{m_{i}}X(w)+i\sum_{i}\lambda_{i}\cdot \bar{\partial}^{n_{i}}X(\bar{w_{i}})\right):\nonumber\\
		\times&\left .\left(\dfrac{\alpha'p\cdot q}{2(z-w)}+\sum_{i}\dfrac{\alpha'm_{i}!q\cdot \zeta_{i}}{2(z-w)^{m_{i}+1}}+iq\cdot \partial X(z)  \right)\left(\dfrac{\alpha'p\cdot q}{2(\bar{z}-\bar{w})}+\sum_{i}\dfrac{\alpha'n_{i}!q\cdot \lambda_{i}}{2(\bar{z}-\bar{w})^{n_{i}+1}}+iq\cdot \bar{\partial} X(\bar{z})  \right)
		\right|_{\text{multilinear in }  i\zeta_{i},i\lambda_{i} }.
		\end{align}
		In the last line the third term in the first bracket does not give the pole  $\dfrac{1}{z-w} $.
		The product of the second term in the first bracket or the first and second term in the second bracket does not give the factor $ |z-w|^{-2} $.
		\begin{align}
		&-\lim_{\xi \to 0}\dfrac{1}{2}h_{\mu\nu}\partial_{\xi}^{\mu}\partial_{\xi}^{\nu}
		|z-w|^{\alpha'p\cdot (q+\xi)}
		\left( -\dfrac{\alpha'}{2}\sum_{i}\dfrac{(n_{i}-1)!(q+\xi)\cdot \lambda_{i}}{(\bar{z}-\bar{w})^{n_{i}}} \right) \nonumber\\
		\times&i(q+\xi)\cdot X(z):\exp\left(ip\cdot X(w)+i\sum_{i}\zeta_{i}\cdot
		\partial_{w}^{m_{i}}X(w)+i\sum_{i}\lambda_{i}\cdot \bar{\partial}^{n_{i}}X(\bar{w_{i}})\right):\nonumber\\
		\times&\left .\left [\left(\dfrac{\alpha'p\cdot q}{2(z-w)}\right)\left(\dfrac{\alpha'p\cdot q}{2(\bar{z}-\bar{w})}+\sum_{i}\dfrac{\alpha'n_{i}!q\cdot \lambda_{i}}{2(\bar{z}-\bar{w})^{n_{i}+1}}+iq\cdot \bar{\partial} X(\bar{z})  \right)+\left (\sum_{i}\dfrac{\alpha'm_{i}!q\cdot \zeta_{i}}{2(z-w)^{m_{i}+1}}\right )iq\cdot \bar{\partial}X(\bar{z})\right ]
		\right|_{\text{multilinear in }  i\zeta_{i},i\lambda_{i} }\nonumber\\
		=&-\lim_{\xi \to 0}\dfrac{1}{2}h_{\mu\nu}\partial_{\xi}^{\mu}\partial_{\xi}^{\nu}
		|z-w|^{\alpha'p\cdot (q+\xi)}\exp\left(ip\cdot X(w)+i\sum_{i}\zeta_{i}\cdot
		\partial_{w}^{m_{i}}X(w)+i\sum_{i}\lambda_{i}\cdot \bar{\partial}^{n_{i}}X(\bar{w_{i}})\right)\nonumber\\
		\times&\left( -\dfrac{\alpha'}{2}\sum_{i}\dfrac{(n_{i}-1)!(q+\xi)\cdot \lambda_{i}}{(\bar{z}-\bar{w})^{n_{i}}} \right) 
		i(q+\xi)\cdot X(z)\nonumber\\
		\times&\left [\left(\dfrac{\alpha'p\cdot q}{2(z-w)}\right)\left\{ \left(\dfrac{\alpha'p\cdot q}{2(\bar{z}-\bar{w})}+iq\cdot \bar{\partial} X(\bar{z})  \right)  \right .+\left (\sum_{j\neq i}\dfrac{\alpha'n_{j}!q\cdot \lambda_{j}}{2(\bar{z}-\bar{w})^{n_{j}+1}}\right )\right\} \nonumber\\
		&\left .+\left .\left(\sum_{i}\dfrac{\alpha'm_{i}!q\cdot \zeta_{i}}{2(z-w)^{m_{i}+1}}\right)iq\cdot \bar{\partial} X(\bar{z}) i(q+\xi)\cdot X(z)    \right ]
		\right|_{\text{multilinear in }  i\zeta_{i},i\lambda_{i} }\nonumber\\
		=&-\lim_{\xi \to 0}\dfrac{1}{2}h_{\mu\nu}\partial_{\xi}^{\mu}\partial_{\xi}^{\nu}
		: \dfrac{2\pi}{\alpha'p\cdot (q+\xi)}
		\exp\left(ip\cdot X(w)+i\sum_{i}\zeta_{i}\cdot
		\partial_{w}^{m_{i}}X(w)+i\sum_{i}\lambda_{i}\cdot \bar{\partial}^{n_{i}}X(\bar{w_{i}})\right)
		\left( -\dfrac{\alpha'}{2}\sum_{i}(q+\xi)\cdot \lambda_{i}(n_{i}-1)! \right) 
		\nonumber\\
		\times&\left [\left(\dfrac{\alpha'p\cdot q}{2}\right)
		\left \{\left(\underline{\underline{\dfrac{\alpha'p\cdot q}{2}\dfrac{i(q+\xi)\cdot \bar{\partial}^{n_{i}}X(w)}{n_{i}!}}}+\uwave{\sum_{k}^{n_{i}-1}\dfrac{iq\cdot \bar{\partial}^{n_{i}-k}X i(q+\xi)\cdot \bar{\partial}^{k}X(w)}{k!(n_{i}-k-1)!}}  \right) \right .\right .\nonumber\\
		&\left .\left.\left . +\uwave{\left (\sum_{j\neq i}\dfrac{\alpha'n_{j}!q\cdot \lambda_{j}}{2}\right )\dfrac{i(q+\xi)\cdot \bar{\partial}^{n_{i}+n_{j}}X(\bar{w})}{(n_{i}+n_{j})!}} \right \}
		+\underline{\left (\sum_{i}\dfrac{\alpha'm_{i}!q\cdot \zeta_{i}}{2}\right )\dfrac{i(q+\xi)\cdot \partial^{m_{i}}X}{m_{i}!}\dfrac{iq\cdot \bar{\partial}^{n_{i}}X}{(n_{i}-1)!}}
		\right ]
		\right|_{\text{multilinear in }  i\zeta_{i},i\lambda_{i} }.\label{eq.(0,1;1)}
		\end{align}
		
		\item \uline{(1,0;1) terms}\\
		This result is given by replacing  $ \lambda_{i}, n_{i}, \bar{X} $ with $ \zeta_{i}, m_{i} X $ in eq.(\ref{eq.(0,1;1)}).
		\begin{align}
		&-\lim_{\xi \to 0}\dfrac{1}{2}h_{\mu\nu}\partial_{\xi}^{\mu}\partial_{\xi}^{\nu}
		: \dfrac{2\pi}{\alpha'p\cdot (q+\xi)}
		\exp\left(ip\cdot X(w)+i\sum_{i}\zeta_{i}\cdot
		\partial_{w}^{m_{i}}X(w)+i\sum_{i}\lambda_{i}\cdot \bar{\partial}^{n_{i}}X(\bar{w})\right)
		\left( -\dfrac{\alpha'}{2}\sum_{i}(q+\xi)\cdot \zeta_{i}(m_{i}-1)! \right) 
		\nonumber\\
		\times&\left [\left(\dfrac{\alpha'p\cdot q}{2}\right)\left \{\left(\uuline{\dfrac{\alpha'p\cdot q}{2}\dfrac{i(q+\xi)\cdot \partial^{m_{i}}X(w)}{m_{i}!}} +\uwave{\sum_{k}^{m_{i}-1}\dfrac{iq\cdot \partial^{m_{i}-k}X i(q+\xi)\cdot \partial^{k}X(w)}{k!(m_{i}-k-1)!} } \right) \right .\right .\nonumber\\
		&\left .\left.\left . +\uwave{\left (\sum_{j\neq i}\dfrac{\alpha'm_{j}!q\cdot \zeta_{j}}{2}\right )\dfrac{i(q+\xi)\cdot \partial^{m_{i}+m_{j}}X(\bar{w})}{(m_{i}+m_{j})!}} \right \}
		+\uline{\left (\sum_{i}\dfrac{\alpha'n_{i}!q\cdot \lambda_{i}}{2}\right )\dfrac{i(q+\xi)\cdot \bar{\partial}^{n_{i}}X}{n_{i}!}\dfrac{iq\cdot \partial^{m_{i}}X}{(m_{i}-1)!}}  \right ].
		\right|_{\text{multilinear in }  i\zeta_{i},i\lambda_{i} }\label{eq.(1,0;1)}
		\end{align}

		\item \uline{(0,0;2) terms}
		\begin{align}
		&-\lim_{\xi \to 0}\dfrac{1}{2}h_{\mu\nu}\partial_{\xi}^{\mu}\partial_{\xi}^{\nu}
		|z-w|^{\alpha'p\cdot (q+\xi)}\dfrac{i(q+\xi)\cdot X(z)i(q+\xi)\cdot X(z)}{2}\nonumber\\
		\times&:\exp\left(ip\cdot X(w)+i\sum_{i}\zeta_{i}\cdot
		\partial_{w}^{m_{i}}X(w)+i\sum_{i}\lambda_{i}\cdot \bar{\partial}^{n_{i}}X(\bar{w_{i}})\right):\nonumber\\
		\times&\left .\left(\dfrac{\alpha'p\cdot q}{2(z-w)}+\sum_{i}\dfrac{\alpha'm_{i}!q\cdot \zeta_{i}}{2(z-w)^{m_{i}}}+iq\cdot \partial X(z)  \right)\left(\dfrac{\alpha'p\cdot q}{2(\bar{z}-\bar{w})}+\sum_{i}\dfrac{\alpha'n_{i}!q\cdot \lambda_{i}}{2(\bar{z}-\bar{w})^{n_{i}}}+iq\cdot \bar{\partial} X(\bar{z})  \right)
		\right|_{\text{multilinear in }  i\zeta_{i},i\lambda_{i} }.
		\end{align}
		In the last line the third term in the first or second brackets does not give the pole of $ z-w $.
		\begin{align}
		&-\lim_{\xi \to 0}\dfrac{1}{2}h_{\mu\nu}\partial_{\xi}^{\mu}\partial_{\xi}^{\nu}
		|z-w|^{\alpha'p\cdot (q+\xi)}
		\dfrac{i(q+\xi)\cdot X(z)i(q+\xi)\cdot X(z)}{2}\nonumber\\
		\times&:\exp\left(ip\cdot X(w)+i\sum_{i}\zeta_{i}\cdot
		\partial_{w}^{m_{i}}X(w)+i\sum_{i}\lambda_{i}\cdot \bar{\partial}^{n_{i}}X(\bar{w_{i}})\right):\nonumber\\
		\times&\left .\left(\dfrac{\alpha'p\cdot q}{2(z-w)}+\sum_{i}\dfrac{\alpha'm_{i}!q\cdot \zeta_{i}}{2(z-w)^{m_{i}+1}}\right)\left(\dfrac{\alpha'p\cdot q}{2(\bar{z}-\bar{w})}+\sum_{i}\dfrac{\alpha'n_{i}!q\cdot \lambda_{i}}{2(\bar{z}-\bar{w})^{n_{i}+1}}\right)
		\right|_{\text{multilinear in }  i\zeta_{i},i\lambda_{i} }\nonumber\\
		=&-\lim_{\xi \to 0}\dfrac{1}{4}h_{\mu\nu}\partial_{\xi}^{\mu}\partial_{\xi}^{\nu}\dfrac{2\pi}{\alpha'p\cdot(q+\xi)}
		\left [\uline{\left(\dfrac{\alpha'p\cdot q}{2}\right)^2 :i(q+\xi)\cdot X(w)i(q+\xi)\cdot X(w)}\right .\nonumber\\
		+&\uwave{\left(\dfrac{\alpha'p\cdot q}{2}\right)\left(\sum_{i}\dfrac{\alpha'm_{i}!q\cdot \zeta_{i}}{2}\right)
			\sum_{k=0}^{m_{i}}\dfrac{i(q+\xi)\cdot \partial^{m_{i}-k}X(w)i(q+\xi)\cdot \partial^{k}X(w)}{(m_{i}-k)!k!}}  \nonumber\\
		+&\uwave{\left(\dfrac{\alpha'p\cdot q}{2}\right)\left(\sum_{i}\dfrac{\alpha'n_{i}!q\cdot \lambda_{i}}{2}\right):
			\sum_{k=0}^{n_{i}}\dfrac{i(q+\xi)\cdot\bar{\partial}^{n_{i}-k}X(w)i(q+\xi)\cdot \bar{\partial}^{k}X(w)}{(n_{i}-k)!k!}}  \nonumber\\
		+&\left .\left .\uline{2\left(\sum_{i}\dfrac{\alpha'm_{i}!q\cdot \zeta_{i}}{2}\right)\left(\sum_{i}\dfrac{\alpha'n_{i}!q\cdot \lambda_{i}}{2}\right)
			\dfrac{i(q+\xi)\cdot \partial^{m_{i}} X(w)}{m_{i}!}\dfrac{i(q+\xi)\cdot\bar{\partial}^{n_{i}}X(\bar{w}) }{n_{i}!}}  \right ]
		\right|_{\text{multilinear in }  i\zeta_{i},i\lambda_{i} }.
		\label{eq.(0,0;2)}
		\end{align}

	\end{enumerate}
	\begin{itemize}
		\item First we sum up the simple lines.
		\begin{align}
		&-\lim_{\xi \to 0}\dfrac{1}{4}h_{\mu\nu}\partial_{\xi}^{\mu}\partial_{\xi}^{\nu}
		: \dfrac{2\pi}{\alpha'p\cdot (q+\xi)}
		\exp\left(ip\cdot X(w)+i\sum_{i}\zeta_{i}\cdot
		\partial_{w}^{m_{i}}X(w)+i\sum_{i}\lambda_{i}\cdot \bar{\partial}^{n_{i}}X(\bar{w_{i}})\right)\nonumber\\
		&\left [\left(\dfrac{\alpha'p\cdot q}{2}\right)^2 i(q+\xi)\cdot X(w)i(q+\xi)\cdot X(w)
		\right .\nonumber\\
		&+2\left .\left( -\dfrac{\alpha'}{2}\sum_{i}(q+\xi)\cdot \lambda_{i}(n_{i}-1)! \right) \left (\sum_{i}\dfrac{\alpha'm_{i}!q\cdot \zeta_{i}}{2}\right )\dfrac{i(q+\xi)\cdot \partial^{m_{i}}X}{m_{i}!}\dfrac{iq\cdot \bar{\partial}^{n_{i}}X}{(n_{i}-1)!}\right .\nonumber\\
		&+2\left .\left( -\dfrac{\alpha'}{2}\sum_{i}(q+\xi)\cdot \zeta_{i}(m_{i}-1)! \right) \left (\sum_{i}\dfrac{\alpha'n_{i}!q\cdot \lambda_{i}}{2}\right )\dfrac{i(q+\xi)\cdot \bar{\partial}^{n_{i}}X}{n_{i}!}\dfrac{iq\cdot \partial^{m_{i}}X}{(m_{i}-1)!}\right .\nonumber\\
		&+2\left .\left(-\dfrac{\alpha'}{2}\sum_{i}(q+\xi)\cdot \zeta_{i}\right ) \left (-\dfrac{\alpha'}{2}\sum_{j}(q+\xi)\cdot \lambda_{j} \right)iq\cdot \partial^{m_{i}}X(w) iq\cdot \bar{\partial}^{n_{i}}X(\bar{w})\right .\nonumber\\
		&\left .+2\left .\left(\sum_{i}\dfrac{\alpha'm_{i}!q\cdot \zeta_{i}}{2}\right)\left(\sum_{i}\dfrac{\alpha'n_{i}!q\cdot \lambda_{i}}{2}\right)
		\dfrac{i(q+\xi)\cdot \partial^{m_{i}} X(w)}{m_{i}!}\dfrac{i(q+\xi)\cdot\bar{\partial}^{n_{i}}X(\bar{w}) }{n_{i}!}
		\right ]
		\right|_{\text{multilinear in }  i\zeta_{i},i\lambda_{i} }\nonumber\\
		=&-\lim_{\xi \to 0}\dfrac{1}{4}h_{\mu\nu}\partial_{\xi}^{\mu}\partial_{\xi}^{\nu}
		: 2\pi
		\exp\left(ip\cdot X(w)+i\sum_{i}\zeta_{i}\cdot
		\partial_{w}^{m_{i}}X(w)+i\sum_{i}\lambda_{i}\cdot \bar{\partial}^{n_{i}}X(\bar{w_{i}})\right)\nonumber\\
		&\left[
		\left(\dfrac{\alpha'p\cdot q}{2}\right)^2 \dfrac{-(q\cdot X)^2(p\cdot \xi)^2+2(q\cdot X)(\xi \cdot X)(p\cdot q)(p\cdot \xi)-(\xi\cdot X)^2(p\cdot q)^2}{\alpha'(p\cdot q)^3}\right .\nonumber\\
		&+\left .\left. \dfrac{\alpha'^2}{2}\dfrac{\left(\sum_{i} iq_{a}\xi_{b}\zeta_{ic}\partial^{m_{i}}X_{d}(S^{ab})^{cd} \right)\left(\sum_{i} iq_{a'}\xi_{b'}\lambda_{ic'}\partial^{n_{i}}X_{d'}(S^{a'b'})^{c'd'} \right) }{p\cdot q}
		\right]
		\right|_{\text{multilinear in }  i\zeta_{i},i\lambda_{i} }\nonumber\\
		=&\dfrac{\pi\alpha'h_{\mu\nu}q_{a}q_{b}\left(L^{\mu a}L^{\nu a}+2S^{\mu a}\bar{S}^{\nu b} \right)}{4p\cdot q}	:\partial^{m_{1}}X^{\rho_{1}}\cdots\bar{\partial}^{n_{1}}X^{\sigma_{1}}\cdots\exp\left(ip\cdot X(w)\right ):.
		\end{align}
		
		\item Secondly we sum up the double lines.
		\begin{align}
		&-\lim_{\xi \to 0}\dfrac{1}{2}h_{\mu\nu}\partial_{\xi}^{\mu}\partial_{\xi}^{\nu}
		: \dfrac{2\pi}{\alpha'p\cdot (q+\xi)}
		\exp\left(ip\cdot X(w)+i\sum_{i}\zeta_{i}\cdot
		\partial_{w}^{m_{i}}X(w)+i\sum_{i}\lambda_{i}\cdot \bar{\partial}^{n_{i}}X(\bar{w_{i}})\right)\left(\dfrac{\alpha'p\cdot q}{2}\right)^2\nonumber\\
		&\left .\left [
		\left( -\dfrac{\alpha'}{2}\sum_{i}(q+\xi)\cdot \lambda_{i}(n_{i}-1)! \right) \dfrac{i(q+\xi)\cdot \bar{\partial}^{n_{i}}X(w)}{n_{i}!}+\left( -\dfrac{\alpha'}{2}\sum_{i}(q+\xi)\cdot \zeta_{i}(m_{i}-1)! \right) \dfrac{i(q+\xi)\cdot \partial^{m_{i}}X(w)}{m_{i}!}
		\right ]
		\right|_{\text{multilinear in }  i\zeta_{i},i\lambda_{i} }\nonumber\\
		=&\lim_{\xi \to 0}\dfrac{1}{2}h_{\mu\nu}\partial_{\xi}^{\mu}\partial_{\xi}^{\nu}
		: 2\pi 
		\exp\left(ip\cdot X(w)+i\sum_{i}\zeta_{i}\cdot
		\partial_{w}^{m_{i}}X(w)+i\sum_{i}\lambda_{i}\cdot \bar{\partial}^{n_{i}}X(\bar{w_{i}})\right)\left(\dfrac{\alpha'p\cdot q}{2}\right)^2\nonumber\\
		&\left [
		\sum_{i}\dfrac{
			q\cdot \lambda_{i}iq\cdot \bar{\partial}^{n_{i}}X(w)(p\cdot\xi)^2
			-\xi\cdot \lambda_{i}iq\cdot \bar{\partial}^{n_{i}}X(w)p\cdot\xi p\cdot q
			-q\cdot \lambda_{i}i\xi\cdot \bar{\partial}^{n_{i}}X(w)p\cdot qp\cdot\xi
			+\xi\cdot \lambda_{i}i\xi\cdot \bar{\partial}^{n_{i}}X(w)(p\cdot q)^2}{2n_{i}(p\cdot q)^3}\right .\nonumber\\
		&\left .\left .+\sum_{i} \dfrac{
			q\cdot \zeta_{i}iq\cdot \partial^{m_{i}}X(w)(p\cdot\xi)^2
			-\xi\cdot \zeta_{i}iq\cdot \partial^{m_{i}}X(w)p\cdot\xi p\cdot q
			-q\cdot \zeta_{i}i\xi\cdot \partial^{m_{i}}X(w)p\cdot\xi p\cdot q
			+\xi\cdot \zeta_{i}i\xi\cdot \partial^{m_{i}}X(w)(p\cdot q)^2
		}{2m_{i}(p\cdot q)^3}
		\right ]
		\right|_{\text{multilinear in }  i\zeta_{i},i\lambda_{i} }\nonumber\\
		=&\dfrac{\pi\alpha'^2h_{\mu\nu}}{4p\cdot q}
		\left(\eta^{ac}\eta^{bd}\eta^{\mu e}\eta^{\nu f}-\eta^{\mu c}\eta^{ad}\eta^{be}\eta^{\nu f}-\eta^{ac}\eta^{\mu d}\eta^{e\nu}\eta^{bf}+\eta^{\mu c}\eta^{\nu d}\eta^{ae}\eta^{bf}\right)\nonumber\\
		&\times\left(\sum_{i}\dfrac{iq_{a}q_{b}\zeta_{ic}\partial^{m_{i}}X_{d}p_{e}p_{f}}{m_{i}p\cdot q} 
		+\sum_{i}\dfrac{iq_{a}q_{b}\lambda_{ic}\bar{\partial}^{n_{i}}X_{d}p_{e}p_{f}}{n_{i}p\cdot q} \right)\nonumber\\
		&\times\left.\exp\left(ip\cdot X(w)+i\sum_{i}\zeta_{i}\cdot
		\partial_{w}^{m_{i}}X(w)+i\sum_{i}\lambda_{i}\cdot \bar{\partial}^{n_{i}}X(\bar{w_{i}})\right)
		\right|_{\text{multilinear in }  i\zeta_{i},i\lambda_{i} }\nonumber\\ 
		=&\dfrac{\pi\alpha'^2h_{\mu\nu}}{4p\cdot q}
		\left(\eta^{\mu e}\eta^{ad}(S^{b\nu})^{cf}-\eta^{ae}\eta^{\mu d}(S^{b\nu})^{cf}\right)\left(\sum_{i}\dfrac{iq_{a}q_{b}\zeta_{ic}\partial^{m_{i}}X_{d}p_{e}p_{f}}{m_{i}p\cdot q} 
		+\sum_{i}\dfrac{iq_{a}q_{b}\lambda_{ic}\bar{\partial}^{n_{i}}X_{d}p_{e}p_{f}}{n_{i}p\cdot q} \right)\nonumber\\
		&\times\left .\exp\left(ip\cdot X(w)+i\sum_{i}\zeta_{i}\cdot
		\partial_{w}^{m_{i}}X(w)+i\sum_{i}\lambda_{i}\cdot \bar{\partial}^{n_{i}}X(\bar{w_{i}})\right)
		\right|_{\text{multilinear in }  i\zeta_{i},i\lambda_{i} }\nonumber\\
		=&\dfrac{\pi\alpha'^2h_{\mu\nu}}{4p\cdot q}
		(S^{b\nu})^{cf}(S^{a\mu})^{de}\left(\sum_{i}\dfrac{iq_{a}q_{b}\zeta_{ic}\partial^{m_{i}}X_{d}p_{e}p_{f}}{m_{i}p\cdot q} 
		+\sum_{i}\dfrac{iq_{a}q_{b}\lambda_{ic}\bar{\partial}^{n_{i}}X_{d}p_{e}p_{f}}{n_{i}p\cdot q} \right)\nonumber\\
		&\times\left .\exp\left(ip\cdot X(w)+i\sum_{i}\zeta_{i}\cdot
		\partial_{w}^{m_{i}}X(w)+i\sum_{i}\lambda_{i}\cdot \bar{\partial}^{n_{i}}X(\bar{w_{i}})\right)
		\right|_{\text{multilinear in }  i\zeta_{i},i\lambda_{i} }.
		\end{align}
		\item Finally we sum up the wavy lines.
		\begin{align}
		&-\lim_{\xi \to 0}\dfrac{1}{2}h_{\mu\nu}\partial_{\xi}^{\mu}\partial_{\xi}^{\nu}
		: \dfrac{2\pi}{\alpha'p\cdot (q+\xi)}
		\exp\left(ip\cdot X(w)+i\sum_{i}\zeta_{i}\cdot
		\partial_{w}^{m_{i}}X(w)+i\sum_{i}\lambda_{i}\cdot \bar{\partial}^{n_{i}}X(\bar{w_{i}})\right)\nonumber\\
		&\left[
		\dfrac{1}{2}\left( -\dfrac{\alpha'}{2}\sum_{i}(n_{i}-1)!(q+\xi)\cdot \lambda_{i} \right)\left( -\dfrac{\alpha'}{2}\sum_{j}(n_{j}-1)!(q+\xi)\cdot \lambda_{j} \right)
		\dfrac{\alpha'p\cdot q}{2}\dfrac{iq\cdot \bar{\partial}^{n_{i}+n_{j}}X}{(n_{i}+n_{j}-1)!}\right.\nonumber\\
		&+\dfrac{1}{2}\left( -\dfrac{\alpha'}{2}\sum_{i}(m_{i}-1)!(q+\xi)\cdot \zeta_{i} \right)\left( -\dfrac{\alpha'}{2}\sum_{j}(m_{j}-1)!(q+\xi)\cdot \zeta_{j} \right)
		\dfrac{\alpha'p\cdot q}{2}\dfrac{iq\cdot \partial^{m_{i}+m_{j}}X}{(m_{i}+m_{j}-1)!}\nonumber\\
		&+\left( -\dfrac{\alpha'}{2}\sum_{i}(q+\xi)\cdot \zeta_{i}(m_{i}-1)! \right) 
		\left(\dfrac{\alpha'p\cdot q}{2}\right)\nonumber\\
		&\times \left\{\left(\sum_{k=0}^{m_{i}-1}\dfrac{iq\cdot \partial^{m_{i}-k}X i(q+\xi)\cdot \partial^{k}X(w)}{k!(m_{i}-k-1)!}  \right) 
		+\left (\sum_{j\neq i}\dfrac{\alpha'm_{j}!q\cdot \zeta_{j}}{2}\right )\dfrac{i(q+\xi)\cdot \partial^{m_{i}+m_{j}}X(w)}{(m_{i}+m_{j})!}\right\}\nonumber\\
		&+\left( -\dfrac{\alpha'}{2}\sum_{i}(q+\xi)\cdot \lambda_{i}(n_{i}-1)! \right) 
		\left(\dfrac{\alpha'p\cdot q}{2}\right)\nonumber\\
		&\times \left\{\left(\sum_{k=0}^{n_{i}-1}\dfrac{iq\cdot \bar{\partial}^{n_{i}-k}X i(q+\xi)\cdot \bar{\partial}^{k}X(w)}{k!(n_{i}-k-1)!}  \right) 
		+\left (\sum_{j\neq i}\dfrac{\alpha'n_{j}!q\cdot \lambda_{j}}{2}\right )\dfrac{i(q+\xi)\cdot \bar{\partial}^{n_{i}+n_{j}}X(\bar{w})}{(n_{i}+n_{j})!}\right\}\nonumber\\
		&+\dfrac{1}{2}\left(\dfrac{\alpha'p\cdot q}{2}\right)\left(\sum_{i}\dfrac{\alpha'm_{i}!q\cdot \zeta_{i}}{2}\right)
		\sum_{k=0}^{m_{i}}\dfrac{i(q+\xi)\cdot \partial^{m_{i}-k}X(w)i(q+\xi)\cdot \partial^{k}X(w)}{(m_{i}-k)!k!}  \nonumber\\
		&\left .+\left .\dfrac{1}{2}\left(\dfrac{\alpha'p\cdot q}{2}\right)\left(\sum_{i}\dfrac{\alpha'n_{i}!q\cdot \lambda_{i}}{2}\right)
		\sum_{k=0}^{n_{i}}\dfrac{i(q+\xi)\cdot\bar{\partial}^{n_{i}-k}X(w)i(q+\xi)\cdot \bar{\partial}^{k}X(w)}{(n_{i}-k)!k!} 
		\right]
		\right|_{\text{multilinear in }  i\zeta_{i},i\lambda_{i} }.
		\end{align}
		First we write the terms that include $ X $'s without any derivative. 
		\begin{align}
		&-\lim_{\xi \to 0}\dfrac{1}{2}h_{\mu\nu}\partial_{\xi}^{\mu}\partial_{\xi}^{\nu}
		: \dfrac{2\pi}{\alpha'p\cdot (q+\xi)}
		\exp\left(ip\cdot X(w)+i\sum_{i}\zeta_{i}\cdot
		\partial_{w}^{m_{i}}X(w)+i\sum_{i}\lambda_{i}\cdot \bar{\partial}^{n_{i}}X(\bar{w_{i}})\right)\nonumber\\
		&\left[
		\left( -\dfrac{\alpha'}{2}\sum_{i}(q+\xi)\cdot \zeta_{i}(m_{i}-1)! \right) 
		\left(\dfrac{\alpha'p\cdot q}{2}\right)\left(\dfrac{iq\cdot \partial^{m_{i}}X i(q+\xi)\cdot X(w)}{(m_{i}-1)!}  \right)\right . \nonumber\\
		&+\left( -\dfrac{\alpha'}{2}\sum_{i}(q+\xi)\cdot \lambda_{i}(n_{i}-1)! \right) 
		\left(\dfrac{\alpha'p\cdot q}{2}\right)\left(\dfrac{iq\cdot \bar{\partial}^{n_{i}}X i(q+\xi)\cdot X(w)}{(n_{i}-1)!}  \right) \nonumber\\
		&+\left(\dfrac{\alpha'p\cdot q}{2}\right)\left(\sum_{i}\dfrac{\alpha'm_{i}!q\cdot \zeta_{i}}{2}\right)
		\dfrac{i(q+\xi)\cdot \partial^{m_{i}}X(w)i(q+\xi)\cdot X(w)}{m_{i}!}  \nonumber\\
		&\left .+\left .\left(\dfrac{\alpha'p\cdot q}{2}\right)\left(\sum_{i}\dfrac{\alpha'n_{i}!q\cdot \lambda_{i}}{2}\right)
		\dfrac{i(q+\xi)\cdot\bar{\partial}^{n_{i}}X(w)i(q+\xi)\cdot X(w)}{n_{i}!} 
		\right]
		\right|_{\text{multilinear in }  i\zeta_{i},i\lambda_{i} }\nonumber\\
		=&-\lim_{\xi \to 0}\dfrac{1}{2}h_{\mu\nu}\partial_{\xi}^{\mu}\partial_{\xi}^{\nu}
		\dfrac{2\pi}{\alpha'p\cdot (q+\xi)}
		\exp\left(ip\cdot X(w)+i\sum_{i}\zeta_{i}\cdot
		\partial_{w}^{m_{i}}X(w)+i\sum_{i}\lambda_{i}\cdot \bar{\partial}^{n_{i}}X(\bar{w_{i}})\right)\nonumber\\
		&\left .\times\left(\dfrac{\alpha'p\cdot q}{2}\right) \sum_{i} i(q+\xi)\cdot X(z)\dfrac{\alpha'}{2}\left(iq_{a}\xi_{b}\zeta_{ic}\partial^{m_{i}}X_{d}(S^{ab})^{cd}+iq_{a}\xi_{b}\lambda_{ic}\bar{\partial}^{n_{i}}X_{d}(S^{ab})^{cd}\right)
		\right|_{\text{multilinear in }  i\zeta_{i},i\lambda_{i} }\nonumber\\
		=&\dfrac{\alpha'\pi h_{\mu\nu}}{2p\cdot q}
		\exp\left(ip\cdot X(w)+i\sum_{i}\zeta_{i}\cdot
		\partial_{w}^{m_{i}}X(w)+i\sum_{i}\lambda_{i}\cdot \bar{\partial}^{n_{i}}X(\bar{w_{i}})\right)\nonumber\\
		&\left .\times  iq_{e}L^{\mu e}\left(\sum_{i}iq_{a}\zeta_{ic}\partial^{m_{i}}X_{d}(S^{\nu a})^{cd}+\sum_{i}iq_{a}\lambda_{ic}\bar{\partial}^{n_{i}}X_{d}(S^{\nu a})^{cd}\right)
		\right|_{\text{multilinear in }  i\zeta_{i},i\lambda_{i} }.
		\end{align}
		Secondly we write the terms that change one prefactor to two.
		\begin{align}
		&-\lim_{\xi \to 0}\dfrac{1}{2}h_{\mu\nu}\partial_{\xi}^{\mu}\partial_{\xi}^{\nu}
		: \dfrac{2\pi}{\alpha'p\cdot (q+\xi)}
		\exp\left(ip\cdot X(w)+i\sum_{i}\zeta_{i}\cdot
		\partial_{w}^{m_{i}}X(w)+i\sum_{i}\lambda_{i}\cdot \bar{\partial}^{n_{i}}X(\bar{w_{i}})\right)\nonumber\\
		&\left[
		\left( -\dfrac{\alpha'}{2}\sum_{i}(q+\xi)\cdot \zeta_{i}(m_{i}-1)! \right) 
		\left(\dfrac{\alpha'p\cdot q}{2}\right)\left(\sum_{k=1}^{m_{i}-1}\dfrac{iq\cdot \partial^{m_{i}-k}X i(q+\xi)\cdot \partial^{k}X(w)}{k!(m_{i}-k-1)!}  \right)\right.  \nonumber\\
		&+\left( -\dfrac{\alpha'}{2}\sum_{i}(q+\xi)\cdot \lambda_{i}(n_{i}-1)! \right) 
		\left(\dfrac{\alpha'p\cdot q}{2}\right)\left(\sum_{k=1}^{n_{i}-1}\dfrac{iq\cdot \bar{\partial}^{n_{i}-k}X i(q+\xi)\cdot \bar{\partial}^{k}X(w)}{k!(n_{i}-k-1)!}  \right) \nonumber\\
		&+\dfrac{1}{2}\left(\dfrac{\alpha'p\cdot q}{2}\right)\left(\sum_{i}\dfrac{\alpha'm_{i}!q\cdot \zeta_{i}}{2}\right)
		\sum_{k=1}^{m_{i}-1}\dfrac{i(q+\xi)\cdot \partial^{m_{i}-k}X(w)i(q+\xi)\cdot \partial^{k}X(w)}{(m_{i}-k)!k!}  \nonumber\\
		&\left .+\left .\dfrac{1}{2}\left(\dfrac{\alpha'p\cdot q}{2}\right)\left(\sum_{i}\dfrac{\alpha'n_{i}!q\cdot \lambda_{i}}{2}\right)
		\sum_{k=1}^{n_{i}-1}\dfrac{i(q+\xi)\cdot\bar{\partial}^{n_{i}-k}X(w)i(q+\xi)\cdot \bar{\partial}^{k}X(w)}{(n_{i}-k)!k!} 
		\right]
		\right|_{\text{multilinear in }  i\zeta_{i},i\lambda_{i} }.
		\end{align}
		We can check the following equation easily.
		\begin{align}
		\sum_{k=1}^{m_{i}-1}\dfrac{(m_{i}-1)!}{k!(m_{i}-k-1)!}\partial^{m_{i}-k}X \partial^{k}X
		=\sum_{k=1}^{m_{i}-1}\dfrac{m_{i}!}{2k!(m_{i}-k)!}\partial^{m_{i}-k}X\partial^{k}X
		\end{align}
		By using this equation we can simplify the above equation as follows:
		\begin{align}
		&-\lim_{\xi \to 0}\dfrac{1}{2}h_{\mu\nu}\partial_{\xi}^{\mu}\partial_{\xi}^{\nu}
		\dfrac{2\pi}{\alpha'p\cdot (q+\xi)}\left(\dfrac{\alpha'p\cdot q}{2}\right)
		\exp\left(ip\cdot X(w)+i\sum_{i}\zeta_{i}\cdot
		\partial_{w}^{m_{i}}X(w)+i\sum_{i}\lambda_{i}\cdot \bar{\partial}^{n_{i}}X(\bar{w_{i}})\right)\nonumber\\
		&\times\left[
		\left(\sum_{k=1}^{m_{i}-1}\dfrac{\alpha'(m_{i}-1)! i(q+\xi)\cdot \partial^{k}X(w)}{2k!(m_{i}-k-1)!}  \right)\left((q\cdot \zeta_{i})i(q+\xi)\cdot \partial^{m_{i}-k}X-(q+\xi)\cdot\zeta_{i} iq\cdot \partial^{m_{i}-k}X  \right)  \right.  \nonumber\\
		&\left .+\left .\left(\sum_{k=1}^{n_{i}-1}\dfrac{\alpha'(n_{i}-1)! i(q+\xi)\cdot \bar{\partial}^{k}X(w)}{2k!(n_{i}-k-1)!}  \right)\left((q\cdot \lambda_{i})i(q+\xi)\cdot \bar{\partial}^{n_{i}-k}X-(q+\xi)\cdot\lambda_{i} iq\cdot \bar{\partial}^{n_{i}-k}X  \right) 
		\right]
		\right|_{\text{multilinear in }  i\zeta_{i},i\lambda_{i} }\nonumber\\
		=&-\lim_{\xi \to 0}\dfrac{1}{2}h_{\mu\nu}\partial_{\xi}^{\mu}\partial_{\xi}^{\nu}
		2\pi\left(\dfrac{\alpha'p\cdot q}{2}\right)
		\exp\left(ip\cdot X(w)+i\sum_{i}\zeta_{i}\cdot
		\partial_{w}^{m_{i}}X(w)+i\sum_{i}\lambda_{i}\cdot \bar{\partial}^{n_{i}}X(\bar{w_{i}})\right)\nonumber\\
		&\times\left[
		\left(\sum_{k=1}^{m_{i}-1}\dfrac{\alpha'(m_{i}-1)! }{2k!(m_{i}-k-1)!}  \right)
		\left(\dfrac{i\xi\cdot\partial^{k}Xp\cdot q-iq\cdot \partial^{k}X(p\cdot\xi)}{(p\cdot q)^2}\right)
		\left(q_{a}\xi_{b}\zeta_{ic}\partial^{m_{i}-k}X_{d}(S^{ab})^{cd}  \right)  \right.  \nonumber\\
		&\left .+\left .\left(\sum_{k=1}^{n_{i}-1}\dfrac{\alpha'(n_{i}-1)! }{2k!(n_{i}-k-1)!}  \right)
		\left(\dfrac{i\xi\cdot\bar{\partial}^{k}Xp\cdot q-iq\cdot \bar{\partial}^{k}X(p\cdot\xi)}{(p\cdot q)^2}\right)
		\left(q_{a}\xi_{b}\lambda_{ic}\bar{\partial}^{n_{i}-k}X_{d}(S^{ab})^{cd}  \right) 
		\right]
		\right|_{\text{multilinear in }  i\zeta_{i},i\lambda_{i} }\nonumber\\
		=&-\dfrac{\pi\alpha'h_{\mu\nu}}{2p\cdot q}
		\left[
		\left(\sum_{k=1}^{m_{i}-1}\dfrac{(m_{i}-1)! }{k!(m_{i}-k-1)!}  \right)
		\left(q_{a}p_{c}\partial^{k}X_{d}(iS^{a\mu})^{cd}  \right)
		\left(q_{a}\zeta_{ic}\partial^{m_{i}-k}X_{d}(S^{a\nu})^{cd}  \right)  \right.  \nonumber\\
		&\left .+\left .\left(\sum_{k=1}^{n_{i}-1}\dfrac{(n_{i}-1)! }{k!(n_{i}-k-1)!}  \right)
		\left(q_{a}p_{c}\bar{\partial}^{k}X_{d}(iS^{a\mu})^{cd}  \right) 
		\left(q_{a}\lambda_{ic}\bar{\partial}^{n_{i}-k}X_{d}(S^{a\nu})^{cd}  \right) 
		\right]
		\right|_{\text{multilinear in }  i\zeta_{i},i\lambda_{i} }.
		\end{align}
		
		Finally we write the terms that change two prefactors to one.
		\begin{align}
		&-\lim_{\xi \to 0}\dfrac{1}{2}h_{\mu\nu}\partial_{\xi}^{\mu}\partial_{\xi}^{\nu}
		: \dfrac{2\pi}{\alpha'p\cdot (q+\xi)}
		\exp\left(ip\cdot X(w)+i\sum_{i}\zeta_{i}\cdot
		\partial_{w}^{m_{i}}X(w)+i\sum_{i}\lambda_{i}\cdot \bar{\partial}^{n_{i}}X(\bar{w_{i}})\right)\nonumber\\
		&\left[
		\dfrac{1}{2}\left( -\dfrac{\alpha'}{2}\sum_{i}(n_{i}-1)!(q+\xi)\cdot \lambda_{i} \right)\left( -\dfrac{\alpha'}{2}\sum_{j}(n_{j}-1)!(q+\xi)\cdot \lambda_{j} \right)
		\dfrac{\alpha'p\cdot q}{2}\dfrac{iq\cdot \bar{\partial}^{n_{i}+n_{j}}X}{(n_{i}+n_{j}-1)!}\right.\nonumber\\
		&+\dfrac{1}{2}\left( -\dfrac{\alpha'}{2}\sum_{i}(m_{i}-1)!(q+\xi)\cdot \zeta_{i} \right)\left( -\dfrac{\alpha'}{2}\sum_{j}(m_{j}-1)!(q+\xi)\cdot \zeta_{j} \right)
		\dfrac{\alpha'p\cdot q}{2}\dfrac{iq\cdot \partial^{m_{i}+m_{j}}X}{(m_{i}+m_{j}-1)!}\nonumber\\
		&+\left( -\dfrac{\alpha'}{2}\sum_{i}(q+\xi)\cdot \zeta_{i}(m_{i}-1)! \right) 
		\left(\dfrac{\alpha'p\cdot q}{2}\right)
		\left (\sum_{j\neq i}\dfrac{\alpha'm_{j}!q\cdot \zeta_{j}}{2}\right )\dfrac{i(q+\xi)\cdot \partial^{m_{i}+m_{j}}X(w)}{(m_{i}+m_{j})!}\nonumber\\
		&\left .+\left .\left( -\dfrac{\alpha'}{2}\sum_{i}(q+\xi)\cdot \lambda_{i}(n_{i}-1)! \right) 
		\left(\dfrac{\alpha'p\cdot q}{2}\right)
		\left (\sum_{j\neq i}\dfrac{\alpha'n_{j}!q\cdot \lambda_{j}}{2}\right )\dfrac{i(q+\xi)\cdot \bar{\partial}^{n_{i}+n_{j}}X(\bar{w})}{(n_{i}+n_{j})!}
		\right]
		\right|_{\text{multilinear in }  i\zeta_{i},i\lambda_{i} }\nonumber
		\end{align}
		
		\begin{align}
		=&-\lim_{\xi \to 0}\dfrac{1}{2}h_{\mu\nu}\partial_{\xi}^{\mu}\partial_{\xi}^{\nu}
		: \dfrac{2\pi}{\alpha'p\cdot (q+\xi)}
		\exp\left(ip\cdot X(w)+i\sum_{i}\zeta_{i}\cdot
		\partial_{w}^{m_{i}}X(w)+i\sum_{i}\lambda_{i}\cdot \bar{\partial}^{n_{i}}X(\bar{w_{i}})\right)\nonumber\\
		&\left[
		\sum_{i,j,i\neq j}
		\dfrac{\alpha'^2(n_{i}-1)!(n_{j}-1)!}{4(n_{i}+n_{j}-1)!}
		\dfrac{\alpha'p\cdot q}{2}(q+\xi)\cdot\lambda_{i}
		\dfrac{n_{j}\left( \xi\cdot \lambda_{j}iq\cdot \bar{\partial}^{n_{i}+n_{j}}X-q\cdot\lambda_{j}i\xi\cdot \bar{\partial}^{n_{i}+n_{j}}X\right )+(i\leftrightarrow j)}{2(n_{i}+n_{j})}\right.\nonumber\\
		&\left .\left .+\sum_{i,j,i\neq j}
		\dfrac{\alpha'^2(m_{i}-1)!(m_{j}-1)!}{4(m_{i}+m_{j}-1)!}
		\dfrac{\alpha'p\cdot q}{2}(q+\xi)\cdot\zeta_{i}
		\dfrac{m_{j}\left (\xi\cdot \zeta_{j}iq\cdot \partial^{m_{i}+m_{j}}X-q\cdot\zeta_{j}i\xi\cdot \partial^{m_{i}+m_{j}}X\right )+(i\leftrightarrow j)}{2(m_{i}+m_{j})}
		\right]
		\right|_{\text{multilinear in }  i\zeta_{i},i\lambda_{i} }\nonumber\\
		=&-\dfrac{\pi\alpha'^2}{8p\cdot q}h_{\mu\nu}
		\exp\left(ip\cdot X(w)+i\sum_{i}\zeta_{i}\cdot
		\partial_{w}^{m_{i}}X(w)+i\sum_{i}\lambda_{i}\cdot \bar{\partial}^{n_{i}}X(\bar{w_{i}})\right)\nonumber\\
		&\left[
		\sum_{i,j,i\neq j}
		\dfrac{(n_{i}-1)!(n_{j}-1)!}{(n_{i}+n_{j}-1)!}
		\dfrac{q_{a}\lambda_{ic}p_{d}(S^{\mu a})^{cd}n_{j}q_{e}\lambda_{jg}\bar{\partial}^{n_{i}+n_{j}}X_{h}(S^{\nu e})^{gh}+(i\leftrightarrow j)}{(n_{i}+n_{j})}\right.\nonumber\\
		&\left .\left .+\sum_{i,j,i\neq j}
		\dfrac{(m_{i}-1)!(m_{j}-1)!}{(m_{i}+m_{j}-1)!}
		\dfrac{q_{a}\zeta_{ic}p_{d}(S^{\mu a})^{cd}m_{j}q_{e}\zeta_{jg}\partial^{m_{i}+m_{j}}X_{h}(S^{\nu e})^{gh}+(i\leftrightarrow j)}{(n_{i}+n_{j})}
		\right]
		\right|_{\text{multilinear in }  i\zeta_{i},i\lambda_{i} }.
		\end{align}
	\end{itemize}
	Thus we can summarize all the results as follows:
	\begin{align}\label{eq;gravitonformula}
	&\dfrac{\pi\alpha'h_{\mu\nu}}{4p\cdot q}
	\left[q_{a}q_{b}\left (L^{\mu a}L^{\nu b}+2S^{\mu a}\bar{S}^{\nu b}\right )
	+2q_{a}L^{\mu a}\left(q_{b}S^{b\nu}+q_{b}\bar{S}^{b\nu}  \right)
	\right]
	\partial^{m_{1}}X\cdots\bar{\partial}^{n_{1}}X\cdots\exp\left(ip\cdot X(w)\right )\nonumber\\
	+&\dfrac{\pi\alpha'^2h_{\mu\nu}}{4p\cdot q}
	\sum_{i}(S^{\mu a})^{ed}(S^{b\nu})^{\rho_{i}f}\dfrac{q_{a}q_{b}\partial^{m_{i}}X_{d}p_{e}p_{f}}{m_{i}} 
	\cdots\underbrace{\partial^{m_{i}}}X\cdots\bar{\partial}^{n_{1}}X\cdots\exp\left(ip\cdot X(w)\right )\nonumber\\
	+&\dfrac{\pi\alpha'^2h_{\mu\nu}}{4p\cdot q}\sum_{i}\left(\eta^{\mu e}\eta^{ad}(S^{b\nu})^{\sigma_{i}f}+\eta^{ae}\eta^{\mu d}(S^{b\nu})^{\sigma_{i}f}\right)
	\dfrac{q_{a}q_{b}\bar{\partial}^{n_{i}}X_{d}p_{e}p_{f}}{n_{i}}  
	\partial^{m_{1}}X\cdots\underbrace{\bar{\partial}^{n_{i}}X}\cdots\exp\left(ip\cdot X(w)\right )\nonumber\\
	-&\dfrac{i\pi\alpha'h_{\mu\nu}}{2p\cdot q}
	\left(\sum_{k=1}^{m_{i}-1}\dfrac{(m_{i}-1)! }{k!(m_{i}-k-1)!}  \right)
	\left(q_{a}p_{c}\partial^{k}X_{d}(S^{a\mu})^{cd}  \right)
	\left(q_{b}\partial^{m_{i}-k}X_{d}(S^{b\nu})^{\rho_{i}d}  \right)
	\cdots\underbrace{\partial^{m_{i}}X}\cdots\bar{\partial}^{n_{1}}X\cdots\exp\left(ip\cdot X(w)\right )\nonumber\\
	-&\dfrac{i\pi\alpha'h_{\mu\nu}}{2p\cdot q}
	\left(\sum_{k=1}^{n_{i}-1}\dfrac{(n_{i}-1)! }{k!(n_{i}-k-1)!}  \right)
	\left(q_{a}p_{c}\bar{\partial}^{k}X_{d}(S^{a\mu})^{cd}  \right) 
	\left(q_{a}\bar{\partial}^{n_{i}-k}X_{d}(S^{a\nu})^{\sigma_{i}d}  \right) 
	\partial^{m_{1}}X\cdots\underbrace{\bar{\partial}^{n_{i}}X}\cdots\exp\left(ip\cdot X(w)\right )\nonumber\\
	+&\dfrac{\pi\alpha'^2h_{\mu\nu}}{8p\cdot q}
	\left[
	\sum_{i,j}
	\dfrac{(n_{i}-1)!(n_{j}-1)!}{(n_{i}+n_{j}-1)!}
	\dfrac{q_{a}p_{d}(S^{\mu a})^{\sigma_{i}d}n_{j}q_{e}i\bar{\partial}^{n_{i}+n_{j}}X_{h}(S^{\nu e})^{\sigma_{j}h}+(i\leftrightarrow j)}{(n_{i}+n_{j})}\right]
	\partial^{m_{1}}X\cdots\underbrace{\bar{\partial}^{n_{i}}X}\underbrace{\bar{\partial}^{n_{j}}X}\cdots\exp\left(ip\cdot X(w)\right )\nonumber\\
	+&\dfrac{\pi\alpha'^2h_{\mu\nu}}{8p\cdot q}\left [\sum_{i,j}
	\dfrac{(m_{i}-1)!(m_{j}-1)!}{(m_{i}+m_{j}-1)!}
	\dfrac{q_{a}p_{d}(S^{\mu a})^{\rho_{i}d}m_{j}q_{e}i\partial^{m_{i}+m_{j}}X_{h}(S^{\nu e})^{\rho_{j}h}+(i\leftrightarrow j)}{(m_{i}+m_{j})}
	\right]
	\cdots\underbrace{\partial^{m_{i}}X}\underbrace{\partial^{m_{j}}X}\cdots\bar{\partial}^{n_{1}}X\cdots\exp\left(ip\cdot X(w)\right ).
	\end{align}

\section{Soft theorem for B field}\label{sec;Bfield}
\quad As in the previous sections we derive the general formula for the soft B field theorem.
The contraction between the B field and the hard vertex operator is as follows:
\begin{align}
&\lim_{\xi,\omega \to 0}h_{\mu\nu}\partial_{\xi}^{\mu}\partial_{\omega}^{\nu}\partial_{z}\partial_{\bar{z}}:\exp\left(iq\cdot X(z)+i\xi\cdot X(z')+
i\omega\cdot \bar{\partial}X(\bar{z}' \right):\nonumber\\
&\left .\times:\exp\left(ip\cdot X(w)+i\sum_{i}\zeta_{i}\cdot \partial_{w}^{m_{i}}X(w)+i\sum_{i}\lambda_{i}\cdot \bar{\partial}^{n_{i}}X(\bar{w_{i}})\right):
\right|_{\text{multilinear in }  i\zeta_{i},i\lambda_{i} }\nonumber\\
=&\lim_{\xi,\omega \to 0}h_{\mu\nu}\partial_{\xi}^{\mu}\partial_{\omega}^{\nu}\partial_{z}
:|z-w|^{\alpha'p\cdot q}|z'-w|^{\alpha'p\cdot \xi}\exp\left (-\dfrac{\alpha'p\cdot\omega}{2(\bar{z}'-\bar{w})}\right )\nonumber\\
\times&\exp\left(-\dfrac{\alpha'}{2}\sum_{i}\dfrac{(m_{i}-1)!q\cdot \zeta_{i}}{(z-w)^{m_{i}}} -\dfrac{\alpha'}{2}\sum_{i}\dfrac{(n_{i}-1)!q\cdot \lambda_{i}}{(\bar{z}-\bar{w})^{n_{i}}} \right) \exp\left(-\dfrac{\alpha'}{2}\sum_{i}\dfrac{(m_{i}-1)!\xi\cdot \zeta_{i}}{(z'-w)^{m_{i}}} -\dfrac{\alpha'}{2}\sum_{i}\dfrac{(n_{i}-1)!\xi\cdot \lambda_{i}}{(\bar{z'}-\bar{w})^{n_{i}}} \right) \nonumber\\
\times&\exp\left(\dfrac{\alpha'}{2}\sum_{i}\dfrac{n_{i}!\omega\cdot\lambda_{i}}{(\bar{z'}-\bar{w})^{n_{i}+1}} \right)\nonumber\\
\times&\left .:\exp\left(iq\cdot X(z)+i\xi\cdot X(z')+i\omega\cdot \bar{\partial}X(\bar{z}')+ip\cdot X(w)+i\sum_{i}\zeta_{i}\cdot
\partial_{w}^{m_{i}}X(w)+i\sum_{i}\lambda_{i}\cdot \bar{\partial}^{n_{i}}X(\bar{w_{i}})\right):
\right|_{\text{multilinear in }  i\zeta_{i},i\lambda_{i} }\nonumber\\
=&\lim_{\xi \to 0}h_{\mu\nu}\partial_{\xi}^{\mu}
: |z-w|^{\alpha'p\cdot (q+\xi)}\nonumber\\
\times&\exp\left(-\dfrac{\alpha'}{2}\sum_{i}\dfrac{(m_{i}-1)!(q+\xi)\cdot \zeta_{i}}{(z-w)^{m_{i}}} -\dfrac{\alpha'}{2}\sum_{i}\dfrac{(n_{i}-1)!(q+\xi)\cdot \lambda_{i}}{(\bar{z}-\bar{w})^{n_{i}}} \right) \nonumber\\
\times&: \left(i(q+\xi)\cdot X(z)+ip\cdot X(w)+i\sum_{i}\zeta_{i}\cdot
\partial_{w}^{m_{i}}X(w)+i\sum_{i}\lambda_{i}\cdot \bar{\partial}^{n_{i}}X(\bar{w_{i}})\right):\nonumber\\
\times&\left .\left(\dfrac{\alpha'p\cdot q}{2(z-w)}+\sum_{i}\dfrac{\alpha'm_{i}!q\cdot \zeta_{i}}{2(z-w)^{m_{i}+1}}+iq\cdot \partial X(z)  \right)
\left(\dfrac{-\alpha'p^{\nu}}{2(\bar{z}-\bar{w})}+\dfrac{\alpha'n_{i}!\lambda_{i}^{\nu}}{2(\bar{z}-\bar{w})^{n_{i}+1}}+i\bar{\partial}X^{\nu}(\bar{z}) \right)
\right|_{\text{multilinear in }  i\zeta_{i},i\lambda_{i} }
.\label{eq.bcont}
\end{align}
We look at the powers of $ z-w $. Because the last line in eq.(\ref{eq.bcont}) is first-order in q,  it does not affect through subleading order unless the z integration yields the singular behavior in q. Therefore we take only the coefficients of  $ |z-w|^{\alpha'p\cdot (q+\xi)-2} $.

\begin{enumerate}
	\item \uline{The 0-th order terms in $ (q+\xi) $}\\
	\begin{align}
	&\lim_{\xi \to 0}h_{\mu\nu}\partial_{\xi}^{\mu}
	|z-w|^{\alpha'p\cdot (q+\xi)}
	\exp\left(ip\cdot X(w)+i\sum_{i}\zeta_{i}\cdot
	\partial_{w}^{m_{i}}X(w)+i\sum_{i}\lambda_{i}\cdot \bar{\partial}^{n_{i}}X(\bar{w_{i}})\right):\nonumber\\
	\times&\left .\left(\dfrac{\alpha'p\cdot q}{2(z-w)}+\sum_{i}\dfrac{\alpha'm_{i}!q\cdot \zeta_{i}}{2(z-w)^{m_{i}+1}}+iq\cdot \partial X(z)  \right)
	\left(\dfrac{-\alpha'p^{\nu}}{2(\bar{z}-\bar{w})}+\dfrac{\alpha'n_{i}!\lambda_{i}^{\nu}}{2(\bar{z}-\bar{w})^{n_{i}+1}}+i\bar{\partial}X^{\nu}(\bar{z}) \right)
	\right|_{\text{multilinear in }  i\zeta_{i},i\lambda_{i} }
	\end{align}
	Only the product of the first terms in each bracket yields the factor $ |z-w|^{-2} $.
	\begin{align}
	&\left .h_{\mu\nu}
	\dfrac{-2\pi p^{\mu}}{\alpha'(p\cdot q)^2}
	\exp\left(ip\cdot X(w)+i\sum_{i}\zeta_{i}\cdot
	\partial_{w}^{m_{i}}X(w)+i\sum_{i}\lambda_{i}\cdot \bar{\partial}^{n_{i}}X(\bar{w_{i}})\right)
	\dfrac{-\alpha'^2 p\cdot qp^{\nu}}{4}\right|_{\text{multilinear in }  i\zeta_{i},i\lambda_{i} }=0.
	\end{align}
	We have used the antisymmetry of the tensor $ h_{\mu\nu} $. The leading B field soft theorem does not exist.
	
	\item \uline{The first order terms in $ (q+\xi) $}\\
	\begin{enumerate}
		\item (0,0;1) terms\\
		\begin{align}
		&\lim_{\xi \to 0}h_{\mu\nu}\partial_{\xi}^{\mu}
		|z-w|^{\alpha'p\cdot (q+\xi)}
		\exp\left(ip\cdot X(w)+i\sum_{i}\zeta_{i}\cdot
		\partial_{w}^{m_{i}}X(w)+i\sum_{i}\lambda_{i}\cdot \bar{\partial}^{n_{i}}X(\bar{w_{i}})\right):\nonumber\\
		\times&\left .i(q+\xi)\cdot X(z)\left(\dfrac{\alpha'p\cdot q}{2(z-w)}+\sum_{i}\dfrac{\alpha'm_{i}!q\cdot \zeta_{i}}{2(z-w)^{m_{i}+1}}+iq\cdot \partial X(z)  \right)
		\left(\dfrac{-\alpha'p^{\nu}}{2(\bar{z}-\bar{w})}+\dfrac{\alpha'n_{i}!\lambda_{i}^{\nu}}{2(\bar{z}-\bar{w})^{n_{i}+1}}+i\bar{\partial}X^{\nu}(\bar{z}) \right)
		\right|_{\text{multilinear in }  i\zeta_{i},i\lambda_{i} }.
		\end{align}
		We write only the terms that have the factor $ |z-w|^{\alpha'p\cdot (q+\xi)-2} $.
		\begin{align}
		&\lim_{\xi \to 0}h_{\mu\nu}\partial_{\xi}^{\mu}
		\dfrac{2\pi}{\alpha'p\cdot (q+\xi)}
		\exp\left(ip\cdot X(w)+i\sum_{i}\zeta_{i}\cdot
		\partial_{w}^{m_{i}}X(w)+i\sum_{i}\lambda_{i}\cdot \bar{\partial}^{n_{i}}X(\bar{w_{i}})\right):\nonumber\\
		\times&\left .\left (-i(q+\xi)\cdot X(w)\dfrac{\alpha'p\cdot q}{2}\dfrac{\alpha'p^{\nu}}{2}
		+\sum_{i}i(q+\xi)\cdot \bar{\partial}^{n_{i}}X(w)\dfrac{\alpha'p\cdot q}{2}\dfrac{\alpha'\lambda_{i}^{\nu}}{2}
		-i(q+\xi)\cdot \partial^{m_{i}}X(w)\sum_{i}\dfrac{\alpha'q\cdot \zeta_{i}}{2}\dfrac{\alpha'p^{\nu}}{2}
	    \right)
	    \right|_{\text{multilinear in }  i\zeta_{i},i\lambda_{i} }\nonumber\\
	    =&\lim_{\xi \to 0}\dfrac{\pi\alpha'h_{\mu\nu}}{2}\partial^{\mu}_{\xi}\exp\left(ip\cdot X(w)+i\sum_{i}\zeta_{i}\cdot
	    \partial_{w}^{m_{i}}X(w)+i\sum_{i}\lambda_{i}\cdot \bar{\partial}^{n_{i}}X(\bar{w_{i}})\right)\nonumber\\
	    \times&\left(\dfrac{-(i\xi\cdot X p\cdot q-iq\cdot X p\cdot \xi)p^{\nu}}{p\cdot q}
	    +\sum_{i}\left (i\xi\cdot \bar{\partial}^{n_{i}}X(w)\lambda_{i}^{\nu}-\dfrac{
	    iq\cdot\bar{\partial}^{n_{i}}X(w)p\cdot\xi\lambda_{i}^{\nu}}{p\cdot q}\right )\right .\nonumber\\
	    &\left .\left .-\sum_{i}q\cdot \zeta_{i}p^{\nu}\left (\dfrac{i\xi\cdot\partial^{m_{i}}X(w)}{p\cdot q}-\dfrac{iq\cdot\partial^{m_{i}}X(w)p\cdot\xi}{(p\cdot q)^2}
	    \right)\right )
	    \right|_{\text{multilinear in }  i\zeta_{i},i\lambda_{i} }\nonumber\\
	    =&\dfrac{\pi\alpha'h_{\mu\nu}}{2}\exp\left(ip\cdot X(w)+i\sum_{i}\zeta_{i}\cdot
	    \partial_{w}^{m_{i}}X(w)+i\sum_{i}\lambda_{i}\cdot \bar{\partial}^{n_{i}}X(\bar{w_{i}})\right)\nonumber\\
	    \times&\left .\left(-i\dfrac{q_{a}L^{\mu a}p^{\nu}}{p\cdot q}
	    +\sum_{i}\left (i\bar{\partial}^{n_{i}}X^{\mu}(w)\lambda_{i}^{\nu}-\dfrac{
	    iq\cdot\bar{\partial}^{n_{i}}X(w)p^{\mu}\lambda_{i}^{\nu}}{p\cdot q}\right )
        -\sum_{i}q\cdot \zeta_{i}p^{\nu}\dfrac{i\partial^{m_{i}}X^{\mu}(w)}{p\cdot q}
	    \right)
	    \right|_{\text{multilinear in }  i\zeta_{i},i\lambda_{i} }
		\end{align}
		
		\item (1,0;0) terms\\
		\begin{align}
		&\lim_{\xi \to 0}h_{\mu\nu}\partial_{\xi}^{\mu}
		|z-w|^{\alpha'p\cdot (q+\xi)}
		\exp\left(ip\cdot X(w)+i\sum_{i}\zeta_{i}\cdot
		\partial_{w}^{m_{i}}X(w)+i\sum_{i}\lambda_{i}\cdot \bar{\partial}^{n_{i}}X(\bar{w_{i}})\right)\nonumber\\
		\times&\left(-\dfrac{\alpha'}{2}\sum_{i}\dfrac{(m_{i}-1)!(q+\xi)\cdot \zeta_{i}}{(z-w)^{m_{i}}}\right )\left(\dfrac{\alpha'p\cdot q}{2(z-w)}+\sum_{i}\dfrac{\alpha'm_{i}!q\cdot \zeta_{i}}{2(z-w)^{m_{i}+1}}+iq\cdot \partial X(z)  \right)\nonumber\\
		\times&\left .\left(\dfrac{-\alpha'p^{\nu}}{2(\bar{z}-\bar{w})}+\dfrac{\alpha'n_{i}!\lambda_{i}^{\nu}}{2(\bar{z}-\bar{w})^{n_{i}+1}}+i\bar{\partial}X^{\nu}(\bar{z}) \right)
		\right|_{\text{multilinear in }  i\zeta_{i},i\lambda_{i} }.
		\end{align}
		We write only the terms that have the factor $ |z-w|^{\alpha'p\cdot (q+\xi)-2} $.
		\begin{align}
		&\lim_{\xi \to 0}h_{\mu\nu}\partial_{\xi}^{\mu}
		\dfrac{2\pi}{\alpha'p\cdot (q+\xi)}
		\exp\left(ip\cdot X(w)+i\sum_{i}\zeta_{i}\cdot
		\partial_{w}^{m_{i}}X(w)+i\sum_{i}\lambda_{i}\cdot \bar{\partial}^{n_{i}}X(\bar{w_{i}})\right)\nonumber\\
		\times&\left .\left(-\dfrac{\alpha'}{2}\sum_{i}(q+\xi)\cdot \zeta_{i}\right )iq\cdot \partial^{m_{i}} X(w) 
		\dfrac{-\alpha'p^{\nu}}{2}
		\right|_{\text{multilinear in }  i\zeta_{i},i\lambda_{i} }\nonumber\\
		=&\left .\sum_{i}\dfrac{\pi\alpha'h_{\mu\nu}p^{\nu}iq\cdot\partial^{m_{i}} X(w)\zeta_{i}^{\mu}}{2p\cdot q}
		\exp\left(ip\cdot X(w)+i\sum_{i}\zeta_{i}\cdot
		\partial_{w}^{m_{i}}X(w)+i\sum_{i}\lambda_{i}\cdot \bar{\partial}^{n_{i}}X(\bar{w_{i}})\right)
		\right|_{\text{multilinear in }  i\zeta_{i},i\lambda_{i} }
		\end{align}
		\item (0,1;0) terms\\
		\begin{align}
		&\lim_{\xi \to 0}h_{\mu\nu}\partial_{\xi}^{\mu}
		|z-w|^{\alpha'p\cdot (q+\xi)}
		\exp\left(ip\cdot X(w)+i\sum_{i}\zeta_{i}\cdot
		\partial_{w}^{m_{i}}X(w)+i\sum_{i}\lambda_{i}\cdot \bar{\partial}^{n_{i}}X(\bar{w_{i}})\right)\nonumber\\
		\times&\left(-\dfrac{\alpha'}{2}\sum_{i}\dfrac{(n_{i}-1)!(q+\xi)\cdot \lambda_{i}}{(\bar{z}-\bar{w})^{n_{i}}} \right)
		\left(\dfrac{\alpha'p\cdot q}{2(z-w)}+\sum_{i}\dfrac{\alpha'm_{i}!q\cdot \zeta_{i}}{2(z-w)^{m_{i}+1}}+iq\cdot \partial X(z)  \right)\nonumber\\
		&\left .\left(\dfrac{-\alpha'p^{\nu}}{2(\bar{z}-\bar{w})}+\dfrac{\alpha'n_{i}!\lambda_{i}^{\nu}}{2(\bar{z}-\bar{w})^{n_{i}+1}} +i\bar{\partial}X^{\nu}(\bar{z})\right)
		\right|_{\text{multilinear in }  i\zeta_{i},i\lambda_{i} }.
		\end{align}
		We write only the terms that have the factor $ |z-w|^{\alpha'p\cdot (q+\xi)-2} $.
		\begin{align}
		&\lim_{\xi \to 0}h_{\mu\nu}\partial_{\xi}^{\mu}
		\dfrac{2\pi}{\alpha'p\cdot(q+\xi)}
		\exp\left(ip\cdot X(w)+i\sum_{i}\zeta_{i}\cdot
		\partial_{w}^{m_{i}}X(w)+i\sum_{i}\lambda_{i}\cdot \bar{\partial}^{n_{i}}X(\bar{w_{i}})\right)\nonumber\\
		\times&\left .\left(-\dfrac{\alpha'}{2}\sum_{i}(q+\xi)\cdot \lambda_{i} \right)
		\dfrac{\alpha'p\cdot q}{2}
		i\bar{\partial}^{n_{i}}X^{\nu}(\bar{w})
		\right|_{\text{multilinear in }  i\zeta_{i},i\lambda_{i} }\nonumber\\
		=&\left .\dfrac{\pi\alpha'h_{\mu\nu}i\bar{\partial}^{n_{i}}X^{\nu}(\bar{w})}{2}
		\sum_{i}\left(-\lambda_{i}^{\mu}+\dfrac{q\cdot\lambda_{i}p^{\mu}}{p\cdot q}\right)
		\exp\left(ip\cdot X(w)+i\sum_{i}\zeta_{i}\cdot
		\partial_{w}^{m_{i}}X(w)+i\sum_{i}\lambda_{i}\cdot \bar{\partial}^{n_{i}}X(\bar{w_{i}})\right)
		\right|_{\text{multilinear in }  i\zeta_{i},i\lambda_{i} }.
		\end{align}
	\end{enumerate}
    By summing up these results, we obtain
	\begin{align}
	&\dfrac{\pi\alpha'h_{\mu\nu}}{2}
	\left[
	\dfrac{-iq_{a}L^{\mu a}p^{\nu}}{p\cdot q}
	-\sum_{i}\dfrac{ip^{\nu}q_{a}\lambda_{ic}\bar{\partial}^{n_{i}}X_{d}(S^{\mu a})^{cd}}{p\cdot q}
	+\sum_{i}\dfrac{ip^{\nu}q_{a}\zeta_{ic}\partial^{m_{i}}X_{d}(S^{\mu a})^{cd}}{p\cdot q}-i(S^{\mu\nu})^{cd}\lambda_{ic}\bar{\partial}^{n_{i}}X_{d}
	\right]\nonumber\\
	&\times \left .\exp\left(ip\cdot X(w)+i\sum_{i}\zeta_{i}\cdot
	\partial_{w}^{m_{i}}X(w)+i\sum_{i}\lambda_{i}\cdot \bar{\partial}^{n_{i}}X(\bar{w_{i}})\right)
	\right|_{\text{multilinear in }  i\zeta_{i},i\lambda_{i} }.\label{eq.subleading soft b field holomorphic}
	\end{align}
	
	On the other hand, we can rewrite the vertex in the alternative form of eq.(\ref{eq.soft B field vertex}).
	\begin{align}
&:\partial X^{\mu}(z)\bar{\partial}X^{\nu}\exp\left(iq\cdot X(z,\bar{z})\right):\nonumber\\
=&\bar{\partial}:\partial X^{\nu}(z,\bar{z})X^{\mu}(\bar{z})\exp\left(iq\cdot X(z,\bar{z})\right):
-:X^{\nu}(z,\bar{z})\partial X^{\mu}(z,\bar{z})\bar{\partial_{z}} \exp\left(iq\cdot X(z,\bar{z})\right):.
	\end{align}
	This contribution is given by changing $ \mu,\lambda_{i},\bar{\partial}^{n_{i}} $ with $ \nu,\zeta_{i},\partial^{m_{i}} $ in the square bracket in eq.(\ref{eq.subleading soft b field holomorphic}).
	\begin{align}
	&\dfrac{\pi\alpha'h_{\mu\nu}}{2}
	\left[
	\dfrac{-iq_{a}L^{\nu a}p^{\mu}}{p\cdot q}
	+\sum_{i}\dfrac{ip^{\mu}q_{a}\lambda_{ic}\bar{\partial}^{n_{i}}X_{d}(S^{\nu a})^{cd}}{p\cdot q}
	-\sum_{i}\dfrac{ip^{\mu}q_{a}\zeta_{ic}\partial^{m_{i}}X_{d}(S^{\nu a})^{cd}}{p\cdot q}+i(S^{\mu\nu})^{cd}\zeta_{ic}\partial^{m_{i}}X_{d}
	\right]\nonumber\\
	&\times \left .\exp\left(ip\cdot X(w)+i\sum_{i}\zeta_{i}\cdot
	\partial_{w}^{m_{i}}X(w)+i\sum_{i}\lambda_{i}\cdot \bar{\partial}^{n_{i}}X(\bar{w_{i}})\right)
	\right|_{\text{multilinear in }  i\zeta_{i},i\lambda_{i} }.\label{eq.subleading soft b field antiholomorphic}
	\end{align}
	Thus averaging these two results, eq.(\ref{eq.subleading soft b field holomorphic}) and eq.(\ref{eq.subleading soft b field antiholomorphic}), we obtain
	\begin{align}
    &\dfrac{\pi\alpha'h_{\mu\nu}}{2}
    \left[
    -\sum_{i}\dfrac{ip^{\nu}q_{a}\lambda_{ic}\bar{\partial}^{n_{i}}X_{d}(S^{\mu a})^{cd}-(\mu\leftrightarrow\nu)}{p\cdot q}
    +\sum_{i}\dfrac{ip^{\nu}q_{a}\zeta_{ic}\partial^{m_{i}}X_{d}(S^{\mu a})^{cd}-(\mu\leftrightarrow\nu)}{p\cdot q}\right .\nonumber\\
    &\left .-\dfrac{1}{2}i(S^{\mu\nu})^{cd}\lambda_{ic}\bar{\partial}^{n_{i}}X_{d}
    +\dfrac{1}{2}i(S^{\mu\nu})^{cd}\zeta_{ic}\partial^{m_{i}}X_{d}
    \right]\nonumber\\
    &\times \left .\exp\left(ip\cdot X(w)+i\sum_{i}\zeta_{i}\cdot
    \partial_{w}^{m_{i}}X(w)+i\sum_{i}\lambda_{i}\cdot \bar{\partial}^{n_{i}}X(\bar{w_{i}})\right)
    \right|_{\text{multilinear in }  i\zeta_{i},i\lambda_{i} }\nonumber\\
    =&\dfrac{-i\pi\alpha'h_{\mu\nu}}{2}
    \left(\dfrac{p^{\nu}q_{a}S^{\mu a}-(\mu\leftrightarrow\nu)}{p\cdot q} 
    -\dfrac{p^{\nu}q_{a}\bar{S}^{\mu a}-(\mu\leftrightarrow\nu)}{p\cdot q}
    +\dfrac{1}{2}\left(S^{\mu\nu}-\bar{S}^{\mu\nu}\right)
    \right)\nonumber\\
    &\times :\partial^{m_{1}}X^{\rho_{1}}\cdots \bar{\partial}^{n_{1}}X^{\sigma_{1}}\cdots \exp\left(ip\cdot X\right) :.
	\end{align}
\end{enumerate}

\section{Derivation of eq.(\ref{eq.soft graviton theorem for massive 2}) from Ward identity}\quad \label{section;from Ward identity}

	We can derive the soft theorem eq.(\ref{eq.soft graviton theorem for massive 2}) by using the Ward identity as in the case of field theory.
	First, we calculate the coefficients of $ |z-w|^{-2} $ in the OPE of a graviton and the massive particle.
	This corresponds to the three-point function of a graviton, the massive particle and the intermediate particle in the left diagram in fig.(\ref{fig:diagram}).
	By using the on shell conditions $ p^2=(p+q)^2=-\dfrac{4}{\alpha'}$, $q^2=0 $ and ignoring $ \mathcal{O}(q^3) $ terms which is irrelevant through subsubleading order, we obtain 
	
	\begin{align}
	&\left .:\partial X^{\mu}(z)\bar{\partial}X^{\nu}(\bar{z})\exp(iq\cdot X(z))::\partial X^{\rho_{1}}\partial X^{\rho_{2}}\bar{\partial}X^{\sigma_{1}}\bar{\partial}X^{\sigma_{2}}\exp(ip\cdot X(w)):\right |_{|z-w|^{-2}}\nonumber\\
	=&-\dfrac{\alpha'^2 p^{\mu}p^{\nu}}{4}V^{\rho_{1}\rho_{2}\sigma_{1}\sigma_{2}}(w;p+q)\nonumber\\
	+&\dfrac{\alpha'^2}{4}\left\{
	i(p^{\mu}\eta^{\nu\sigma_{1}}+p^{\nu}\eta^{\mu\sigma_{1}})iq_{a}V^{\rho_{1}\rho_{2}a\sigma_{2}}
	-i(p^{\mu}\delta^{\nu}_{c}+p^{\nu}\delta^{\mu}_{c})iq^{\sigma_{1}}T^{\rho_{1}\rho_{2}c\sigma_{2}}
	+i(\eta^{\mu\rho_{1}}p^{\nu}+\eta^{\nu\rho_{1}}p^{\mu})iq_{a}V^{a\rho_{2}\sigma_{1}\sigma_{2}}
	-i(p^{\nu}\delta^{\mu}_{c}+p^{\mu}\delta^{\nu}_{c})iq^{\rho_{1}}T^{c\rho_{2}\sigma_{1}\sigma_{2}}\right \}\nonumber\\
	+&\dfrac{\alpha'^2}{16}\left\{
	4\alpha'p^{\mu}p^{\nu}iq^{\rho_{1}}iq_{a}V^{a\rho_{2}\sigma_{1}\sigma_{2}}
	+4\alpha'p^{\mu}p^{\nu}iq^{\sigma_{1}}iq_{a}V^{\rho_{1}\rho_{2}a\sigma_{2}}
	+\alpha'i(\eta^{\nu\sigma_{1}}p^{\mu}+\eta^{\mu\sigma_{1}}p^{\nu})iq^{\sigma_{2}}iq_{a}V^{\rho_{1}\rho_{2};a}
	-\alpha'i(p^{\mu}\delta^{\nu}_{c}+p^{\nu}\delta^{\mu}_{c})iq^{\sigma_{1}}iq^{\sigma_{2}}V^{\rho_{1}\rho_{2};c}\right.\nonumber\\
	+&\left .\alpha'i(\eta^{\mu\rho_{1}}p^{\nu}+\eta^{\nu\rho_{1}}p^{\mu})iq^{\rho_{2}}iq_{a}V^{a;\sigma_{1}\sigma_{2}}
	+8(\eta^{\mu\rho_{1}}\eta^{\nu\sigma_{1}}+\eta^{\nu\rho_{1}}\eta^{\mu\sigma_{1}})iq_{a}iq_{b}V^{a\rho_{2}b\sigma_{2}}
	-8(\eta^{\mu\rho_{1}}\delta^{\nu}_{c}+\eta^{\nu\rho_{1}}\delta^{\mu}_{c})iq^{\sigma_{1}}iq_{a}V^{a\rho_{2}c\sigma_{2}}\right .\nonumber\\
	&\left .-\alpha'i(p^{\mu}\delta^{\nu}_{c}+p^{\nu}\delta^{\mu}_{c})iq^{\rho_{1}}iq^{\rho_{2}}V^{c;\sigma_{1}\sigma_{2}}
	-8(\eta^{\nu\sigma_{1}}\delta^{\mu}_{c}+\eta^{\mu\sigma_{1}}\delta^{\nu}_{c})iq^{\rho_{1}}iq_{a}V^{c\rho_{2}a\sigma_{2}}
	+4iq^{\rho_{1}}iq^{\sigma_{1}}(V^{\mu\rho_{2}\nu\sigma_{2}}+V^{\nu\rho_{2}\nu\sigma_{2}})
	\right\}(w;p+q)+\mathcal{O}(q^3),
	\end{align}
	where $ V^{\rho_{1}\rho_{2}\sigma_{1}\sigma_{2}}(w;p+q)=\partial X^{\rho_{1}}\partial X^{\rho_{2}} \bar{\partial}X^{\sigma_{1}}\bar{\partial}X^{\sigma_{2}}\exp(i(p+q)\cdot X)(w) $, $ V^{\rho_{1}\rho_{2};a}(w;p+q)=\partial X^{\rho_{1}}\partial X^{\rho_{2}} \bar{\partial}^2X^{a}\exp(i(p+q)\cdot X)(w)  $.
	Here the polarization tensors of the graviton and massive hard particle, $ h_{\mu\nu}$ and $A_{\rho_{1}\rho_{2}\sigma_{1}\sigma_{2}} $, are omitted.
	By multiplying the propagator in fig.(\ref{fig:diagram}), we can obtain the terms proportional to $ \dfrac{1}{p\cdot q} $ in the soft theorem. 
	
	We write the right diagram in fig.(\ref{fig:diagram}) as $ h_{\mu\nu}N^{\mu\nu}_{n}(q;p_{1},\cdots,p_{n}) $, which does not have the factor $ \dfrac{1}{p\cdot q} $.
	Then the amplitude becomes
	\begin{align}
	&M_{n+1}(q;p,\cdots,p_{n})\nonumber\\
	=&-\dfrac{\alpha'^2h_{\mu\nu}p^{\mu}p^{\nu}}{8p\cdot q}T_{n}^{\rho_{1}\rho_{2}\sigma_{1}\sigma_{2}}(p+q,\cdots)\nonumber\\
	-&\dfrac{\alpha'^2h_{\mu\nu}}{8p\cdot q}
	\left [
	(p^{\mu}\eta^{\nu\sigma_{1}}+p^{\nu}\eta^{\mu\sigma_{1}})q_{a}T_{n}^{\rho_{1}\rho_{2}a\sigma_{2}}
	-(p^{\mu}\delta^{\nu}_{c}+p^{\nu}\delta^{\mu}_{c})q^{\sigma_{1}}T_{n}^{\rho_{1}\rho_{2}c\sigma_{2}}
	+(\eta^{\mu\rho_{1}}p^{\nu}+\eta^{\nu\rho_{1}}p^{\mu})q_{a}T_{n}^{a\rho_{2}\sigma_{1}\sigma_{2}}
	-(p^{\nu}\delta^{\mu}_{c}+p^{\mu}\delta^{\nu}_{c})q^{\rho_{1}}T_{n}^{c\rho_{2}\sigma_{1}\sigma_{2}}
	\right ](p+q,\cdots)\nonumber\\
	-&\dfrac{\alpha'^2h_{\mu\nu}}{32p\cdot q}
	\left[
	4\alpha'p^{\mu}p^{\nu}q^{\rho_{1}}q_{a}T_{n}^{a\rho_{2}\sigma_{1}\sigma_{2}}
	+4\alpha'p^{\mu}p^{\nu}q^{\sigma_{1}}q_{a}T_{n}^{\rho_{1}\rho_{2}a\sigma_{2}}
	+\alpha'(\eta^{\nu\sigma_{1}}p^{\mu}+\eta^{\mu\sigma_{1}}p^{\nu})q^{\sigma_{2}}q_{a}iT_{n}^{\rho_{1}\rho_{2};a}
	-\alpha'(p^{\mu}\delta^{\nu}_{c}+p^{\nu}\delta^{\mu}_{c})q^{\sigma_{1}}q^{\sigma_{2}}iT_{n}^{\rho_{1}\rho_{2};c}\right .\nonumber\\
	+&\alpha'(\eta^{\mu\rho_{1}}p^{\nu}+\eta^{\nu\rho_{1}}p^{\mu})q^{\rho_{2}}q_{a}iT_{n}^{a;\sigma_{1}\sigma_{2}}
	+8(\eta^{\mu\rho_{1}}\eta^{\nu\sigma_{1}}+\eta^{\nu\rho_{1}}\eta^{\mu\sigma_{1}})q_{a}q_{b}T_{n}^{a\rho_{2}b\sigma_{2}}
	-8(\eta^{\mu\rho_{1}}\delta^{\nu}_{c}+\eta^{\nu\rho_{1}}\delta^{\mu}_{c})q^{\sigma_{1}}q_{a}T_{n}^{a\rho_{2}c\sigma_{2}}\nonumber\\
	-&\left .\alpha'(p^{\nu}\delta^{\mu}_{c}+p^{\mu}\delta^{\nu}_{c})q^{\rho_{1}}q^{\rho_{2}}iT_{n}^{c;\sigma_{1}\sigma_{2}}
	-8(\eta^{\nu\sigma_{1}}\delta^{\mu}_{c}+\eta^{\mu\sigma_{1}}\delta^{\nu}_{c})q^{\rho_{1}}q_{a}T_{n}^{c\rho_{2}a\sigma_{2}}
	+8q^{\rho_{1}}q^{\sigma_{1}}(T_{n}^{\mu\rho_{2}\nu\sigma_{2}}+T_{n}^{\nu\rho_{2}\mu\sigma_{2}})
	\right](p+q,\cdots)+\mathcal{O}(q^2)\nonumber\\
	+&h_{\mu\nu}N^{\mu\nu}_{n}(q;p,\cdots),
	\end{align}
	where $ T_{n}^{\rho_{1}\rho_{2}\sigma_{1}\sigma_{2}}(p,\cdots,p_{n}) $ is the result of removing the polarization tensor $ A_{\rho_{1}\rho_{2}\sigma_{1}\sigma_{2}} $ from the original scattering amplitude $ M_{n}(p,\cdots, p_{n}) $ : $ M_{n}(p,\cdots, p_{n})=A_{\rho_{1}\rho_{2}\sigma_{1}\sigma_{2}}T_{n}^{\rho_{1}\rho_{2}\sigma_{1}\sigma_{2}}(p,\cdots,p_{n}) $ or  $ T^{\rho_{1}\rho_{2}\sigma_{1}\sigma_{2}}_{n}(p,\cdots)=\int d^2w\cdots \langle :V^{\rho_{1}\rho_{2}\sigma_{1}\sigma_{2}}(w;p):\cdots \rangle  $.
	
	Next, we expand the Ward identity, $ \left. M_{n+1}(q;p,\cdots)\right|_{h_{\mu\nu}\rightarrow q_{\mu}}=0 $, by the soft momentum q. 
	\begin{enumerate}
		\item $ \mathcal{O}(q^0)  $
		\begin{align}
		0=
		-\dfrac{\alpha'^2q_{\mu}p^{\mu}p^{\nu}}{8p\cdot q}T_{n}^{\rho_{1}\rho_{2}\sigma_{1}\sigma_{2}}(p,\cdots)
		=-\alpha'^2p^{\nu}T^{\rho_{1}\rho_{2}a\sigma_{2}}(p,\cdots).
		\end{align}
		When we sum up for all the hard vertices, the right hand side becomes zero by momentum conservation.
		
		\item $ \mathcal{O}(q) $
		\begin{align}
		q_{\mu}N^{\mu\nu}(q=0;p_{1},\cdots,p_{n})
		=&\dfrac{\alpha'^2p^{\nu}}{8}q_{a}\partial_{p}^{a}T_{n}^{\rho_{1}\rho_{2}\sigma_{1}\sigma_{2}}(p,\cdots )\nonumber\\
		&+\dfrac{\alpha'^2q_{\mu}}{8p\cdot q}
		\left [
		(p\cdot q\eta^{\nu\sigma_{1}}+p^{\nu}q^{\sigma_{1}})q_{a}T_{n}^{\rho_{1}\rho_{2}a\sigma_{2}}
		-(p\cdot q\delta^{\nu}_{c}+p^{\nu}q_{c}) q^{\sigma_{1}}T_{n}^{\rho_{1}\rho_{2}c\sigma_{2}}\right .\nonumber\\
		&\left .+(q^{\rho_{1}}p^{\nu}+p\cdot q\eta^{\nu\rho_{1}})q_{a}T_{n}^{a\rho_{2}\sigma_{1}\sigma_{2}}
		-(p^{\nu}q_{c}+p\cdot q\delta^{\nu}_{c})q^{\rho_{1}}T_{n}^{c\rho_{2}\sigma_{1}\sigma_{2}}
		\right ](p,\cdots)\nonumber\\
		=&\dfrac{\alpha'^2p^{\nu}}{8}q_{a}\partial_{p}^{a}T_{n}^{\rho_{1}\rho_{2}\sigma_{1}\sigma_{2}}(p,\cdots )\nonumber\\
		&+\dfrac{\alpha'^2}{8}
		\left [
		\eta^{\nu\sigma_{1}}q_{a}T_{n}^{\rho_{1}\rho_{2}a\sigma_{2}}
		-q^{\sigma_{1}}T_{n}^{\rho_{1}\rho_{2}\nu\sigma_{2}}
		\eta^{\nu\rho_{1}}q_{a}T_{n}^{a\rho_{2}\sigma_{1}\sigma_{2}}
		-q^{\rho_{1}}T_{n}^{\nu\rho_{2}\sigma_{1}\sigma_{2}}
		\right ](p,\cdots ).
		\end{align}
		Therefore, for soft graviton we obtain
		\begin{align}
		N^{\mu\nu}(q=0;p,\cdots,p_{n})
		=\dfrac{\alpha'^2}{16}(p^{\nu}\partial_{p}^{\mu}+p^{\mu}\partial_{p}^{\nu})T_{n}^{\rho_{1}\rho_{2}\sigma_{1}\sigma_{2}}(p,\cdots )
		\end{align}
		
		\item $ \mathcal{O}(q^2) $
		\begin{align}
		q_{a}q_{\mu}\partial_{q}^{a}N^{\mu\nu}(q=0;p_{1}\cdots,p_{n})
		=&\dfrac{\alpha'^2q_{\mu}p^{\mu}p^{\nu}}{8p\cdot q}\dfrac{q_{a}q_{b}}{2}\partial_{p}^{a}\partial_{p}^{b}T_{n}^{\rho_{1}\rho_{2}\sigma_{1}\sigma_{2}}(p,\cdots )\nonumber\\
		&+\dfrac{\alpha'^2p\cdot q}{8p\cdot q}\left(
		\eta^{\nu\sigma_{1}}q_{a}-\delta^{\nu}_{a}q^{\sigma_{1}}
		\right)q_{b}\partial_{p}^{b}T_{n}^{\rho_{1}\rho_{2}a\sigma_{2}}(p,\cdots )\nonumber\\
		&+\dfrac{\alpha'^2p\cdot q}{8p\cdot q}\left(
		\eta^{\nu\rho_{1}}q_{a}-\delta^{\nu}_{a}q^{\rho_{1}}
		\right)q_{b}\partial_{p}^{b}T_{n}^{a\rho_{2}\sigma_{1}\sigma_{2}}(p,\cdots )\nonumber\\
		&+\dfrac{\alpha'^3}{32p\cdot q}\left(
		4p\cdot qp^{\nu}q^{\rho_{1}}q_{a}T_{n}^{a\rho_{2}\sigma_{1}\sigma_{2}}
		+4p\cdot qp^{\nu}q^{\sigma_{1}}q_{a}T_{n}^{\rho_{1}\rho_{2}a\sigma_{2}}\right .\nonumber\\
		&\left .+p\cdot q\eta^{\nu\sigma_{1}}q^{\sigma_{2}}q_{a}iT_{n}^{\rho_{1}\rho_{2};a}
		-p\cdot qq^{\sigma_{1}}q^{\sigma_{2}}iT_{n}^{\rho_{1}\rho_{2};\nu}
		+p\cdot q\eta^{\nu\rho_{1}}q^{\rho_{2}}q_{a}iT_{n}^{a;\sigma_{1}\sigma_{2}}
		-p\cdot qq^{\rho_{1}}q^{\rho_{2}}iT_{n}^{\nu;\sigma_{1}\sigma_{2}}
		\right)	(p,\cdots)
		\end{align}
		We can determine the symmetric part $ \dfrac{\partial_{q}^{a}N^{\mu\nu}+\partial_{q}^{\mu}N^{a\nu}}{2} $ from the above equation.
		\begin{align}
		\dfrac{\partial_{q}^{a}N^{\mu\nu}+\partial_{q}^{\mu}N^{a\nu}}{2}
		=&\dfrac{\alpha'^2p^{\nu}}{16}\partial_{p}^{a}\partial_{p}^{\mu}T^{\rho_{1}\rho_{2}\sigma_{1}\sigma_{2}}_{n}(p\cdots,)\nonumber\\
		+&\dfrac{\alpha'^2}{16}(S^{\nu\mu})^{\sigma_{1}}_{\ c}\partial_{p}^{a}T_{n}^{\rho_{1}\rho_{2}c\sigma_{2}}(p\cdots,)
		+\dfrac{\alpha'^2}{16}(S^{\nu a})^{\sigma_{1}}_{\ c}\partial_{p}^{\mu}T_{n}^{\rho_{1}\rho_{2}c\sigma_{2}}(p\cdots,)\nonumber\\
		+&\dfrac{\alpha'^2}{16}(S^{\nu\mu})^{\rho_{1}}_{\ c}\partial_{p}^{a}T_{n}^{c\rho_{2}\sigma_{1}\sigma_{2}}(p\cdots,)
		+\dfrac{\alpha'^2}{16}(S^{\nu a})^{\rho_{1}}_{\ c}\partial_{p}^{\mu}T_{n}^{c\rho_{2}\sigma_{1}\sigma_{2}}(p\cdots,)\nonumber\\
		+&\dfrac{\alpha'^3}{64}\left[
		4p^{\nu}(\eta^{\rho_{1}\mu}\delta^{a}_{c}+\eta^{\rho_{1}a}\delta^{\mu}_{c})T_{n}^{c\rho_{2}\sigma_{1}\sigma_{2}}
		+4p^{\nu}(\eta^{\sigma_{1}\mu}\delta^{a}_{c}+\eta^{a\sigma_{1}}\delta^{\mu}_{c})T_{n}^{\rho_{1}\rho_{2}c\sigma_{2}}\right .\nonumber\\
		+&\eta^{\nu\sigma_{1}}(\eta^{\mu\sigma_{2}}\delta^{a}_{c}+\eta^{a\sigma_{2}}\delta^{\mu}_{c})iT_{n}^{\rho_{1}\rho_{2};c}
		-(\eta^{\mu\sigma_{1}}\eta^{a\sigma_{2}}+\eta^{\mu\sigma_{2}}\eta^{a\sigma_{1}})iT_{n}^{\rho_{1}\rho_{2};\nu}\nonumber\\
		+&\left .\eta^{\nu\rho_{1}}(\eta^{\mu\rho_{2}}\delta^{a}_{c}+\eta^{a\rho_{2}}\delta^{\mu}_{c})iT_{n}^{c;\sigma_{1}\sigma_{2}}
		-(\eta^{\mu\rho_{1}}\eta^{a\rho_{2}}+\eta^{\mu\rho_{2}}\eta^{a\rho_{1}})iT_{n}^{\nu;\sigma_{1}\sigma_{2}}
		\right](p,\cdots)\nonumber\\
		\equiv&f^{a\mu\nu}.\label{eq.symm}
		\end{align}
		On the other hand, by exchanging the indexes $ \mu$ and $\nu$ in the above equation we obtain
		\begin{align}
		\dfrac{\partial_{q}^{a}N^{\mu\nu}+\partial_{q}^{\nu}N^{a\mu}}{2}
		=f^{a\nu\mu}.
		\end{align}
		By taking difference between the above two equations, we obtain
		\begin{align}
		\dfrac{\partial_{q}^{\mu}N^{a\nu}-\partial_{q}^{\nu}N^{a\mu }}{2}(q=0;p_{1},\cdots,p_{n})
		=f^{a\mu\nu}-f^{a\nu\mu}
		\end{align}
		By changing the indexes $ \mu\rightarrow a,\nu\rightarrow\mu$ and $a\rightarrow\nu $, we obtain 
		\begin{align}
		\dfrac{\partial_{q}^{a}N^{\nu\mu}-\partial_{q}^{\mu}N^{\nu a}}{2}(q=0;p_{1},\cdots,p_{n})
		=&f^{\nu a \mu}-f^{\nu \mu a}.\label{eq.antisymm}
		\end{align}

		Then we add eq.(\ref{eq.symm}) and eq.(\ref{eq.antisymm}),
		\begin{align}
		\partial_{q}^{a}N^{\mu\nu}(q=0;p_{1},\cdots,p_{n})
		=&f^{a\mu\nu}+f^{\nu a \mu}-f^{\nu \mu a}\nonumber\\
		=&\dfrac{\alpha'^2}{16}\left \{(p^{\nu}\partial_{p}^{\mu}+p^{\mu}\partial_{p}^{\nu})\partial_{p}^{a}-p^{a}\partial_{p}^{\mu}\partial_{p}^{\nu}\right \}T^{\rho_{1}\rho_{2}\sigma_{1}\sigma_{2}}_{n}(p\cdots,)\nonumber\\
		+&\dfrac{\alpha'^2}{8}\left[
		(S^{\nu a})^{\sigma_{1}}_{\ c}\partial_{p}^{\mu}T_{n}^{\rho_{1}\rho_{2}c\sigma_{2}}
		+(S^{\mu a})^{\sigma_{1}}_{\ c}\partial_{p}^{\nu}T_{n}^{\rho_{1}\rho_{2}c\sigma_{2}}
		+(S^{\nu a})^{\rho_{1}}_{\ c}\partial_{p}^{\mu}T_{n}^{c\rho_{2}\sigma_{1}\sigma_{2}}
		+(S^{\mu a})^{\rho _{1}}_{\ c}\partial_{p}^{\nu}T_{n}^{c\rho_{2}\sigma_{1}\sigma_{2}}
		\right]
		(p\cdots,)\nonumber\\
		+&\dfrac{\alpha'^3}{64}\left[
		\left\{4\eta^{\rho_{1}a}(p^{\mu}\delta^{\nu}_{c}+p^{\nu}\delta^{\mu}_{c})
		+4(\eta^{\rho_{1}\mu}p^{\nu}+\eta^{\rho_{1}\nu}p^{\mu})\delta^{a}_{c}
		-4p^{a}(\eta^{\rho_{1}\mu}\delta^{\nu}_{c}+\eta^{\rho_{1}\nu}\delta^{\mu}_{c})
		\right\}T_{n}^{c\rho_{2}\sigma_{1}\sigma_{2}}\right .\nonumber\\
		&\left.+\left\{4\eta^{\sigma_{1}a}(p^{\mu}\delta^{\nu}_{c}+p^{\nu}\delta^{\mu}_{c})
		+4(\eta^{\sigma_{1}\mu}p^{\nu}+\eta^{\sigma_{1}\nu}p^{\mu})\delta^{a}_{c}
		-4p^{a}(\eta^{\sigma_{1}\mu}\delta^{\nu}_{c}+\eta^{\nu\sigma_{1}}\delta^{\mu}_{c})
		\right\}T_{n}^{\rho_{1}\rho_{2}c\sigma_{2}}\right .\nonumber\\
		&+\left\{4\eta^{\mu\sigma_{1}}\eta^{\nu\sigma_{2}}\delta^{a}_{c}
		-2\eta^{a\sigma_{2}}(\eta^{\mu\sigma_{1}}\delta^{\nu}_{c}
		+\eta^{\nu\sigma_{1}}\delta^{\mu}_{c})\right\}iT_{n}^{\rho_{1}\rho_{2};c}\nonumber\\
		&+\left.\left\{4\eta^{\mu\rho_{1}}\eta^{\nu\rho_{2}}\delta^{a}_{c}
		-2\eta^{a\rho_{2}}(\eta^{\mu\rho_{1}}\delta^{\nu}_{c}
		+\eta^{\nu\rho_{1}}\delta^{\mu}_{c})\right\}iT_{n}^{c;\sigma_{1}\sigma_{2}}
		\right](p,\cdots).
		\end{align}

	\end{enumerate}
	If we substitute this, the subsubleading soft theorem becomes
	\begin{align}
	\left .M_{n+1}(p,\cdots,q)\right |_{\text{subsubleading}}=&-\dfrac{\alpha'^2q_{a}q_{b}p^{\mu}p^{\nu}}{16p\cdot q}\partial_{p}^{a}\partial_{p}^{b}T_{n}(p,\cdots)\nonumber\\
	&-\dfrac{\alpha'^2p^{\mu}q_{a}q_{b}}{4p\cdot q}(\eta^{\nu\sigma_{1}}\delta^{a}_{c}-\eta^{a\sigma_{1}}\delta^{\nu}_{c})\partial_{p}^{b}T_{n}^{\rho_{1}\rho_{2}c\sigma_{2}}(p,\cdots)
	-\dfrac{\alpha'^2p^{\nu}q_{a}q_{b}}{4p\cdot q}(\eta^{\mu\rho_{1}}\delta^{a}_{c}-\eta^{a\rho_{1}}\delta^{\mu}_{c})\partial_{p}^{b}T_{n}(p,\cdots )\nonumber\\
	&-\dfrac{\alpha'^2}{16p\cdot q}
	\left[
	-2\alpha'p^{\mu}p^{\nu}q^{\rho_{1}}q_{a}T_{n}^{a\rho_{2}\sigma_{1}\sigma_{2}}
	-2\alpha'p^{\mu}p^{\nu}q^{\sigma_{1}}q_{a}T_{n}^{\rho_{1}\rho_{2}a\sigma_{2}}
	+\alpha'p^{\mu}\eta^{\nu\sigma_{1}}q^{\sigma_{2}}q_{a}iT_{n}^{\rho_{1}\rho_{2};a}\right .\nonumber\\
	&\left .-\alpha'p^{\mu}q^{\sigma_{1}}q^{\sigma_{2}}iT_{n}^{\rho_{1}\rho_{2};\nu}
	+\alpha'\eta^{\mu\rho_{1}}p^{\nu}q^{\rho_{2}}q_{a}iT_{n}^{a;\sigma_{1}\sigma_{2}}\right .\nonumber\\
	&\left .+8\eta^{\mu\rho_{1}}\eta^{\nu\sigma_{1}}q_{a}q_{b}iT_{n}^{a\rho_{2}b\sigma_{2}}
	-8\eta^{\mu\rho_{1}}q^{\sigma_{1}}q_{a}T_{n}^{a\rho_{2}\nu\sigma_{2}}
	-\alpha'p^{\nu}q^{\rho_{1}}q^{\rho_{2}}iT_{n}^{\mu;\sigma_{1}\sigma_{2}}
	-8\eta^{\nu\sigma_{1}}q^{\rho_{1}}q_{a}T_{n}^{\mu\rho_{2}a\sigma_{2}}
	+8q^{\rho_{1}}q^{\sigma_{2}}T_{n}^{\mu\rho_{2}\nu\sigma_{2}}
	\right](p,\cdots)\nonumber\\
	&+\dfrac{\alpha'^3}{16}\left \{(p^{\nu}\partial_{p}^{\mu}+p^{\mu}\partial_{p}^{\nu})\partial_{p}^{a}-p^{a}\partial_{p}^{\mu}\partial_{p}^{\nu}\right \}T^{\rho_{1}\rho_{2}\sigma_{1}\sigma_{2}}_{n}(p\cdots,)\nonumber\\
	&+\dfrac{\alpha'^3}{8}\left[
	(S^{\nu a})^{\sigma_{1}}_{\ c}\partial_{p}^{\mu}T_{n}^{\rho_{1}\rho_{2}c\sigma_{2}}
	+(S^{\mu a})^{\sigma_{1}}_{\ c}\partial_{p}^{\nu}T_{n}^{\rho_{1}\rho_{2}c\sigma_{2}}
	+(S^{\nu a})^{\rho_{1}}_{\ c}\partial_{p}^{\mu}T_{n}^{c\rho_{2}\sigma_{1}\sigma_{2}}
	+(S^{\mu a})^{\rho_{1}}_{\ c}\partial_{p}^{\nu}T_{n}^{c\rho_{2}\sigma_{1}\sigma_{2}}
	\right]
	(p\cdots,)\nonumber\\
	&+\dfrac{\alpha'^3}{64}\left[
	\left\{4\eta^{\rho_{1}a}(p^{\mu}\delta^{\nu}_{c}+p^{\nu}\delta^{\mu}_{c})
	+4(\eta^{\rho_{1}\mu}p^{\nu}+\eta^{\rho_{1}\nu}p^{\mu})\delta^{a}_{c}
	-4p^{a}(\eta^{\rho_{1}\mu}\delta^{\nu}_{c}+\eta^{\rho_{1}\nu}\delta^{\mu}_{c})
	\right\}T_{n}^{c\rho_{2}\sigma_{1}\sigma_{2}}\right .\nonumber\\
	&\left.+\left\{4\eta^{\sigma_{1}a}(p^{\mu}\delta^{\nu}_{c}+p^{\nu}\delta^{\mu}_{c})
	+4(\eta^{\sigma_{1}\mu}p^{\nu}+\eta^{\sigma_{1}\nu}p^{\mu})\delta^{a}_{c}
	-4p^{a}(\eta^{\sigma_{1}\mu}\delta^{\nu}_{c}+\eta^{\nu\sigma_{1}}\delta^{\mu}_{c})
	\right\}T_{n}^{\rho_{1}\rho_{2}c\sigma_{2}}\right .\nonumber\\
	&+\left\{4\eta^{\mu\sigma_{1}}\eta^{\nu\sigma_{2}}\delta^{a}_{c}
	-2\eta^{a\sigma_{2}}(\eta^{\mu\sigma_{1}}\delta^{\nu}_{c}
	+\eta^{\nu\sigma_{1}}\delta^{\mu}_{c})\right\}iT_{n}^{\rho_{1}\rho_{2};c}\nonumber\\
	&+\left.\left\{4\eta^{\mu\rho_{1}}\eta^{\nu\rho_{2}}\delta^{a}_{c}
	-2\eta^{a\rho_{2}}(\eta^{\mu\rho_{1}}\delta^{\nu}_{c}
	+\eta^{\nu\rho_{1}}\delta^{\mu}_{c})\right\}iT_{n}^{c;\sigma_{1}\sigma_{2}}
	\right](p,\cdots).\nonumber\\
	=&\dfrac{\alpha'^2q_{a}q_{b}h_{\mu\nu}}{16p\cdot q}
	\left [
	\left(
	J^{\mu a}J^{\nu b}-S^{\mu a}S^{\nu b}-\bar{S^{\mu a}}\bar{S^{\nu b}}
	\right)
	T_{n}(p,\cdots)
	\right ]^{\rho_{1}\rho_{2}\sigma_{1}\sigma_{2}}\nonumber\\
	&+\dfrac{\alpha'^3q_{a}q_{b}h_{\mu\nu}}{32p\cdot q}\left [
	(S^{\mu a}p)^{\rho_{1}}
	p_{c}\left(S^{\nu b}T_{n}
	\right)^{c\rho_{2}\sigma_{1}\sigma_{2}}
	+(\bar{S}^{\mu a}p)^{\sigma_{1}}
	p_{c}\left(\bar{S}^{\nu b}T_{n}
	\right)^{\rho_{1}\rho_{2}c\sigma_{2}}
	\right ](p,\cdots)\nonumber\\
	&-\dfrac{\alpha'^3q_{a}q_{b}h_{\mu\nu}}{16p\cdot q}
	\left [
	\left(S^{\nu b}p\right)^{\rho_{2}}\left(S^{\mu a}iT\right)^{\rho_{1};\sigma_{1}\sigma_{2}}
	+\left(\bar{S}^{\nu b}p\right)^{\sigma_{2}}\left(S^{\mu a}iT\right)^{\rho_{1}\rho_{2};\sigma_{1}}
	\right ](p,\cdots)
	\label{eq.subsubleading soft theorem}
	\end{align}
	This is the same result as eq.(\ref{eq.soft graviton theorem for massive}) up to the overall factor.

\end{document}